\documentclass[a4paper,11pt]{article}
\pdfoutput=1 
\usepackage{jheppub} 
\usepackage[T1]{fontenc} 
\usepackage{multirow}
\usepackage{enumitem}
\usepackage{mathtools}

\usepackage{jheppub}

{\end{list}}
\newcounter{enumct}

\renewcommand{\d}{\mathrm{d}}

\newcommand{\g}{{\mathrm g}}
\newcommand{\q}{{\mathrm q}}
\newcommand{\qbar}{\bar{\mathrm q}}
\renewcommand{\t}{{\mathrm t}}
\newcommand{\W}{{\mathrm W}}
\newcommand{\Z}{{\mathrm Z^0}}

\newcommand{\as}{\alpha_{\mathrm{s}}}

\newcommand{\pT}{p_{\perp}}
\newcommand{\pTv}{\boldsymbol{p_{\perp}}}
\newcommand{\kT}{k_{\perp}}
\newcommand{\kTv}{\boldsymbol{k_{\perp}}}
\newcommand{\kTvp}{\boldsymbol{k}^+_{\boldsymbol\perp}}
\newcommand{\kTvm}{\boldsymbol{k}^-_{\boldsymbol\perp}}
\newcommand{\yp}{\boldsymbol{y}_+}
\newcommand{\ym}{\boldsymbol{y}_-}
\newcommand{\z}{\boldsymbol{z}}
\newcommand{\pTs}{p^2_{\perp}}

\newcommand{\dP}{\d\mathcal{P}}
\newcommand{\dPh}{\d\hat{\mathcal{P}}}
\newcommand{\y}{\boldsymbol{y}}

\newcommand{\Herwig}{\textsf{Herwig}}
\newcommand{\dShower}{\textsf{dShower}}
\newcommand{\GeV}{\mathrm{GeV}}

\newcommand{\PyEight}{\textsc{Pythia 8}}
\newcommand{\OL}{\textsc{OpenLoops 2}}
\newcommand{\hxs}{\hat{\sigma}}
\newcommand{\xsDPS}{\sigma^{\mathrm{DPS}}}
\newcommand{\xsSPS}{\sigma^{\mathrm{SPS}}}
\newcommand{\xsSub}{\sigma^{\mathrm{sub}}}
\newcommand{\xsOne}{\sigma^{\mathrm{1v1,pt}}}
\newcommand{\xsTot}{\sigma^{\mathrm{tot}}}
\newcommand{\FSplPT}{F^\mathrm{spl,pt}}
\newcommand{\LQCD}{\Lambda_\mathrm{QCD}}
\newcommand{\mZ}{M_{\mathrm{Z}}}
\newcommand{\mZZ}{m_{\mathrm{ZZ}}}
\newcommand{\Ei}{\mathrm{Ei}}
\newcommand{\dSh}{\mathbf{S}_2}
\newcommand{\sSh}{\mathbf{S}_1}
\newcommand{\tdSh}{\widetilde{\mathbf{S}}_2}

\newcommand{\Eq}{Equation\,}
\newcommand{\Eqs}{Equations\,}
\newcommand{\Sec}{\mathrm{Section}}

\newcommand{\Fig}{\mathrm{Figure}}
\newcommand{\Figs}{\mathrm{Figures}}
\newcommand{\Tab}{\mathrm{Table}}

\newcommand{\widest}{\ensuremath{ccc}}
\newcommand{\newWidth}[1]{\makebox[\widthof{\widest}]{\ensuremath{#1}}}

\begin{document}
\sloppy


\title{Combining single and double parton \\ scatterings in a parton shower}

\author[a]{Baptiste Cabouat}

\author[b]{and Jonathan R.~Gaunt}

\emailAdd{baptiste.cabouat@manchester.ac.uk}
\emailAdd{jonathan.richard.gaunt@cern.ch}

\affiliation[a]{School of Physics and Astronomy, Schuster Building, Oxford Road, \\University of Manchester, Manchester M13 9PL, United Kingdom}
\affiliation[b]{CERN Theory Department, 1211 Geneva 23, Switzerland}

\abstract{Double parton scattering (DPS) processes in which there is a perturbative \mbox{``$1\to2$''} splitting in both protons overlap with loop corrections to single parton scattering (SPS). Any fundamental theoretical treatment of DPS needs to address this double-counting issue. In this paper, we augment our Monte-Carlo simulation of DPS, \textsf{dShower}, to be able to generate kinematic distributions corresponding to the combination SPS+DPS without double counting. To achieve this, we formulate a fully-differential version of the subtraction scheme introduced in Diehl et al. (JHEP 06 (2017) 083). A shower is attached to the subtraction term, and this is combined with the \textsf{dShower} DPS shower along with the usual SPS shower. We perform a proof-of-concept study of this new algorithm in the context of $\mathrm{Z}^0\mathrm{Z}^0$ production. Once the subtraction term is included, we verify that the results do not depend strongly on the artificial ``DPS-SPS demarcation'' scale $\nu$. As part of the development of the new algorithm, we improve the kinematics of the $1\to2$ splitting in the DPS shower (and subtraction term), allowing the daughter partons to have a relative transverse momentum. Several reasonable choices for the transverse profile in the $1\to2$ splitting are studied. We find that many kinematic distributions are not strongly affected by the choice, although we do observe some differences in the region where the transverse momenta of both bosons are small.
}

 \preprint{\begin{flushright}
MAN/HEP/2020/007\\
CERN-TH-2020-114\\
MCnet-20-18\\
 \end{flushright}
 }
 

\arxivnumber{2008.01442}
 
\maketitle












\section{Introduction}

Double parton scattering (DPS) is where one has two separate hard parton-parton collisions in the same  proton-proton collision, producing two sets of final states that we shall denote by $A$ and $B$. In terms of the total cross section for the production of $A+B$, DPS is formally a power suppressed effect compared to the usual single parton  scattering (SPS) mechanism \cite{Politzer:1980me, Paver:1982yp, Paver:1983hi}. However, DPS populates the final-state phase space in a different way to SPS, with the result that when making more-differential measurements, DPS can play an important role, and there are various regions of phase space where DPS contributes at the same level as SPS. One generic example is the region where the transverse momenta of both $A$ and $B$ are small \cite{Diehl:2011tt, Diehl:2011yj}, and for many processes (such as double J/$\Psi$ production \cite{Aaij:2016bqq}), another is the region where $A$ and $B$ are widely separated in rapidities. For certain processes where the SPS mechanism is suppressed by small or multiple coupling constants, DPS can compete with SPS even at the level of the total cross section -- a well known example is same-sign WW production \cite{Kulesza:1999zh, Gaunt:2010pi}. The importance of DPS relative to SPS increases with collider energy (as lower momentum fractions are probed, where the population of partons is greater), such that DPS is more relevant at the Large Hadron Collider (LHC) than at any previous collider, and will be yet-more relevant at any future higher-energy proton-proton collider. DPS can also be an important effect in proton-nucleus and nucleus-nucleus collisions, with certain contributions to DPS rising more quickly with the nucleon number $A$ than SPS does \cite{Strikman:2001gz, Frankfurt:2004kn, Blok:2012jr, Strikman:2010bg, Blok:2017alw, Alvioli:2019kcy,dEnterria:2013mrp, Calucci:2013pza, Cattaruzza:2004qb, dEnterria:2012jam, Cazaroto:2013fua, dEnterria:2014lwk, Helenius:2019uge, Blok:2019fgg, PhysRevD.101.054036, Fedkevych:2019ofc} (for a review, see \cite{dEnterria:2017yhd}). Finally, DPS reveals information about the proton structure that is not accessible via any SPS process -- namely, correlations between partons. For all of these reasons, the experimental measurement of DPS contributions to various processes at the LHC, and the ability to make corresponding theoretical predictions of these contributions, is of great interest and importance.

The simplest and crudest approach to make theoretical predictions for DPS is to assume that two partons entering a DPS process from a given proton are uncorrelated to one another. This leads to the ``pocket formula'', in which the DPS cross section for $A+B$ is computed as the product of SPS cross sections for $A$ and $B$, divided by a geometrical prefactor $\sigma_{\mathrm{eff}}$. Here, the kinematics of the final state $A+B$ in DPS events is simply that obtained by overlaying SPS $A$ and $B$ events. The simulations of DPS (and more general multiple parton interactions, MPI) in general-purpose event generators such as $\Herwig$ \cite{Butterworth:1996zw, Borozan:2002fk, Bahr:2008dy, Bahr:2008spa, Bahr:2008pv, Bellm:2015jjp, Bellm:2017bvx, Bellm:2019icn, Bellm:2019zci}, \textsc{Pythia} \cite{Sjostrand:1987su, Sjostrand:2004ef, Sjostrand:2004pf, Sjostrand:2006za, Corke:2010yf, Corke:2011yy, Sjostrand:2014zea, Sjostrand:2017cdm} and \textsc{Sherpa}  \cite{Schumann:2007mg,Gleisberg:2008ta,Bothmann:2019yzt} (in particular, the \textsc{AMISIC++} model \cite{Alekhin:2005dx}) are fundamentally based on the pocket-formula picture. These Monte-Carlo simulations are key tools in experimental extractions of DPS, precisely because many such extractions rely on the different kinematic ``shapes'' of DPS $(A,B)$ and SPS $A+B$ events, and Monte-Carlo generators provide fully-differential predictions of these shapes (for both SPS and DPS). The number of kinematic distributions used to extract the DPS contribution in past analyses ranges from two in the ATLAS and CMS extractions of DPS in \mbox{W + 2 jets} \cite{Aad:2013bjm, Chatrchyan:2013xxa}, three or four in the ATLAS and CMS extractions in the four-jet process \cite{Chatrchyan:2013qza, Khachatryan:2016rjt, Aaboud:2016dea}, to eleven in the recent CMS extraction in same-sign WW \cite{Sirunyan:2019zox}.

The pocket-formula picture of DPS cannot be the complete one, however, and over the past few years a complete theoretical framework for the description of DPS in Quantum Chromodynamics (QCD) has been developed \cite{Gaunt:2009re, Ryskin:2011kk, Blok:2010ge, Blok:2011bu, Gaunt:2011xd, Diehl:2011yj, Manohar:2012jr, Manohar:2012pe, Diehl:2015bca, Diehl:2017kgu, Vladimirov:2017ksc, Buffing:2017mqm, Diehl:2018wfy} (see \cite{Diehl:2017wew, Blok:2017alw, Gaunt:2018eix} for reviews). One key aspect is that in QCD, the two partons entering the DPS process from a proton can have a common origin in a single parton splitting perturbatively into two (the ``$1 \to 2$ splitting'') \cite{Shelest:1982dg, Diehl:2011yj, Blok:2011bu, Gaunt:2011xd}. Treating this splitting appropriately requires a formalism in which the transverse separation between the partons $\y$ is taken into account.\footnote{Bold symbols are used for two-dimensional vectors in the plane perpendicular to the beam axis.}  Inclusion of the $1 \to 2$ splitting leads to potential double counting issues; most notably, the process in which one has a $1 \to 2$ splitting in both protons overlaps with a loop correction to SPS (see $\Fig$ \ref{Fig:yZero}). The DPS description is clearly more appropriate at large $y=|\y|$, whilst the SPS one is appropriate at smaller $y \sim 1/Q_h$, with $Q_h$ the hard scale. A QCD framework that consistently incorporates the $1 \to 2$ splittings in DPS and overcomes the double counting issues was developed by M. Diehl, JRG and K. Sch\"{o}nwald \cite{Diehl:2017kgu}, and will be referred to here as the DGS framework. The first core aspect of this framework is that the DPS cross section is written in terms of $\y$-dependent double parton density functions (dPDFs), which are integrated over $\y$ down to a cut-off $\sim 1/\nu$. The parameter $\nu$ is an unphysical scale, taken to be of order $Q_h$. The second core aspect of the framework is the inclusion of a ``subtraction term'' into the total cross section for the production of $A+B$ (in addition to the DPS and SPS terms), which cancels the dependence on $\nu$ order-by-order in the strong coupling $\as$, as well as ensuring that the total cross section smoothly interpolates between the DPS description at large $y$ and the SPS description at small $y$, as is intuitively appropriate.

Other effects also exist beyond the pocket-formula picture. The dPDFs should be ``aware'' of the constraints associated with the finite number of valence quarks in the proton (and the fact that its composition is uud) and the fact that the momentum of all partons should add up to the proton momentum. Formally this information is encoded in the number and momentum sum rules for the dPDFs \cite{Gaunt:2009re, Diehl:2018kgr, Gaunt:2012, Blok:2013bpa, Golec-Biernat:2015aza, Diehl:2020xyg}, which place non-trivial constraints on their structure. The MPI model in \PyEight\, in fact takes account of number and momentum sum-rule constraints in an approximate way, by ordering the interactions in scale and rescaling  the PDFs following each hard interaction \cite{Sjostrand:2004pf}. In addition to this, there can be non-perturbative correlations between the parton momentum fractions and $y$ in the dPDFs, and correlations in spin, colour and flavour between partons \cite{Mekhfi:1985dv, Diehl:2011yj} (for a review, see \cite{Kasemets:2017vyh}). All of these types of effects can lead to differences in the DPS rate and/or DPS shapes (for examples of these, see \cite{Gaunt:2010pi, Blok:2013bpa, Gaunt:2014rua, Echevarria:2015ufa, Ceccopieri:2017oqe, Cao:2017bcb, Cotogno:2018mfv, Cabouat:2019gtm, Cotogno:2020iio, Fedkevych:2020cmd}), where effects on the DPS shapes are particularly important with regards to the correct experimental extraction of DPS contributions.

In light of this, there is a need for an improved approach to generate event-level DPS predictions that goes beyond the pocket formula and, ideally, is based on the full QCD framework of \cite{Diehl:2017kgu}. One possible approach involves reweighting events generated by an existing Monte-Carlo generator; this approach has been used to incorporate certain $1 \to 2$ splitting effects \cite{Blok:2015afa, Blok:2015rka} and the effect of quark spin correlations \cite{Cotogno:2018mfv, Cotogno:2020iio}. In our work, we have chosen to take a different approach, building a whole new Monte-Carlo simulation of DPS from the ground up based on the DGS framework, which we believe to be advantageous in terms of flexibility, ease of use, and future development. We refer to this algorithm as \dShower. In a previous work \cite{Cabouat:2019gtm} we developed a parton-shower description of the DPS term, with proper account of the $\y$ dependence and $1\to 2$ splitting effects, and a cut-off on the $\y$ integral $\sim 1/\nu \sim 1/Q_h$. That is, we recast the first core aspect of the DGS framework into a parton-shower description. The goal of the present work is to do the same for the second core aspect of the DGS framework, and develop a parton shower that can generate both DPS and SPS events without double counting. This requires a formulation of the DGS subtraction scheme at the fully-differential level, with an appropriate parton shower for all terms. In order to achieve this goal, we adapt techniques used in the matching of fixed next-to-leading-order (NLO) computations to the parton shower \cite{Bengtsson:1986hr, Seymour:1994we, Seymour:1994df, Miu:1998ju, Lonnblad:1995ex, Frixione:2007vw, Nason:2004rx, Catani:2001cc, Lonnblad:2001iq, Mrenna:2003if, Frixione:2002ik, Frixione:2010wd, Frederix:2012ps, Frederix:2020trv}. Also in that context, there is a potential double counting issue (for example, between the real-emission process in the NLO fixed-order process and the first emission in the shower), and a subtraction scheme is needed to remove this double counting.

The paper is organised as follows. In $\Sec$ \ref{Sec:dShowerrecap} we present a brief review of the DGS framework, along with an overview of the key features of the DPS shower that we developed in \cite{Cabouat:2019gtm}. $\Sec$ \ref{Sec:impSub} describes in detail our implementation of the DGS subtraction scheme at the differential level in the parton shower. As part of this procedure, we alter one aspect of the DPS shower from its formulation in \cite{Cabouat:2019gtm}; whereas previously the $1 \to 2$ splitting occurred with the two daughter partons having no transverse momentum relative to the parent, we now add the possibility for the daughter partons to have a relative transverse momentum $\kTv \sim 1/y$ drawn from a distribution $g(\kTv,y)$. This is beneficial in terms of being able to construct a subtraction term that cancels both the DPS at small $y$ and the SPS at large $y$ at the differential level, and yields a more realistic DPS description at large $y$. We construct the algorithm in the context of on-shell vector-boson pair production ($\Z\Z$, $\W^+\W^-$), where the SPS $\g\g \to \Z\Z/\W^+\W^-$ loop corrections overlapping with DPS are known \cite{Glover:1988rg, Dicus:1987dj, Binoth:2005ua, Binoth:2006mf} (in fact, up to the next-to-leading order \cite{Caola:2015psa, Caola:2015rqy}). Extension of this procedure to more complex processes is in principle straightforward. 

In $\Sec$ \ref{Sec:numerics} we present numerical results from the algorithm in the context of on-shell $\Z\Z$ production. Our purpose here is not to perform a full phenomenological study of $\Z\Z$ production, but rather to study the behaviour and performance of the algorithm. Thus, in this proof-of-concept study we include only the $\mathcal{O}(\as^2)$ $\g\g \to \Z\Z$ loop-induced process in the SPS piece, and divide this contribution by $10$ -- this is to boost the relative importance of DPS and reduce the Monte-Carlo statistics needed to obtain distinguishable DPS effects. We perform the important validation check that the subtraction term cancels the $\nu$ dependence of the DPS term, and investigate the effect of various sensible choices for the profile $g(\kTv,y)$ in the DPS term (with corresponding choices in the subtraction term). We also show that in several distributions we see a difference in the SPS+DPS results compared to the SPS results alone, in the context of this toy study.

Finally, in $\Sec$ \ref{Sec:summary}, we conclude and discuss potential future directions.

\section{Review of the $\dShower$ algorithm} \label{Sec:dShowerrecap} 

In this section, a review of the algorithm proposed in \cite{Cabouat:2019gtm} is given. This algorithm is based on the QCD framework developed by M. Diehl, JRG and K. Sch\"{o}nwald \cite{Diehl:2017kgu} (DGS framework) whose main features are gathered in the following. This section also  introduces the subtraction scheme presented in \cite{Diehl:2017kgu} that addresses the double-counting issue mentioned in the introduction.

\subsection{The DGS framework}
\label{Sec:DGS}

In a proton-proton collision happening at a centre-of-mass energy of $\sqrt{s}$, the total cross section for the production of a final state $A+B$ via a process involving two separate hard interactions \mbox{$ij\to A$} and \mbox{$kl\to B$} is given by the factorisation formula\footnote{This formula is derived under the so-called ``collinear factorisation'' approach. The partons are considered to be parallel to the incoming proton beams in the expressions of the partonic cross sections. In the PDFs, the transverse momenta of the incoming partons are integrated over.} \cite{Gaunt:2009re, Diehl:2011yj, Diehl:2015bca, Diehl:2017kgu, Vladimirov:2017ksc, Buffing:2017mqm, Diehl:2018wfy}

\begin{equation}
\begin{split}
\xsDPS_{(A,B)}(s)=\frac{1}{1+\delta_{AB}}\sum_{i,j,k,l}&\int\d x_1\,\d x_2\,\d x_3\,\d x_4\,\hxs_{ij\to A}(x_1x_2s,\mu^2)\,\hxs_{kl\to B}(x_3x_4s,\mu^2) \\
&\times\int \d^2\y\,\Phi^2(y\nu)\,F_{ik}(x_1,x_3,\y,\mu^2)\,F_{jl}(x_2,x_4,\y,\mu^2).
\end{split}
\label{eq:XSDPS}
\end{equation}

\begin{figure}[t!] 
\centering
\includegraphics[width=0.5\textwidth]{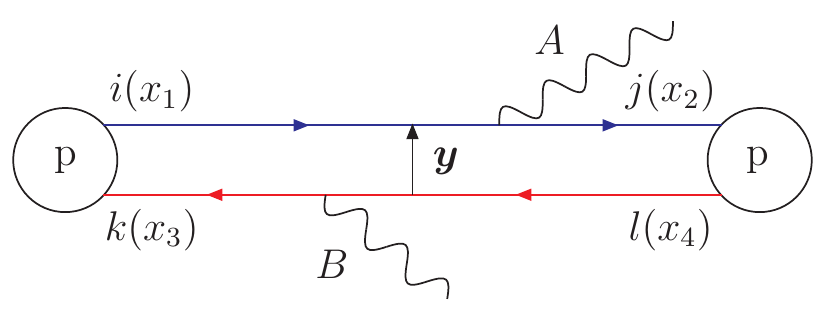}
\caption{Sketch of a DPS at a pp collider leading to the production of the final state $A+B$. The transverse distance $\y$ between the partons is represented.}
\label{Fig:XSDPS}
\end{figure}

\noindent See $\Fig$ \ref{Fig:XSDPS} for an illustration of a DPS process. Here, $\hxs_{ij\to A}$ and $\hxs_{kl\to B}$ are the parton-level cross sections for the subprocesses $ij \to A$ and $kl \to B$. The symmetry factor in front of the sum is equal to one half if $A = B$ and to unity otherwise. The functions $F_{ij}(x_1,x_2,\y,\mu^2)$ are the $\y$-dependent dPDFs; note that in this work we will only consider the case in which the two hard scatters are at equal scales, such that there is only one scale $\mu^2$ in the dPDFs. A dPDF is proportional to the joint probability (or, more specifically, the number density) of finding two partons of flavours $i$ and $j$ within the same proton with longitudinal momentum fractions $x_1$ and $x_2$ when those partons participate in two different hard interactions characterised by the same scale $Q_h$ \cite{Diehl:2011yj}. The evolution of the dPDFs with respect to the factorisation scale $\mu$ is described by the homogeneous double DGLAP equations \cite{Diehl:2011yj, Diehl:2017kgu}. It is customary to choose $\mu\sim Q_h$. The impact parameter $\y$ gives the relative distance between the two partons.

For small values of $y$, the dominant behaviour of the dPDFs can be expressed in terms of the single PDFs (sPDFs) and a perturbative $1\to2$ splitting kernel. At leading order (LO) in the strong coupling $\as$, this perturbative splitting expression reads \cite{Diehl:2011yj}

\begin{equation}
\FSplPT_{ij}(x_1,x_2,\y,\mu^2)=\frac{1}{\pi y^2}\frac{f_k(x_1+x_2,\mu^2)}{x_1+x_2}\,\frac{\as(\mu^2)}{2\pi}\,P_{k\to i+j}\left(\frac{x_1}{x_1+x_2}\right).
\label{eq:SplPT}
\end{equation}

\noindent This expression includes the effects of the $1\to2$ splitting mechanism presented in the introduction. More precisely, it takes into account the fact that the pair of partons $ij$ can originate from the perturbative splitting of a parton $k$ with longitudinal momentum fraction $x_1+x_2$. The flavour $k$ is uniquely determined by the flavours $i$ and $j$ for LO QCD splittings.  If there is no flavour $k$ such that the branching $k \to i+j$ is allowed, because of colour or flavour considerations, then the perturbative splitting expression for the pair $ij$ is equal to zero. This small-$y$ expression involves the unregularised splitting kernel $P_{k\to i+j}(z)$ (see e.g.~\cite{Gaunt:2009re}) and the sPDF $f_k$ of parton $k$, which gives the probability of probing such a flavour $k$ at the scale $\mu$.  

In \cite{Diehl:2017kgu}, the $\y$-dependent dPDFs are modelled as the sum of an intrinsic part and a splitting part. The evolutions of both components as a function of $\mu$ are given by the (homogeneous) double DGLAP equations. For the intrinsic part, the initial condition for the evolution is a product of sPDFs multiplied by a phase-space factor and a Gaussian in~$y$. The starting scale for the evolution is chosen to be $\mu_0\simeq\LQCD$, where $\LQCD\sim 1\,\GeV$ is the typical non-perturbative scale of QCD. In contrast, the input for the evolution of the splitting part of the dPDFs, is the perturbative splitting expression given in \Eq (\ref{eq:SplPT}) (multiplied by a Gaussian factor that suppresses this expression for $y \gtrsim 1/\LQCD$). The input is then evolved starting from the scale $\mu_y=b_0/y^*$ with $y^*=y/\sqrt{1+y^2/y_\mathrm{max}^2}$, \mbox{$y_\mathrm{max}=0.5\,\GeV^{-1}$}, $b_0=2 e^{-\gamma_E}\simeq1.12$ and $\gamma_E$, the Euler-Mascheroni constant \cite{Diehl:2017kgu}. The scale $\mu_y$ is not simply $1/y$ to avoid the sPDF and the strong coupling present in \Eq (\ref{eq:SplPT}) being evaluated at a scale which is below $\LQCD$ when $y\to+\infty$. Instead, $\mu_y\to b_0/y_\mathrm{max}\simeq2.24\,\GeV$, which is still in the perturbative regime. This construction for the dPDFs ensures that the dominant behaviour of the dPDFs at small $y$ is given by the perturbative splitting expression written in \Eq (\ref{eq:SplPT}), as required.

\begin{figure}[t!] 
\centering
\includegraphics[width=0.45\textwidth]{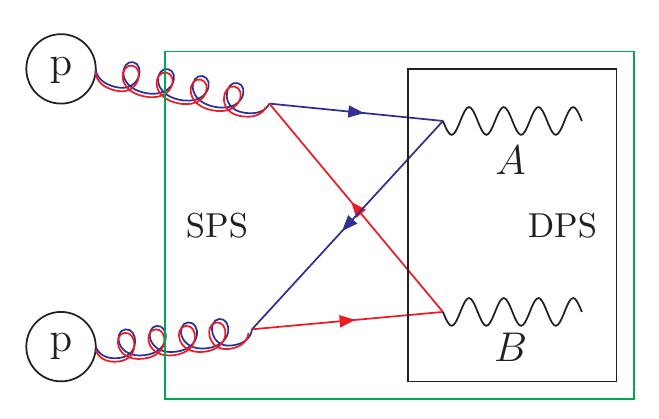}
\caption{Example of a process which can be seen either as a DPS or as an SPS. If the hard process is defined by the black box, then it is a DPS with the two subprocesses $\q\qbar\to A$ and $\q\qbar\to B$. In the case where the hard process is defined by the green box, then one has the SPS $\g\g\to A+B$. The pieces which are not included within the boxes are integrated out inside the PDFs.}
\label{Fig:yZero}
\end{figure}

The function $\Phi(y\nu)$ in \Eq \eqref{eq:XSDPS} is a cut-off at small $y$ values. It regulates the divergence of the DPS cross section which appears when $y\to 0$ (recall the $1/y^2$ behaviour in \Eq (\ref{eq:SplPT})). This power divergence is related to a double-counting issue between SPS and DPS, which is inherent to the $1\to2$ splitting mechanism. More specifically, a DPS process where $1\to2$ splittings occur in both protons (commonly referred to in the literature as a ``1v1'' DPS process) can also be considered as a loop correction to the SPS process. The latter description is actually the more appropriate one at small $y$ where the entire loop process is contained in a small space-time volume. An illustration of this double-counting issue is given in $\Fig$ \ref{Fig:yZero}. In the following, the Heaviside function $\Theta(y\nu-b_0)$ will be used as a cut-off, as was also done in the numerical studies of \cite{Diehl:2017kgu}.

Introducing the cut-off $\Phi(y\nu)$ simply regulates the DPS cross section: it does not solve the double-counting issue. There is double counting between the SPS and DPS contributions for all $y>b_0/\nu$, where the DPS (SPS) term gives a poor description for small (large) $y$ values. The simple sum of SPS and DPS terms has a strong dependence on the unphysical parameter $\nu$. These two related problems are cured by defining the total cross section for the production of a final state $A+B$ as \cite{Diehl:2017kgu}

\begin{equation}
\label{eq:sub}
\xsTot_{A+B}=\xsSPS_{A+B}+\xsDPS_{(A,B)}-\xsSub_{(A,B)},
\end{equation}

\noindent where $\xsSPS_{A+B}$ is the usual total cross section for the production of the final-state $A+B$ via SPS given by the factorisation formula \cite{Bodwin:1984hc, Collins:1985ue, Collins:1988ig, Collins:1989gx} as

\begin{equation}
\xsSPS_{A+B}(s)=\sum_{i,j}\int \d x_1\,\d x_2 \;f_i(x_1,\mu^2)\,f_j(x_2,\mu^2)\,\hxs_{ij\to A+B}(x_1x_2s,\mu^2).
\label{eq:XSSPS}
\end{equation}

\noindent The subtraction term $\xsSub_{(A,B)}$ is the integral over $\y$ of a quantity $\d\xsSub_{(A,B)}/\d^2\y$ that is defined to satisfy $\d\xsSub_{(A,B)}/\d^2\y\simeq\d\xsDPS_{(A,B)}/\d^2\y$ for $y\sim 1/\nu$ and $\d\xsSub_{(A,B)}/\d^2\y\simeq\d\xsSPS_{A+B}/\d^2\y$ for $y\gg 1/\nu$. When the two partons are well separated, the subtraction and SPS terms cancel and one is left with the DPS description which is valid in this region of the phase space. At small $y$, the subtraction and DPS terms cancel and leave the SPS term, which is the appropriate description in this region. Such a scheme removes the double counting and ensures a smooth transition between the SPS and DPS regimes. To achieve this objective in practice, the following form for the subtraction term $\xsSub_{(A,B)}$ is taken\footnote{Note that in \cite{Diehl:2017kgu}, the subtraction term in fact comprises two terms: $\xsOne_{(A,B)}$, which removes double counting between DPS and SPS, and $\sigma_{(A,B)}^{\mathrm{2v1},\mathrm{pt}}$, which removes double counting between DPS and the so-called ``twist 2 $\times$ twist 4'' mechanism. The twist 2 $\times$ twist 4 mechanism and $\sigma_{(A,B)}^{\mathrm{2v1},\mathrm{pt}}$ do not contribute at the leading logarithmic level when we take $\nu \sim Q_h$ (as we do here), and we do not consider them in what follows.}


\begin{equation}
\begin{split}
\xsSub_{(A,B)}(s) = \xsOne_{(A,B)}(s)\equiv&\frac{1}{1+\delta_{AB}}\sum_{i,j,k,l}\int\d x_1\,\d x_2\,\d x_3\,\d x_4\,\hxs_{ij\to A}(x_1x_2s,\mu^2)\,\hxs_{kl\to B}(x_3x_4s,\mu^2) \\
&\times\int \d^2\y\,\Phi^2(y\nu)\,\FSplPT_{ik}(x_1,x_3,\y,\mu^2)\,\FSplPT_{jl}(x_2,x_4,\y,\mu^2).
\end{split}
\label{eq:XS1v1}
\end{equation}

\noindent This term is nothing else but the DPS cross section given by \Eq (\ref{eq:XSDPS}), but with the full dPDFs replaced by their small-$y$ perturbative expressions written in \Eq (\ref{eq:SplPT}). 

Let us briefly sketch how this term satisfies the requirements. At small $y \sim 1/\nu\sim1/Q_h$, the DPS cross section is dominated by the 1v1 term, and there is little room for evolution between $\mu_y$ and $Q_h$, such that \mbox{$\d\xsOne_{(A,B)}/\d^2\y\simeq\d\xsDPS_{(A,B)}/\d^2\y$} and we recover the SPS term in this limit. SPS loop contributions are typically written as an integral over momenta rather than positions, but it is known that at large $y$ the dominant contribution to the SPS loop term has the form of \Eq \eqref{eq:XS1v1} \cite{Gaunt:2011xd, Diehl:2011yj}, such that \mbox{$\d\xsOne_{(A,B)}/\d^2\y\simeq\d\xsSPS_{A+B}/\d^2\y$} and we recover the DPS term.
We will only consider the unpolarised colour-singlet term in the DPS and subtraction cross sections here, for simplicity and because this is typically the dominant contribution to DPS. In this case, at large $y$, we only replace the unpolarised colour-singlet piece of the SPS loop by the DPS description, and all spin/colour/flavour interference/correlation contributions remain described by the SPS term.

Since the DPS and subtraction terms coincide in the vicinity of the cut-off $y=b_0/\nu$, up to higher order terms in $\as$, the leading dependence of the two terms on $\nu$ is the same, and cancels out. Using the change of variables $u=y\nu$, one can show that this leading behaviour is $\propto\nu^2$:
\begin{equation}
\label{eq:nuDep}
\int_{0}^{+\infty}\frac{\d^2\y}{y^4}\,\Phi^2(y\nu)=2\pi\nu^2\int_{0}^{+\infty}\frac{\d u}{u^3}\,\Phi^2(u).
\end{equation}

 In later sections, the implementation of this subtraction scheme within a parton-shower algorithm as well as a numerical example of this implementation will be presented. A key aspect of this implementation will be the cancellation of the $\nu$ dependence of the DPS and subtraction terms, as in \Eq (\ref{eq:sub}), albeit now at the differential level.

\subsection{The $\dShower$ algorithm}
\label{Sec:dSh}

The aim of the algorithm proposed in \cite{Cabouat:2019gtm} is to simulate exclusive parton-level DPS events. The starting point is to select two hard scatters with their respective kinematics according to the DPS cross section introduced in \Eq (\ref{eq:XSDPS}). A value for $\y$ is also sampled according to the cross section. After that, the two hard scatters are evolved simultaneously using a variant of the usual parton-shower algorithms. In particular, the evolution of the initial-state partons which are initiating the two hard scatters is guided by the $\y$-dependent dPDFs presented in the previous section. More precisely, consider a pair of partons of flavours $i$ and $j$ belonging to the same proton with momentum fraction $x_1$ and $x_2$ and participating in two different hard interactions characterised by the same hard scale $Q_h$. The probability that this pair remains resolved during a backward evolution from the scale $Q_h^2$ down to a lower scale $Q^2$ and then appears as coming either from the pair $i'j$ or the pair $ij'$ is  \cite{Cabouat:2019gtm}

\begin{equation}
\dP_{ij}=\dPh_{ij}\exp\left(-\int_{Q^2}^{Q_h^2}\dPh_{ij}\right),
\label{eq:dPijSud}
\end{equation}

\noindent with

\begin{equation}
\label{eq:dPij}
\begin{split}
\dPh_{ij}=\frac{\d Q^2}{Q^2}&\left(\sum_{i'}\int_{x_1}^{1-x_2}\frac{\d x_1'}{x_1'}\,\frac{\as}{2\pi}\,P_{i'\to i}\left(\frac{x_1}{x_1'}\right)\,\frac{F_{i'j}(x_1',x_2,\y,Q^2)}{F_{ij}(x_1,x_2,\y,Q^2)} \right . \\
& \left . + \sum_{j'}\int_{x_2}^{1-x_1}\frac{\d x_2'}{x_2'}\,\frac{\as}{2\pi}\,P_{j'\to j}\left(\frac{x_2}{x_2'}\right)\,\frac{F_{ij'}(x_1,x_2',\y,Q^2)}{F_{ij}(x_1,x_2,\y,Q^2)} \right).
\end{split}
\end{equation}

\noindent By iterating \Eq (\ref{eq:dPijSud}), QCD emissions are attached to the incoming partons and their effects are consistently included. Once an emission has occurred at a scale $Q_\mathrm{emi} < Q_h$, the evolution is carried on, but with starting scale $Q_\mathrm{emi}$ instead of $Q_h$. The algorithm stops when the evolution scale $Q$ reaches $\LQCD$.

The algorithm described in \cite{Cabouat:2019gtm} also includes the possibility that the two incoming partons inside the same proton may be resolved into a single parton. This phenomenon, referred to as ``merging'', aims to give a geometrical picture of the backward evolution of the system that is consistent with the $1\to2$ splitting mechanism. The merging procedure proceeds as follows. At the scale $Q=\mu_y\simeq1/y$, the backward evolution gets frozen and the merging happens with a probability given by

\begin{equation}
p_\mathrm{Mrg}=\frac{F_{ij}^\mathrm{spl}(x_1,x_2,\y,\mu_y^2)}{F_{ij}(x_1,x_2,\y,\mu_y^2)},
\end{equation}

\noindent where $F_{ij}^\mathrm{spl}$ is the splitting part of the full dPDF $F_{ij}$, which is obtained as explained in the previous section. In the case where the merging does not happen, then the evolution of the pair $ij$ is carried on as before, but with the term corresponding to the $1\to2$ splitting mechanism removed from the expression of the full dPDF (i.e. the splitting part is omitted and only the intrinsic one remains). In the case where the merging happens, the two partons $i$ and $j$ are merged into a single parton $k$ with momentum fraction $x_1+x_2$. The evolution of this single parton $k$ from the scale $\mu_y$ down to the non-perturbative scale $\LQCD$ is carried on using the conventional one-parton branching algorithm.  For the whole procedure to work, one needs to have $Q_h \geqslant 1/y$.  With our choice for the cut-off $\Phi(y\nu)$, this can be ensured at the cost of requiring that $\nu \leqslant Q_h$. This is one of the limitations of the algorithm. In order to be able to include the case $\nu>Q_h$, one would need to combine forward and backward evolutions, which is beyond the scope of this work.

In the procedure introduced in \cite{Cabouat:2019gtm}, the merging of the two partons $i$ and $j$ happens at zero transverse momentum. More precisely, the four-momenta $p_i$ and $p_j$ of partons $i$ and $j$ after the merging occurred are aligned with the beam axis in the laboratory frame. It will be seen in a later section how one can modify the kinematics such that $p_i$ and $p_j$ get a non-vanishing transverse momentum during the merging procedure.

\section{Implementation of the subtraction scheme}
\label{Sec:impSub}

As mentioned previously, there is a potential double counting issue between DPS processes in which there is a $1\to2$ splitting in both protons (referred to as 1v1 events), and loop corrections to SPS. The subtraction scheme introduced by the DGS framework removes the double counting in the physical quantity -- the cross section for the production of $A+B$ via both DPS and SPS -- via the master formula, \Eq (\ref{eq:sub}). This equation is written at the inclusive level. However, we require a subtraction scheme that can be implemented in a parton-shower framework where the DPS part is generated using the $\dShower$ algorithm, such that we can simulate full events for the combination of SPS and DPS without double counting. This subtraction scheme must be formulated at the fully-differential level, and its construction will be detailed below.

We note that more-differential formulations of the DGS framework do exist -- in particular a formulation differential in the transverse momenta of the two produced systems $A$ and $B$ was obtained in \cite{Buffing:2017mqm}. The framework constructed in that paper can be used to resum logarithms of the transverse momenta $\pT$ over the hard scale $Q_h$ to, in principle, arbitrary accuracy. However, in this formulation, the DPS and subtraction terms have different $\y$ values in amplitude and conjugate (termed $\yp$ and $\ym$), and there are further terms associated with interference between DPS and SPS. These features are necessary in the full all-order framework with transverse-momentum dependence. However, such features do not appear to be amenable to a probabilistic parton-shower treatment (and some kind of amplitude-level parton branching framework \cite{Nagy:2014mqa, Nagy:2017ggp, Martinez:2018ffw, Forshaw:2019ver} would presumably be needed). In this work we take a simpler approach, neglecting DPS/SPS interference, having only a single value of $\y$ in the DPS and subtraction terms, and making the most ``physically reasonable'' choices of transverse-momentum profiles $g(\kTv,y)$ in the $1 \to 2 $ splitting (to be discussed shortly). Our treatment should be sufficient to achieve (at least) leading logarithmic accuracy for a broad set of observables, and represents the best we can achieve in the context of a probabilistic approach.

In $\Sec$ \ref{Sec:alg}, the subtraction term at the differential level will be constructed by combining the cross section $\xsSub_{(A,B)}=\xsOne_{(A,B)}$ with a shower algorithm. As suggested by \Eq \eqref{eq:XS1v1} itself, the subtraction term is ``SPS-like'' in terms of the shower (there is only one parton in each proton) so the shower algorithm will be the usual one-parton branching one. The kinematics of the subtraction term, which results from this combination, should match the SPS one for large $y$ whereas it should coincide with the DPS one for small $y$. In order to best satisfy both requirements, and following the spirit of the DGS subtraction approach, we decide to assign to the subtraction term the same kinematics as the one generated by the $\dShower$ algorithm for a 1v1 event where no QCD emissions occurred before the merging phase, which is forced to happen at a scale $\sim Q_h$. Such DPS events are referred to as ``1v1,pt'' events in the following. 

The cancellation between the subtraction term and DPS at small $y$ occurs essentially by definition. In the implementation of the $\dShower$ algorithm, the DPS events corresponding to small $y\sim1/\nu\sim1/Q_h$ are 1v1 events, where $1\to2$ splittings occur in both protons. These splittings occur very close in scale to $Q_h$ such that there is little room for emissions above the scale $\mu_y\sim\nu\sim Q_h$ of the $1\to2$ splittings. At small $y$ and for $\nu\sim Q_h$, 1v1,pt events are indistinguishable from 1v1 events (up to small corrections), and thus the subtraction term matches the DPS one.

\begin{figure}[t!] 
\centering
\includegraphics[width=0.3\textwidth]{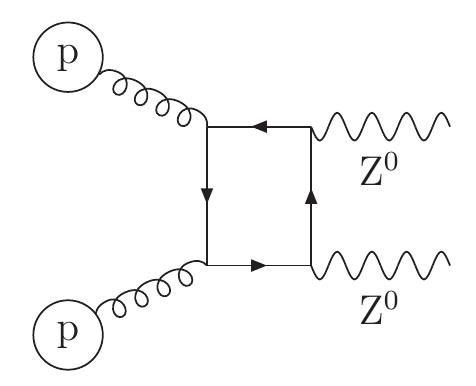}
\hspace{25pt}
\includegraphics[width=0.3\textwidth]{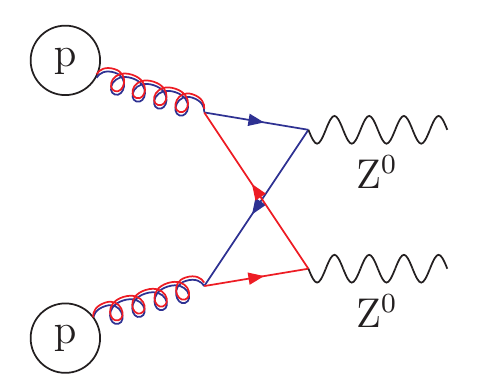} \\
(a) \hspace{175pt} (b)
\caption{Examples of graphs contributing to the loop-induced $\g\g\to\Z\Z$ process. The graph in (b) has the same topology as a 1v1,pt event.}
\label{Fig:ZZprod}
\end{figure}

At large $y$ values, the kinematics of the subtraction term needs to be equivalent to the SPS kinematics (to be more specific, the unpolarised colour-singlet contribution to SPS). In the following, $\Z\Z$ production is used as an illustration. Here, for the SPS process, we will consider only the $\mathcal{O}(\as^2)$ loop-induced process initiated by a pair of gluons, see $\Fig$ \ref{Fig:ZZprod}, since this is the contribution that overlaps with DPS (i.e. has a large-$y$ tail). It is also gauge invariant and well-defined on its own. The topology of the only graph in the loop-induced contribution that has a large-$y$ tail is the one in $\Fig$ \ref{Fig:ZZprod}b, such that the topologies of SPS and 1v1,pt events match. The choice to start the shower with a forced double merging at a scale $\sim Q_h$ for all $y$ in 1v1,pt events ensures that the shower starting scales match between the SPS and 1v1,pt (and thus subtraction) terms at large $y$. On the other hand, with the current version of the $\dShower$ algorithm, a reasonable kinematic match between the subtraction and SPS terms at large $y$ cannot be achieved. The kinematics of the loop-induced process leads at LO to bosons that have a non-vanishing transverse momentum with respect to the beam axis, even without the shower. In contrast, the equivalent topology obtained with a DPS 1v1,pt event gives bosons which are produced along the beam axis at LO, since partons are merged with zero relative transverse momenta. In $\Sec$ \ref{Sec:Mrg}, an improved merging kinematics for the DPS (and subtraction) term will be proposed such that it follows more closely the SPS kinematics at large $y$. This will yield an improved description at large $y$ overall -- the cancellation between SPS and the subtraction term will be more complete, and the mergings in the remaining DPS term, which are then dressed by QCD emissions with $\dShower$, will have more realistic kinematics.

\subsection{Merging with non-vanishing transverse momentum}
\label{Sec:Mrg}

Before presenting the new kinematics which includes a non-vanishing transverse momentum, the old kinematics developed in \cite{Cabouat:2019gtm} is reviewed in detail.

\subsubsection{The old procedure}

Consider a pair of hard scatters that was evolved from a hard scale $Q_h$ down to the scale $Q=\mu_y$ with the double-parton branching algorithm presented earlier. At this resolution scale, the two incoming partons $i$ and $j$ inside the proton moving along the $+z$ axis in the laboratory frame have momenta $\tilde{p}_{i,j}=\xi_{i,j}(\sqrt{s}/2)(1;0,0,1)$, where the momentum fractions $\xi_{i,j}$ will be referred to as the ``pre-kick'' momentum fractions in the following. The merging happens with a probability equal to $F_{ij}^\mathrm{spl}(\xi_i,\xi_j,\y,\mu_y^2)/F_{ij}(\xi_i,\xi_j,\y,\mu_y^2)$. Before implementing the merging, one needs to apply longitudinal boosts to these partons (and their daughters) in order to recover overall momentum conservation. Indeed, some parton emissions might have been added to the two hard scatters during their common evolution from $Q_h$ down to~$\mu_y$. Adding these emissions breaks momentum conservation since some partons turn into virtual particles. In particular, the partons which are initiating the hard scatters are now space-like and have acquired a transverse momentum by recoiling against the emissions, whereas they used to be light-like and moving along the beam axis. The longitudinal boosts are determined by requiring the invariant mass and the rapidity of each hard scatter to remain as they were before the shower algorithm \cite{Bahr:2008pv, Cabouat:2019gtm}. The longitudinal boosts have the following form

\begin{equation}
\Lambda(\lambda)=\left(\begin{array}{cccc}\mathrm{ch}(\lambda) & 0 & 0 & \mathrm{sh}(\lambda) \\ 0 & \newWidth{1} & 0 & 0 \\ 0 & 0 & \newWidth{1} & 0 \\ \mathrm{sh}(\lambda) & 0 & 0 & \mathrm{ch}(\lambda) \end{array}\right),
\end{equation}

\noindent with 

\begin{equation}
\mathrm{ch}(\lambda) = \frac{\lambda^2+1}{2\lambda}, 
\hspace{80pt}
\mathrm{sh}(\lambda) = \frac{\lambda^2-1}{2\lambda}.
\end{equation}

\noindent The parameter $\lambda$ is the exponential of the rapidity associated to the longitudinal boost. Therefore, a boost with $\lambda\simeq 1$ does not change the initial momenta too much. In practice, if the parton emissions that were added are hard, then $\lambda$ may be larger than unity. After applying the boosts, the two partons $i$ and $j$ extracted from the proton have momenta \mbox{$p_{i,j}=\Lambda(\lambda_{i,j})\,\tilde{p}_{i,j}=\lambda_{i,j}\,\xi_{i,j}(\sqrt{s}/2)(1;0,0,1)$} in the laboratory frame. Since the old procedure does not add any transverse momentum to these latter momenta, the resulting parton after merging has a momentum given by $(\lambda_i\,\xi_i+\lambda_j\,\xi_j)(\sqrt{s}/2)(1;0,0,1)$. The ``post-kick'' momentum fractions $x_i=\lambda_i\,\xi_i$ and $x_j=\lambda_j\,\xi_j$ are usually different from the ``pre-kick'' ones $\xi_i$ and $\xi_j$. This ensures that the emissions prior to the merging phase do not break momentum conservation.

\subsubsection{The new procedure}

With the new procedure, the two partons $i$ and $j$ involved in the merging are now allowed to have a non-vanishing transverse momentum $\kT$. More precisely, before applying the boosts, the momenta in the laboratory frame are defined as

\begin{equation}
\tilde{p}_{i,j} =  \left(E_{i,j};\pm\kT\cos\varphi,\pm\kT\sin\varphi,{p_z}_{i,j}\right),
\end{equation}

\noindent with $\varphi$ some azimuthal angle. The energies and longitudinal components of these two momenta are related to the pre-kick momentum fractions $\xi_{i,j}$ as follows

\begin{equation}
E_{i,j}+{p_z}_{i,j}=\sqrt{s}\,\xi_{i,j}.
\end{equation}

\noindent We also define the virtualities of these momenta as

\begin{equation}
Q^2_{i,j}=-\tilde{p}^2_{i,j}\geqslant0.
\end{equation}

\noindent These relations lead to

\begin{equation}
E_{i,j}=\frac{\sqrt{s}}{2}\xi_{i,j} + \frac{\kT^2-Q^2_{i,j}}{2\sqrt{s}\,\xi_{i,j}},
\hspace{70pt}
{p_z}_{i,j}=\frac{\sqrt{s}}{2}\xi_{i,j} - \frac{\kT^2-Q^2_{i,j}}{2\sqrt{s}\,\xi_{i,j}}.
\end{equation}

One is left with three degrees of freedom: $\kT$, $Q^2_i$ and $Q^2_j$. Momentum conservation gives us one constraint. Indeed, when one sums $\tilde{p}_i$ and $\tilde{p}_j$, one would like to get a light-like momentum along the $+z$ axis. This implies
$E_i+E_j={p_z}_i+{p_z}_j$ which can be rewritten as

\begin{equation}
\kT^2=\frac{\xi_j}{\xi_i+\xi_j}\,Q^2_i+\frac{\xi_i}{\xi_i+\xi_j}\,Q^2_j.
\end{equation}

\noindent Unfortunately, this is the only constraint. Let us now apply the longitudinal boosts that restore overall momentum conservation, as in the old procedure. The two boosted momenta $p_i$ and $p_j$ should now add up to a light-like momentum along the $+z$ axis. Given that the two boosts are in general different ($\lambda_i \neq \lambda_j$), this is possible only if $E_{i,j} = {p_z}_{i,j}$. These two last constraints imply that

\begin{equation}
Q^2_i=Q^2_j=\kT^2.
\label{eq:beta0}
\end{equation}

\noindent With this prescription, the resulting parton after the merging has a light-like momentum moving along the $+z$ axis, as with the old procedure. Partons $i$ and $j$ now have a transverse momentum which will be propagated to the final states by recoil. For $\kT=0$, one recovers exactly the old kinematics. Note that a similar kinematics was proposed in \cite{Sjostrand:2004ef}.

The only remaining degree of freedom is thus $\kT$. Naively, $\kT$ should be a function of three parameters: $\xi_i$, $\xi_j$ and $\mu_y$. Intuitively, one also expects $\kT\sim \mu_y$. This is not enough to fix an expression for $\kT$ and several choices are thus possible. The choice that is made in this work will be presented shortly.

Let us now consider a 1v1,pt event i.e. there are no emissions before the double merging. With this new procedure, after the boosts, the virtual partons involved in the merging inside the proton moving along the $+z$ axis have four-momenta

\begin{equation}
p_{1,2}^+=\left(\lambda_{1,2}^+\,\xi_{1,2}^+\frac{\sqrt{s}}{2};\pm\kTvp,\lambda_{1,2}^+\,\xi_{1,2}^+\frac{\sqrt{s}}{2}\right),
\end{equation}

\noindent whereas the ones moving along the $-z$ axis\footnote{For the proton moving along the $-z$ axis, the boosts that must be applied are $\Lambda(1/\lambda_{1,2}^-)$.} have momenta

\begin{equation}
p_{1,2}^-=\left(\lambda_{1,2}^-\,\xi_{1,2}^-\frac{\sqrt{s}}{2};\pm\kTvm,-\lambda_{1,2}^-\,\xi_{1,2}^-\frac{\sqrt{s}}{2}\right),
\end{equation}

\noindent with $\kTvp$ and $\kTvm$ the transverse momenta generated during the merging procedure. In the case of $\Z\Z$ production, the pre-kick momentum fractions are given by 

\begin{equation}
\xi_{1}^{\pm}=\sqrt{\frac{\mZ^2}{s}}\,e^{\pm Y_1},
\hspace{80pt}
\xi_{2}^{\pm}=\sqrt{\frac{\mZ^2}{s}}\,e^{\pm Y_2},
\label{eq:prekick}
\end{equation}

\noindent with $\mZ$ the $\Z$ mass and $Y_{1,2}$ the rapidities of the bosons in the laboratory frame. According to momentum conservation, the $\Z$ bosons now have momenta

\begin{equation}
p_1^\mathrm{Z}=p_1^++p_1^-,
\hspace{80pt}
p_2^\mathrm{Z}=p_2^++p_2^-.
\end{equation}

\noindent Both bosons thus get a transverse momentum given by ${\boldsymbol{p}_{\boldsymbol\perp}}_{1,2}=\pm\pTv$, with $\pTv=\kTvp+\kTvm$. Therefore, the transverse momenta of the bosons produced in a 1v1,pt event are directly related to the choice of $\kT$ profile made. In such a 1v1,pt event, extra emissions may be attached to the merged system after the merging phase, thus modifying further the transverse-momentum distributions of the bosons. For the purposes of comparing 1v1,pt (i.e. subtraction) and SPS events, those additional emissions are actually not relevant because they lead to the exact same effects in both event types, and in the study in the next part of the section, we will neglect their effect. Since there are no prior emissions before the double merging, the $\lambda$ coefficients can be analytically calculated. One finds that they are all equal to $\sqrt{1+\pTs/\mZ^2}$. The post-kick momentum fractions are thus

\begin{equation}
x_1^{\pm}=\lambda_1^{\pm}\,\xi_1^{\pm}=\sqrt{\frac{\mZ^2+\pTs}{s}}\,e^{\pm Y_1},
\hspace{50pt}
x_2^{\pm}=\lambda_2^{\pm}\,\xi_2^{\pm}=\sqrt{\frac{\mZ^2+\pTs}{s}}\,e^{\pm Y_2},
\label{eq:postkick}
\end{equation}

\noindent  and depend explicitly on $\pT$. They lead to a squared invariant mass of the diboson system equal to

\begin{equation}
\label{eq:mZZ}
\mZZ^2=2(\mZ^2+\pTs)\left(1+\cosh(Y_1-Y_2)\right).
\end{equation}

\subsubsection{Choice of the transverse profile}
\label{Sec:choicepT}

Whatever choice for $\kT$ is made, the kinematics of a 1v1,pt event obtained with this choice should match as closely as possible the SPS kinematics for large $y$ values. In this work, rather than aiming for an exact match, we will adopt a simple choice for the transverse profile in the merging procedure, which should nevertheless reproduce the SPS kinematics at large $y$ reasonably well. More specifically, values for $\kTv$ will be sampled randomly according to the following distribution

\begin{equation}
g(\kTv,y)=\frac{\beta}{\pi}\,y^2\exp\left(-\beta y^2\kT^2\right),
\label{eq:gausskT}
\end{equation}

\noindent which is normalised as

\begin{equation}
\int g(\kTv,y)\,\d^2\kTv=1,
\end{equation}

\begin{figure}[t!] 
\centering
\includegraphics[width=1\textwidth]{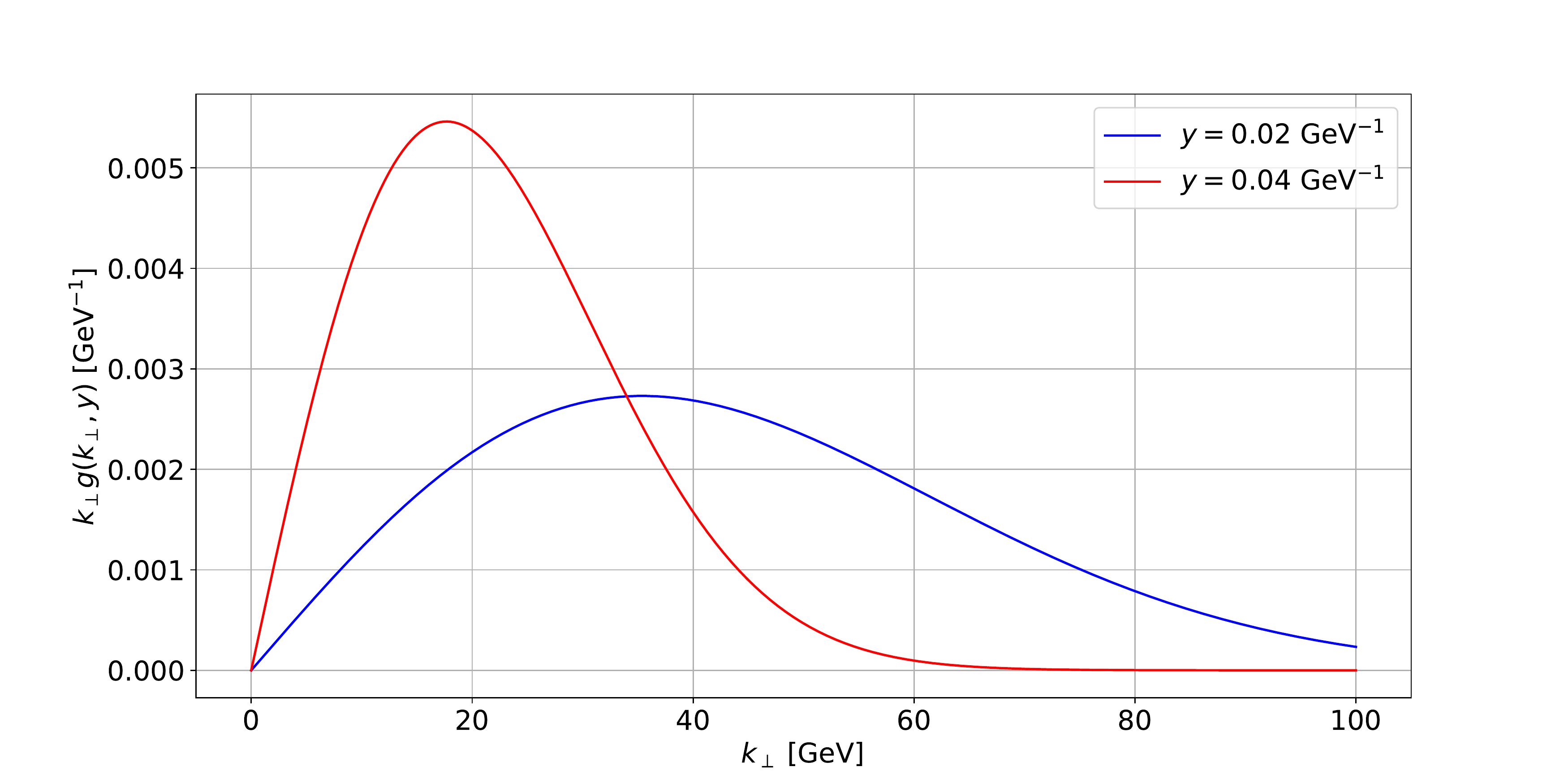}
\caption{Plot of the distribution $\kT\,g(\kTv,y)$ as a function of $\kT$ for $y=0.02\,\GeV^{-1}$ and \mbox{$y=0.04\,\GeV^{-1}$}. Here, $\beta=1$ is used.}
\label{Fig:plotGauss}
\end{figure}

\noindent with $\d^2\kTv = \kT\d\kT\d\varphi=\d\kT^2\d\varphi/2$. $\beta$ is a free parameter of the model that controls the width of the distribution. In the following, $\beta=1$ will be used but the impact of different choices for $\beta$ will be discussed in a later section. The distribution is represented for a few values of $y$ in $\Fig$ \ref{Fig:plotGauss}. One can see that the distribution peaks at $\kT=1/(\sqrt{2\beta}\,y)\simeq\mu_y$, as desired. It will now be shown how this choice leads to a reasonable match between the 1v1,pt events and the SPS events at large $y$ in the case of $\Z\Z$ production.

Loop diagrams are generally computed as integrals over internal momenta rather than positions, and no full result exists for the $\g\g\to\Z\Z$ loops differential in the transverse partonic separation $y$. However, the small-$\pT$ behaviour of the loop-induced process $\g\g\to\Z\Z$ is dominated by the contribution from the region of large $y$ values \cite{Gaunt:2011xd, Diehl:2011yj}. Therefore, if the kinematics of a 1v1,pt event and the SPS one lead to the same behaviour at small $\pT$, then one can state that the two kinematics match to a reasonable degree of accuracy in the large-$y$ region (and thus, that the kinematics of the subtraction and SPS terms also match in the large $y$ region). This can be checked by studying the $\pTv$ distribution of the produced bosons. For the 1v1,pt events, $\pTv$ is defined as the sum of the two vectors $\kTvp$  and $\kTvm$, which are selected according to \Eq (\ref{eq:gausskT}). This quantity is thus distributed according to

\begin{equation}
\label{eq:ypTprofile}
\begin{split}
h(\pTv,y)&=\int\d^2\kTvp\,\d^2\kTvm\, g(\kTvp,y)\,g(\kTvm,y)\,\delta^{(2)}(\kTvp+\kTvm -\pTv) \\
&=\frac{\beta}{2\pi}\,y^2\exp\left(-\beta\,\frac{y^2\pTs}{2}\right),
\end{split}
\end{equation}

\noindent with the following normalisation

\begin{equation}
\int h(\pTv,y)\,\d^2\pTv=1.
\end{equation}

\noindent In the SPS cross section, the $y$ parameter is integrated over. One thus needs to do the same for the 1v1,pt events in order to be able to compare. The 1v1,pt cross section differential in $\pT$ is given by \Eq \eqref{eq:XS1v1}, but with the profile $h(\pTv,y)$ inserted into the $y$ integral. Then the $\pT$ distribution of the bosons obtained for a 1v1,pt event can be estimated to be 

\begin{equation}
\label{eq:pTprofile}
\begin{split}
\int_{0}^{+\infty}\frac{\d^2 \y}{y^4}\,\Phi^2(y\nu)\,h(\pTv,y)&=\pi\int_{b_0^2/\nu^2}^{+\infty}\frac{\d y^2}{y^4}\,h(\pTv,y) \\
&=-\frac{\beta}{2}\,\Ei\left(-\beta\,\frac{b_0^2\,\pTs}{2\nu^2}\right),
\end{split}
\end{equation}

\noindent with $\Ei(x)$ the exponential integral function defined as

\begin{equation}
\Ei(x)=-\int_{-x}^{+\infty}\frac{e^{-t}}{t}\,\d t.
\end{equation}

\noindent In the limit where $\pT \ll \nu$, one gets 

\begin{equation}
\int_{b_0/\nu}^{+\infty}\frac{\d^2 \y}{y^4}\,h(\pTv,y)\sim\frac{\beta}{2}\left(-\log\left(\frac{\pTs}{\nu^2}\right)-\log\left(\frac{\beta b_0^2}{2}\right)-\gamma_E\right).
\label{eq:DPSsing}
\end{equation}

\noindent This is, at least, not too far from the $\log^2(\pTs/\nu^2)$ behaviour one obtains for the $\pT$ spectrum of the loop-induced process for small $\pT$ values \cite{Glover:1988rg, Nagy:2006xy, Gaunt:2011xd}. This behaviour leads to a divergence when $\pT\to0$ referred to as the ``DPS singularity'', since this one also originates from the double counting between SPS and DPS. This singularity is however integrable, meaning that integrating $\log^2(\pTs/\nu^2)$ down to $\pT=0$ yields a finite result. In the case of a 1v1,pt event, the $\log(\pTs/\nu^2)$ behaviour obtained in \Eq (\ref{eq:DPSsing}) leads also to an integrable singularity.

\subsection{Subtraction scheme at the differential level}
\label{Sec:alg}

The new kinematics presented in the previous section was introduced so that the 1v1,pt events and the SPS kinematics lead to similar behaviours at large $y$ values. The subtraction term will then correctly reproduce the DPS one at small $y$ and approximately the SPS one at large $y$, both at the inclusive and differential levels. The objective now is to create a shower algorithm that can simulate event shapes for the combination SPS+DPS without double counting. The procedure which will be presented in the following uses ideas from matching \cite{Bengtsson:1986hr, Seymour:1994we, Seymour:1994df, Miu:1998ju, Lonnblad:1995ex, Frixione:2007vw, Nason:2004rx, Catani:2001cc, Lonnblad:2001iq, Mrenna:2003if, Frixione:2002ik} between NLO matrix elements and parton showers. Similarly to the \textsc{MC@NLO} method \cite{Frixione:2002ik, Frixione:2010wd, Frederix:2012ps, Frederix:2020trv}, we decide to split the cross section for the production of a final state $A+B$  into two terms. More precisely, for any observable $O$, we write symbolically 

\begin{equation}
\frac{\d\xsTot_{A+B}}{\d O}=\sSh(t_{1})\otimes\left[\frac{\d\xsSPS_{A+B}}{\d O}-\frac{\d\xsSub_{(A,B)}}{\d O}\right]+\int\d^2\y\,\dSh(t_{2})\otimes\frac{\d\xsDPS_{(A,B)}}{\d O\,\d^2\y}.
\label{eq:master}
\end{equation}

\noindent This formula is the differential version of \Eq (\ref{eq:sub}). The operators $\sSh$ and $\dSh$ encapsulate the effects of the one-parton and two-parton branching algorithms respectively. In other words, $\sSh$ is the usual shower algorithm whereas $\dSh$ is the $\dShower$ algorithm (including the merging procedure) recalled in $\Sec$ \ref{Sec:dSh}. The quantities $t_1$ and $t_2$ are the starting scales of the shower algorithms. Usually, it is the type of shower algorithm that is implemented that determines which scale should be used. However, they should be related to the hard scales of the corresponding hard scatters. As explained in $\Sec$ \ref{Sec:dSh}, one must impose $t_2\geqslant\nu$. In order to achieve the best matching between DPS and subtraction terms at small $y$, one must take $t_1 = t_2$, as will be discussed later. The two operators $\sSh$ and $\dSh$ are unitary, meaning that they cannot modify the value of the total cross section $\xsTot_{A+B}$, but only the event shapes. One thus has two types of events: SPS-like events (first term of \Eq (\ref{eq:master})) and DPS-like events (second term). For an SPS-like event, there is only one hard scatter and its kinematics is sampled according to $\xsSPS_{A+B}-\xsSub_{(A,B)}$. The event is then showered using the one-parton branching algorithm. The DPS-like events start from two hard scatters whose kinematics are selected according to $\xsDPS_{(A,B)}$. The $\dShower$ algorithm $\dSh$ is then applied to this pair of hard scatters. The DPS-like events include all the contributions to DPS (1v1 contribution as well). Since $\y$ is not an observable, one needs to integrate over it in the second term of \Eq (\ref{eq:master}). $\dSh$ contains an implicit dependence on $y$ due to the way the merging procedure is implemented, recall $\Sec$ \ref{Sec:dSh}. Note that for each term in \Eq (\ref{eq:master}), both the shower and cross section parts can contribute to the total value of $O$.

Let us now explain how \Eq (\ref{eq:master}) is implemented from an algorithmic point of view. The first technical aspect is that the phase spaces for SPS-like and DPS-like events are different. More precisely, in the instance of diboson production via SPS, the kinematics of the diboson system can be parametrised by three non-trivial\footnote{The azimuthal angles are selected according to flat distributions and are omitted here.} variables $\Phi_1=\{Y_1,Y_2,\pTs\}$, with $Y_1$ and $Y_2$ the rapidities of the two bosons and $\pTs$ the transverse momentum squared of the bosons with respect to the beam axis in the laboratory frame. All the relevant kinematic quantities can be derived from these three variables, as illustrated in $\Sec$ \ref{Sec:Mrg}. In the case of two hard scatters, the same rapidities $Y_1$ and $Y_2$ can be used to characterise the kinematics of the two bosons. At LO, the two bosons are produced with zero transverse momenta so there is no need for the variable $\pTs$ in the DPS case. The bosons get a non-vanishing transverse momentum afterwards via the shower algorithm $\dSh$. The phase space for DPS can thus be encapsulated in the variable $\Phi_2=\{Y_1,Y_2,y\}$, with $y$ the impact parameter. Since $\Phi_1\neq\Phi_2$, one has to choose the event type before sampling the kinematics. This can be done with the following algorithm \cite{Frixione:2002ik, Sjostrand:2006za}

\begin{enumerate}
\item \label{enu:step1} Select a random number $R$ uniformly between 0 and 1. If one has \mbox{$R<M_1/(M_1+M_2)$} then the event is an SPS-like one, otherwise it is a DPS-like one.
\item Select a phase-space point $\Phi_i$ according to the distribution $p_i(\Phi_i)$, $i$ being equal to 1 or 2, depending on the event type previously determined. Calculate the corresponding quantity $w_i(\Phi_i)$.
\item Accept the event with a probability given by $w_i(\Phi_i)/M_i$. In the case where the event is rejected, then go back to the first step. If the event is accepted then apply the corresponding shower algorithm $\mathbf{S}_i$.
\end{enumerate}

\noindent Here, the event weight $w_i(\Phi_i)$ is defined for $i=1,2$ as

\begin{equation}
w_i(\Phi_i)=\frac{1}{p_i(\Phi_i)}\,\frac{\d\sigma_i}{\d\Phi_i},
\end{equation}

\noindent with $\sigma_1=\xsSPS_{A+B}-\xsSub_{(A,B)}$ and $\sigma_2=\xsDPS_{(A,B)}$. The functions $p_i(\Phi_i)$ are some positive-definite distributions normalised to unity which are used during the importance-sampling procedure to increase the efficiency of the Monte-Carlo method. The number $M_i$ is defined as the maximum value of the event weight $w_i(\Phi_i)$ over the whole phase space parametrised by $\Phi_i$, thus ensuring that $w_i(\Phi_i)/M_i<1$. On average, the events are generated with the correct weight $\xsTot_{A+B}$ since

\begin{equation}
\begin{split}
 \xsTot_{A+B}=&\int\d\Phi_1\,p_1(\Phi_1)\,(M_1+M_2)\left(\frac{w_1(\Phi_1)}{M_1}\,\frac{M_1}{M_1+M_2}\right) \\
 &+ \int\d\Phi_2\,p_2(\Phi_2)\,(M_1+M_2)\left(\frac{w_2(\Phi_2)}{M_2}\,\frac{M_2}{M_1+M_2}\right),
 \end{split}
\end{equation}

\noindent where the right-hand side of the equation is the sum of two terms: the first one (second one) is the product averaged over the corresponding phase space of the weight associated with an SPS-like (DPS-like) event with the probability to accept this event type. Also, on average, the relative probability to select the event type $i$ is $\sigma_i/\xsTot_{A+B}$, as desired.

The second technical aspect is linked to the fact that the implementation of \Eq (\ref{eq:master}) implies the handling of events with negative weights, as in the \textsc{MC@NLO} procedure. Indeed, for some specific values of $\Phi_1$, it may happen that $w_1(\Phi_1)<0$. The algorithm proposed above can be adapted to account for such cases by accepting the SPS-like events with a probability equal to $|w_1(\Phi_1)|/M_1$ instead of simply $w_1(\Phi_1)/M_1$. In that case, $M_1$ must be defined as the maximum of $|w_1(\Phi_1)|$. When constructing histograms, the SPS-like events with $w_1(\Phi_1)<0$ contribute with a weight $-1$ whereas the ones with $w_1(\Phi_1)>0$ and the DPS-like events are recorded with weight $+1$. Such a procedure ensures that the average weight of an SPS-like event is $\sigma_1$. Indeed, one can write

\begin{equation}
\sigma_1=\int\d\Phi_1\,p_1(\Phi_1)\left(\frac{\d\sigma_1/\d\Phi_1}{|\d\sigma_1/\d\Phi_1|}\,(M_1+M_2)\right)\left(\frac{|w_1(\Phi_1)|}{M_1}\,\frac{M_1}{M_1+M_2}\right),
\end{equation}

\noindent which is the product averaged over the phase space parametrised by $\Phi_1$ of the weight associated with an SPS-like event in the histograms with the probability to accept an SPS-like event. This is similar to what is proposed in the \textsc{MC@NLO} implementation \cite{Frixione:2002ik, Frixione:2010wd, Frederix:2012ps, Frederix:2020trv}. In order for the whole procedure to be working efficiently, the fraction of events with negative weights should not be too large, typically a few percent.

\subsection{The subtraction term}
\label{Sec:subTerm}

\subsubsection{Analytical expression}

Let us now understand how the subtraction term is coupled to the one-parton branching algorithm $\sSh$, as indicated by \Eq (\ref{eq:master}). First of all, the algorithm that implements \Eq (\ref{eq:master}) requires to be able to calculate $\d\xsSub_{(A,B)}/\d\Phi_1$. We recall that $\Phi_1$ includes the variable $\pT$, such that we need a suitable $\pT$ profile for this term. As mentioned in the beginning of this section, we choose to assign to the subtraction term the $\pT$ profile that is generated by the $\dShower$ algorithm for a 1v1,pt event (i.e. a 1v1 event with no QCD  emissions before the merging phase). This latter profile was derived earlier in $\Sec$ \ref{Sec:choicepT} for diboson production. One can thus insert the profile $h(\pTv,y)$ given by \Eq (\ref{eq:ypTprofile}) inside the subtraction term as follows

\begin{equation}
\label{eq:subZZinsert}
\begin{split}
\xsSub_{(A,B)}(s)=\,&\xsOne_{(A,B)}(s)=\frac{1}{1+\delta_{AB}}\sum_{i,j,k,l}\int\d x_1\,\d x_2\,\d x_3\,\d x_4\,\hxs_{ij\to A}(x_1x_2s,\mu^2)\,\hxs_{kl\to B}(x_3x_4s,\mu^2) \\
&\times\int \d^2\y\,\Phi^2(y\nu)\,\FSplPT_{ik}(x_1,x_3,\y,\mu^2)\,\FSplPT_{jl}(x_2,x_4,\y,\mu^2)\int\d^2\pTv\,h(\pTv,y).
\end{split}
\end{equation}

\noindent Plugging \Eq (\ref{eq:SplPT}) into this expression and using the rapidities $Y_i$ of the bosons instead of the momentum fractions $x_i$, one gets, in the case of $\Z\Z$ production

\begin{equation}
\begin{split}
    \xsSub_\mathrm{(Z,Z)}(s)=\,&\frac{\hxs_\mathrm{Z}^2(s)}{2}\,\int_{b_0^2/\nu^2}^{+\infty}\frac{\pi\,\d y^2}{(\pi\,y^2)^2}\int\d Y_1\,\d Y_2\,\frac{f_\g(X_1^+ + X_2^+,\mu^2)}{X_1^+ + X_2^+}\,\frac{f_\g(X_1^- + X_2^-,\mu^2)}{X_1^- + X_2^-}\left(\frac{\as(\mu^2)}{2\pi}\right)^2 \\
    &\times 2\sum_\q c_\q^2\,\,P_{\g\to\q}\left(\frac{X_1^+}{X_1^+ +X_2^+}\right)\,P_{\g\to\q}\left(\frac{X_1^-}{X_1^- +X_2^-}\right)\int\d^2\pTv\,h(\pTv,y),
\end{split}
\label{eq:subZZ}
\end{equation}

\noindent where $\hxs_\mathrm{Z}$ is the partonic cross section for the process $\q\qbar\to\Z$. The $c_\q$ coefficients are the couplings of the $\Z$ with the incoming quarks q and only depend on the flavour of those quarks. The sum over q includes all the quark flavours which are allowed. The factor two in front of that sum accounts for the symmetry between the branchings $\g\to\q\qbar$ and $\g\to\qbar\q$. There is some freedom in choosing which momentum fractions $X$ should be used in the splitting kernels and in the gluon sPDFs $f_\g$: one could use either the pre-kick or the post-kick fractions defined in $\Sec$ \ref{Sec:Mrg}. The scale $\mu$ should be set to the hard scale appropriate to the process, although there are several potential choices. We will come back to this question shortly. Provided the scale $\mu$ does not depend on $y$, we can straightforwardly perform the $y$ integral in \Eq (\ref{eq:subZZ}) analytically, yielding

\begin{equation}
\begin{split}
    \xsSub_\mathrm{(Z,Z)}(s)&=\,\frac{\hxs_\mathrm{Z}^2(s)}{2}\,\frac{2\pi}{\pi^2}\,\int\d Y_1\,\d Y_2\,\frac{f_\g(X_1^+ + X_2^+,\mu^2)}{X_1^+ + X_2^+}\,\frac{f_\g(X_1^- + X_2^-,\mu^2)}{X_1^- + X_2^-}\left(\frac{\as(\mu^2)}{2\pi}\right)^2 \\
    &\times 2\sum_\q c_\q^2\,\,P_{\g\to\q}\left(\frac{X_1^+}{X_1^+ +X_2^+}\right)\,P_{\g\to\q}\left(\frac{X_1^-}{X_1^- +X_2^-}\right)\int\d \pTs\left[-\frac{\beta}{4}\Ei\left(-\beta\,\frac{b_0^2\,\pTs}{2\nu^2}\right)\right].
\end{split}
\label{eq:subZZbis}
\end{equation}

\begin{figure}[t!] 
\centering
\includegraphics[width=1\textwidth]{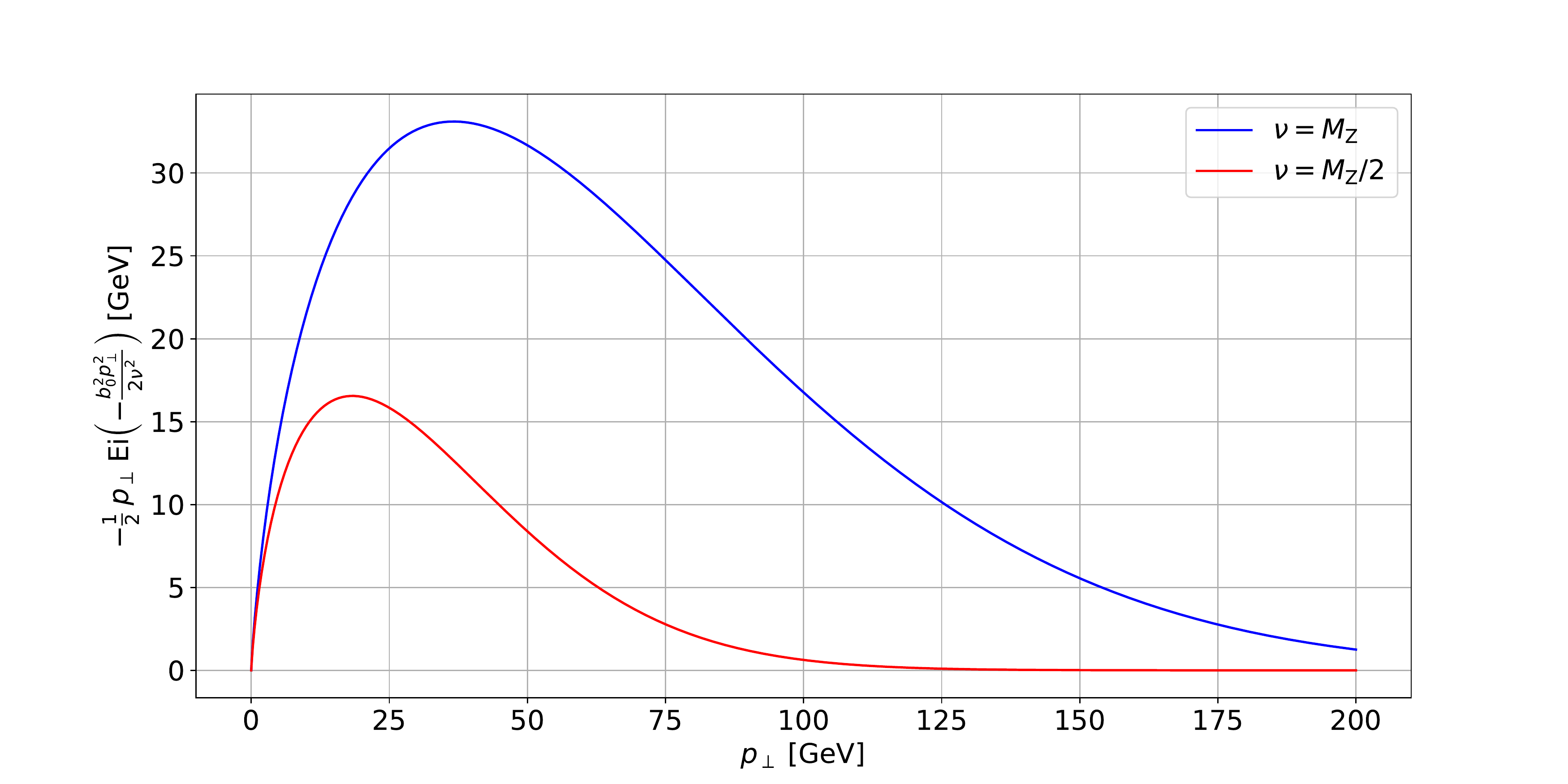}
\caption{$\pT$ profile of the subtraction term for $\nu=\mZ$ and $\nu=\mZ/2$. The area under each curve is equal to $\nu^2/(2b_0^2)$. Here, $\beta=1$ is used.}
\label{Fig:pTSpect}
\end{figure}

\noindent This last expression is what is needed for the implementation of \Eq (\ref{eq:master}). Indeed, the subtraction term is now written as an integral over $\Phi_1$. Inserting the $\pT$ profile does not change the dependence of the subtraction term on $\nu$ since

\begin{equation}
\label{eq:nuNorm}
\int_{0}^{+\infty}\d \pTs\left[-\frac{\beta}{4}\Ei\left(-\beta\,\frac{b_0^2\,\pTs}{2\nu^2}\right)\right]=\frac{\nu^2}{2b_0^2},
\end{equation}

\noindent which is the same dependence as in \Eq (\ref{eq:nuDep}). The $\pT$ profile of the subtraction term is represented in $\Fig$ \ref{Fig:pTSpect} for two values of $\nu$.

\subsubsection{Choices of scales and momentum fractions}
\label{sec:Scales}

Let us now discuss the choice of scale $\mu$ in the subtraction term, as well as the momentum fractions $X_1^\pm$ and $X_2^\pm$. We will also discuss the issues of the choice of renormalisation/factorisation scales in the SPS and DPS terms, which we shall refer to here as $\mu_{\text{SPS}}$ and $\mu_{\text{DPS}}$ respectively, and the choice of shower starting scales $t_1$ and $t_2$ in \Eq \eqref{eq:master}.

Clearly, all renormalisation/factorisation scales should be set to be of the order of the hard scale $Q_h$. But for the SPS, DPS (and subtraction) terms slightly different choices of hard scale may be optimal, even though formally the differences will be beyond the accuracy of the computation. Customary choices for $\mu_{\text{SPS}}$ in the context of $\Z\Z$ production are $\mu_{\text{SPS}}=\mZZ$ \cite{Aad:2015zqe, Khachatryan:2016txa, Aaboud:2016urj,  Sirunyan:2017zjc, Aaboud:2019lgy}, $\mu_{\text{SPS}}=\mZZ/2$ \cite{Aaboud:2018puo, Kallweit:2018nyv, Grazzini:2018owa,  Aaboud:2019lxo, Alioli:2016xab, Caola:2016trd} and $\mu_{\text{SPS}}=\mZ$ \cite{Alioli:2016xab, Chatrchyan:2012sga, Cascioli:2014yka, Grazzini:2015hta}, with $\mZZ$ the invariant mass of the diboson system given by \Eq (\ref{eq:mZZ}). By contrast, for $\Z\Z$ production via DPS one would typically choose $\mu_{\text{DPS}}=\mZ$. At large $y$, the SPS term should predominantly produce the bosons with $\pT \sim 1/y \ll \mZ$, such that at such $y$ values one can drop $\pT$ in dynamic scales like $\mZZ$ and write this as a function of $\mZ$ and the rapidities $Y_i$ alone. To achieve best matching between the subtraction and DPS at small $y$, and subtraction and SPS at large $y$, the optimal choice of $\mu$ in the subtraction term would then be a $y$-dependent choice that tends to $\mu_{\text{DPS}}$ at small $y$, and to $\mu_{\text{SPS}}(\pT=0)$ at large $y$ (this in practice could be implemented via appropriate profile scales \cite{Ligeti:2008ac,Abbate:2010xh,Diehl:2017kgu}). With this choice, one can straightforwardly follow the procedure above up to \Eq \eqref{eq:subZZ} (since the scales are independent of $\pT$), but would no longer be able to perform the $y$ integral analytically to obtain \Eq \eqref{eq:subZZbis}.

An alternative possibility is to choose $\mu$ to either be $\mu_{\text{SPS}}$ (or $\mu_{\text{SPS}}(\pT=0)$) or $\mu_{\text{DPS}}$. In this case the matching between the subtraction term and either DPS or SPS will be degraded at small $y$ or large $y$, where the degradation in matching will be, in general, more observable at small $y$ (since this is the leading-power SPS region). This would favour the choice $\mu = \mu_{\text{DPS}}$ in this case.

Now let us discuss the choice of starting scales $t_i$ for the showers. We set the shower starting scales for the SPS and subtraction terms to be equal ($=t_1$), as written in \Eq (\ref{eq:master}). The reason for this is that then we can treat these terms together as SPS-like events in the algorithm. This in turn minimises the number of events with negative weights -- given that the SPS term is usually much larger that the subtraction term, one is ensured that the combination $\d(\xsSPS_\mathrm{ZZ}-\xsSub_\mathrm{(Z,Z)})/\d\Phi_1$ is positive-definite over a large region of the phase space parametrised by $\Phi_1$. As in the \textsc{MC@NLO} method, a minimal fraction of negative-weight events is desired because, for a given accuracy, the larger the fraction is, the higher the statistics needs to be. If one separates the scales of the SPS and subtraction terms, then one has to split the SPS-like events  into pure SPS events and subtraction counter-events which contribute to the histograms with weight $-1$. This will increase the number of negative weights drastically.

It is in principle possible to choose the shower starting scale to be different from the renormalisation/factorisation scale in each term, although having such a mismatch between the cross section expression and shower is somewhat unnatural. If we want to match the shower starting scale with the renormalisation/factorisation scale, the constraint that the shower starting scales of the SPS and subtraction terms are equal implies that $\mu = \mu_{\text{SPS}}$. If $\mu_{\text{DPS}} \neq \mu_{\text{SPS}}$, this choice is incompatible with $\mu = \mu_{\text{DPS}}$. 

In the $\Z\Z$ production example we study here, we will simply set all renormalisation, factorisation and shower starting scales to $\mZ$. In such a case, where we set $\mu_{\text{DPS}} = \mu_{\text{SPS}}$, we can achieve all desired properties above simultaneously.

Now we discuss which momentum fractions $X$ should be used in the expression of the subtraction term. To achieve the best match between the DPS and subtraction terms at small $y$, the pre-kick fractions $\xi$ constitute a better choice than the post-kick fractions~$x$. Indeed, the DPS cross section uses the pre-kick fractions given by \Eq (\ref{eq:prekick}). Moreover, the post-kick fractions contain an explicit dependence on $\pTs$, see \Eq (\ref{eq:postkick}), which technically prevents us from inserting the integral over $\pTs$ in \Eq \eqref{eq:subZZinsert}. 

\subsubsection{Numerical checks}
\label{sec:subtractionNumerics}

It will now be shown how the subtraction term performs numerically. The first step is to check that the kinematics of the subtraction term is indeed equal to that of a DPS 1v1,pt event. The kinematics corresponding to a 1v1,pt event can be simulated by combining the cross section $\xsOne_\mathrm{(Z,Z)}$ defined by \Eq (\ref{eq:XS1v1}) (with $\mu = \mZ$) with the $\dShower$ algorithm~$\dSh$. By definition, the shower evolution of a 1v1,pt event starts with a forced double merging at $t_2=\mZ$, in contrast with a usual 1v1 event where the merging phase happens at the scale $\mu_y\simeq 1/y$ which is below $t_2$. To highlight this technical difference, the shower algorithm used to shower the 1v1,pt events is denoted by $\tdSh$. Since the evolution of a 1v1,pt event starts directly with the merging phase, there are no emissions before this phase, as mentioned before. Recall that at small $y \sim 1/\nu\sim1/\mZ$ the 1v1,pt DPS term coincides with the full one. The subtraction term in this comparison is simply the corresponding term in \Eq \eqref{eq:master} i.e. the cross section given by \Eq \eqref{eq:subZZbis} coupled with the shower algorithm $\sSh$, with $t_1=\mZ$. In the rest of this section, $\nu=\mZ$ is used. The effects of a variation in $\nu$ are studied in $\Sec$ \ref{Sec:val}. The only differences between the two terms are then the shower algorithm, the way the phase space is sampled (recall that $\Phi_1\neq\Phi_2$) and the choices of scales and momentum fractions.

In the following figures, the two previously described terms \mbox{$\sSh(\mZ)\otimes\d\xsSub_\mathrm{(Z,Z)}/\d O$} and \mbox{$\int\d^2\y\,\tdSh(\mZ)\otimes\d\xsOne_\mathrm{(Z,Z)}/(\d O\,\d^2\y)$} are designated by ``Sub'' and ``1v1,pt'' respectively. The results for $\sqrt{s}=13$ TeV were obtained using the 3-flavour MSTW2008 set of LO sPDFs  \cite{Martin:2009iq, Martin:2010db} and the 3-flavour scheme for $\as$ developed by the same authors \cite{Martin:2009bu}, with \mbox{$\as(\mZ)=0.126$}. Consequently, only the massless u, d and s quarks are allowed in the cross-section formulae and in the showers. We only include three flavours to avoid to have to deal with the different mass thresholds that would add further complications to the problem. The showers are angular ordered and stop when the evolution scale reaches the value of 2 GeV. No cuts are applied to the hard process $\q\qbar\to\Z\otimes\q\qbar\to\Z$. We take $\mZ = 91.188$ GeV.

\begin{figure}[t!] 
\centering
\includegraphics[width=0.45\textwidth]{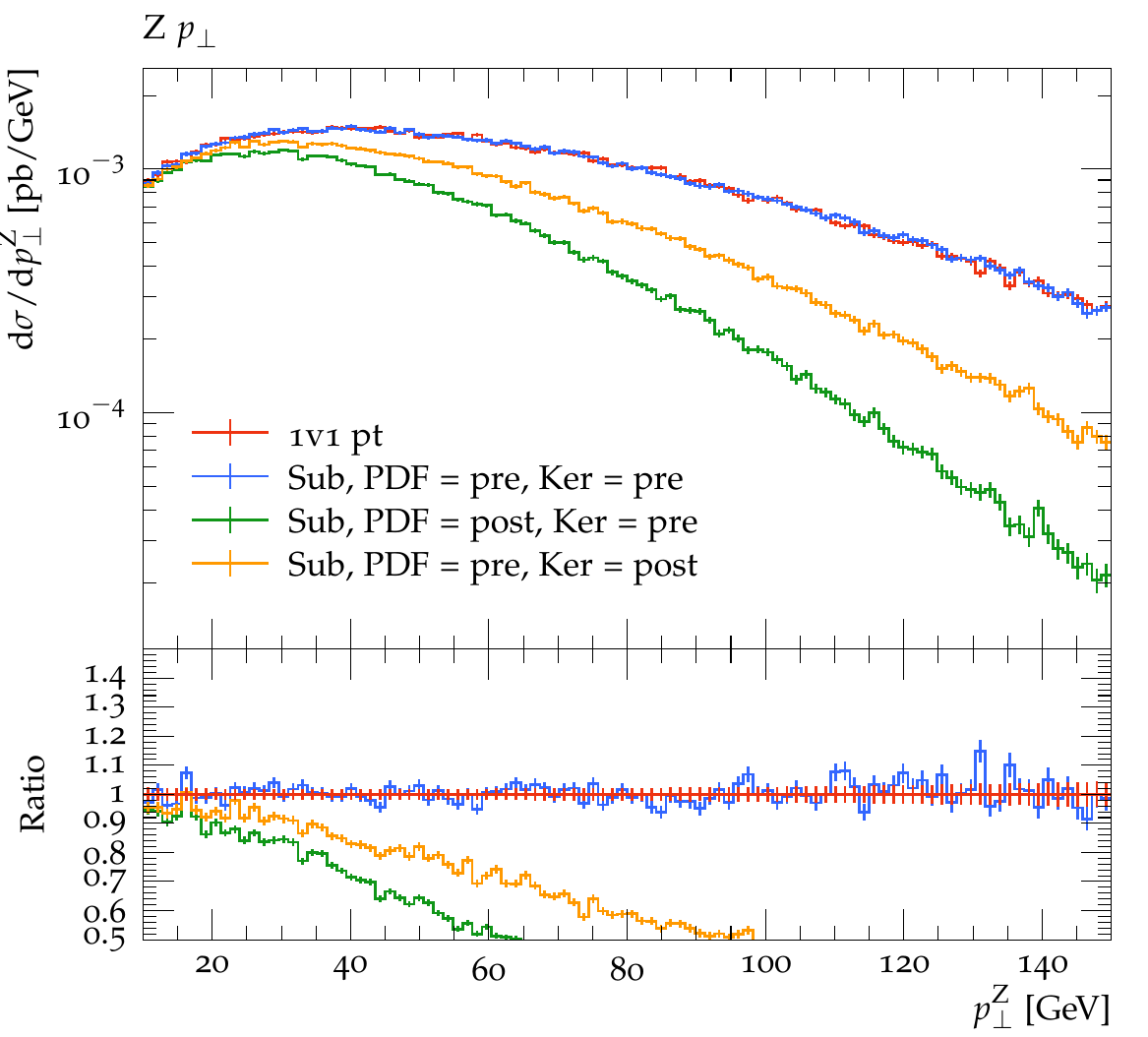}
\includegraphics[width=0.45\textwidth]{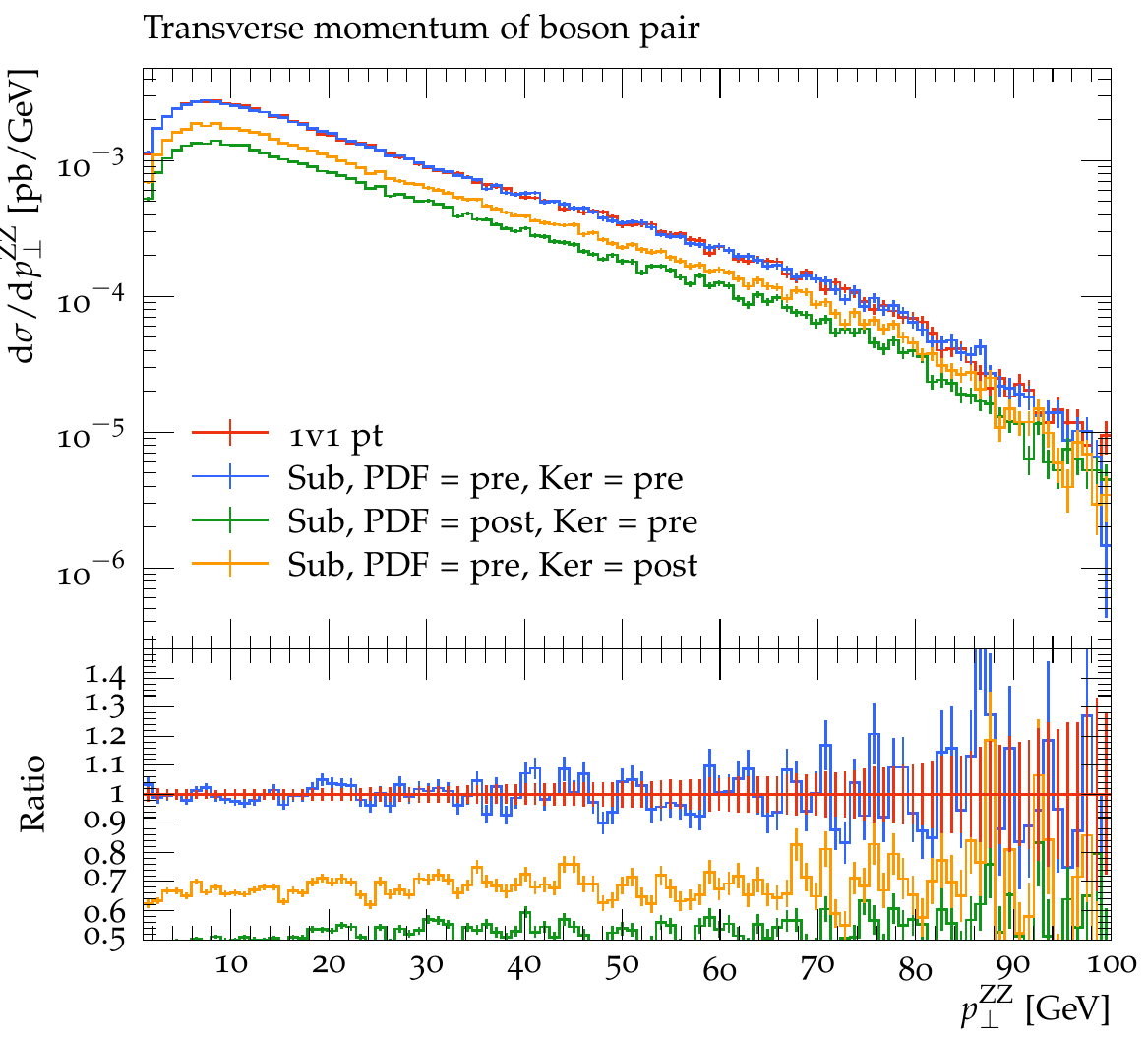} \\
(a) \hspace{220pt} (b)
\caption{(a) Transverse momenta of the $\Z$ bosons and (b) transverse momentum of the $\Z\Z$ pair for different values of the momentum fractions $X$ used in \Eq (\ref{eq:subZZbis}). The label ``PDF'' refers to the momentum fractions used in the gluon PDFs, whereas ``Ker'' labels the fractions in the splitting kernels. In both cases, these fractions are set to be either the pre-kick fractions or the post-kick ones. The scale $\mu$ is set to be equal to $\mZ$. The 1v1,pt setup is the reference in the ratio plots. The histograms are not normalised to unity.}
\label{Fig:subFrac}
\end{figure}

\begin{figure}[t!] 
\centering
\includegraphics[width=0.45\textwidth]{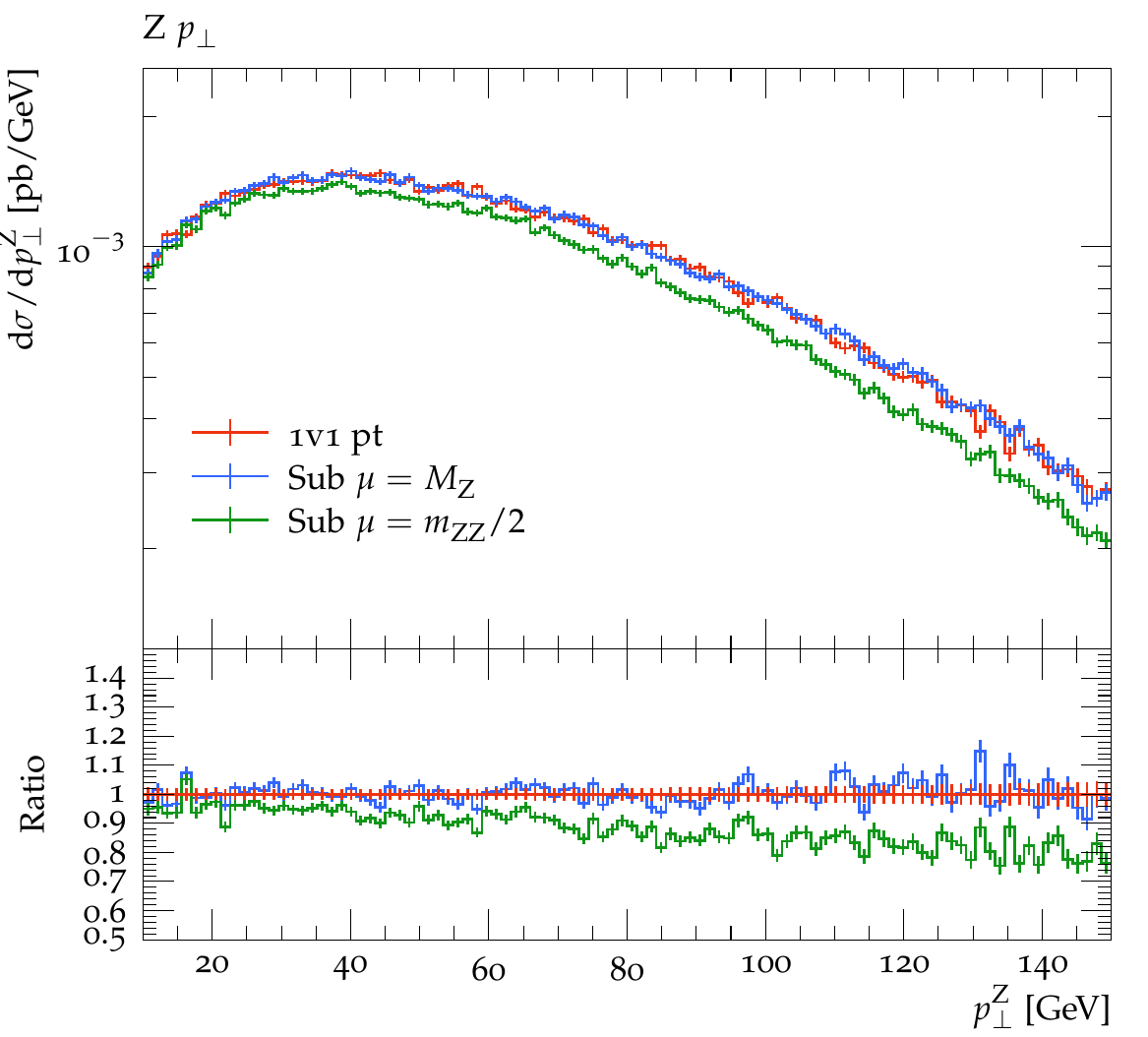}
\includegraphics[width=0.45\textwidth]{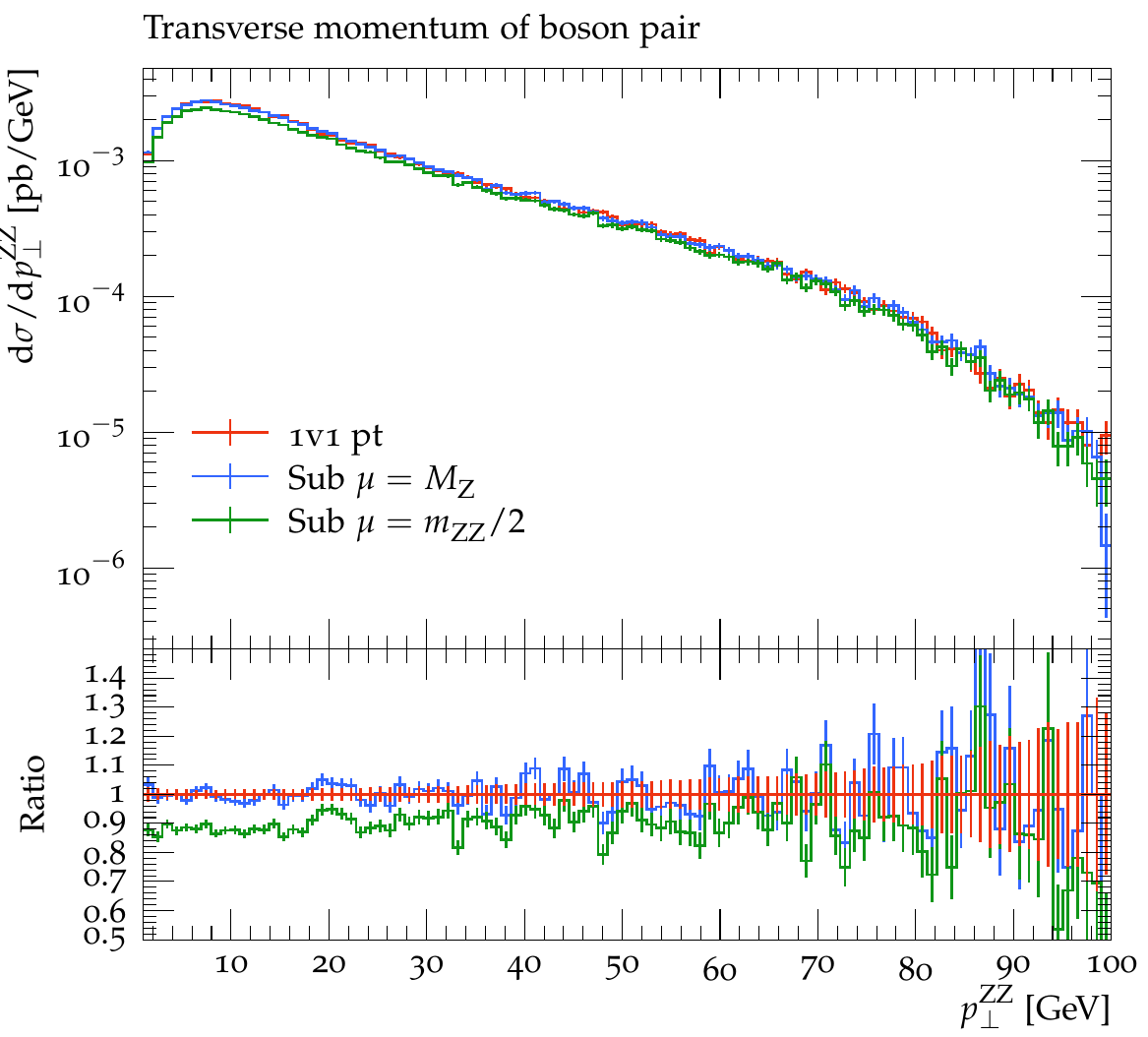} \\
(a) \hspace{220pt} (b)
\caption{(a) Transverse momenta of the $\Z$ bosons and (b) transverse momentum of the $\Z\Z$ pair for different values of the scale $\mu$ used in \Eq (\ref{eq:subZZbis}). The momentum fractions $X$ are set to be equal to the pre-kick fractions. The 1v1,pt setup is the reference in the ratio plots. The histograms are not normalised to unity.}
\label{Fig:subScale}
\end{figure}

In $\Figs$ \ref{Fig:subFrac} and \ref{Fig:subScale}, the histograms of the transverse momenta of the $\Z$ bosons and of the $\Z\Z$ pair are given for several choices of momentum fractions $X$ ($\Fig$ \ref{Fig:subFrac}) and scale~$\mu$ ($\Fig$ \ref{Fig:subScale}). These two histograms give complementary pieces of information since the transverse momenta of the $\Z$ bosons are mostly determined by the cross section whereas the transverse momentum of the $\Z\Z$ pair is particularly sensitive to the shower activity. Indeed, the transverse momentum of the $\Z\Z$ pair must balance that of all the extra parton emissions in order to achieve overall momentum conservation. In all the histograms, the error bars represent the statistical errors due to the use of Monte-Carlo techniques. As motivated above, the choice $\mu=\mZ$ and $X=\xi$ for both the PDFs and splitting kernels leads to the best match between the 1v1,pt and subtraction terms, at least for the presented distributions. With this choice, the subtraction term should reproduce the DPS one at small~$y$, since this latter is equal to the 1v1,pt term in that region.

The second step is to check the large-$y$ region. For $y\gg1/\nu$, the subtraction term should match the unpolarised, colour-singlet part of the SPS loop-induced term. The subtraction term $\sSh(\mZ)\otimes\d\xsSub_\mathrm{(Z,Z)}/\d O$ will now be compared to the loop-induced SPS cross section coupled to the $\sSh(\mZ)$ algorithm. In the region $y\gg 1/\nu$, the choice of scale $\mu$ and fractions $X$ does not matter as much as it does for $y\sim 1/\nu$ because the $\pT$ values are here small and the different choices thus coincide. In this study, our focus will be on comparing the overall shapes of the two terms (particularly at small $\pT \ll \nu$) rather than making precise numerical comparisons between the two -- in any case the magnitudes of the two should not coincide even at low $\pT$, as the full SPS loop-induced term contains additional colour, spin and flavour interference/correlation contributions, that are not contained in our subtraction term. 

In this study, the loop-induced cross section was computed using the matrix-element generator $\OL$ \cite{Cascioli:2011va, Buccioni:2017yxi,  Buccioni:2019sur, vanHameren:2009dr, vanHameren:2010cp}. The factorisation scale and the argument of the strong coupling are set to $\mZ$. In the $\OL$ calculation one has all six quark flavours running inside the loop (with all quarks treated as massless except the top quark), instead of the three massless flavours in the calculation of the subtraction term. However, since we only aim at a rough shape comparison between the SPS and subtraction terms, this mismatch is not critical. We use the same 3-flavour $\as$ in both the SPS and subtraction terms. In the SPS calculation, we use the default values for the Higgs and top masses, $M_\mathrm{H}=125$ GeV and $M_\t=172$ GeV. 

\begin{figure}[t!] 
\centering
\includegraphics[width=0.55\textwidth]{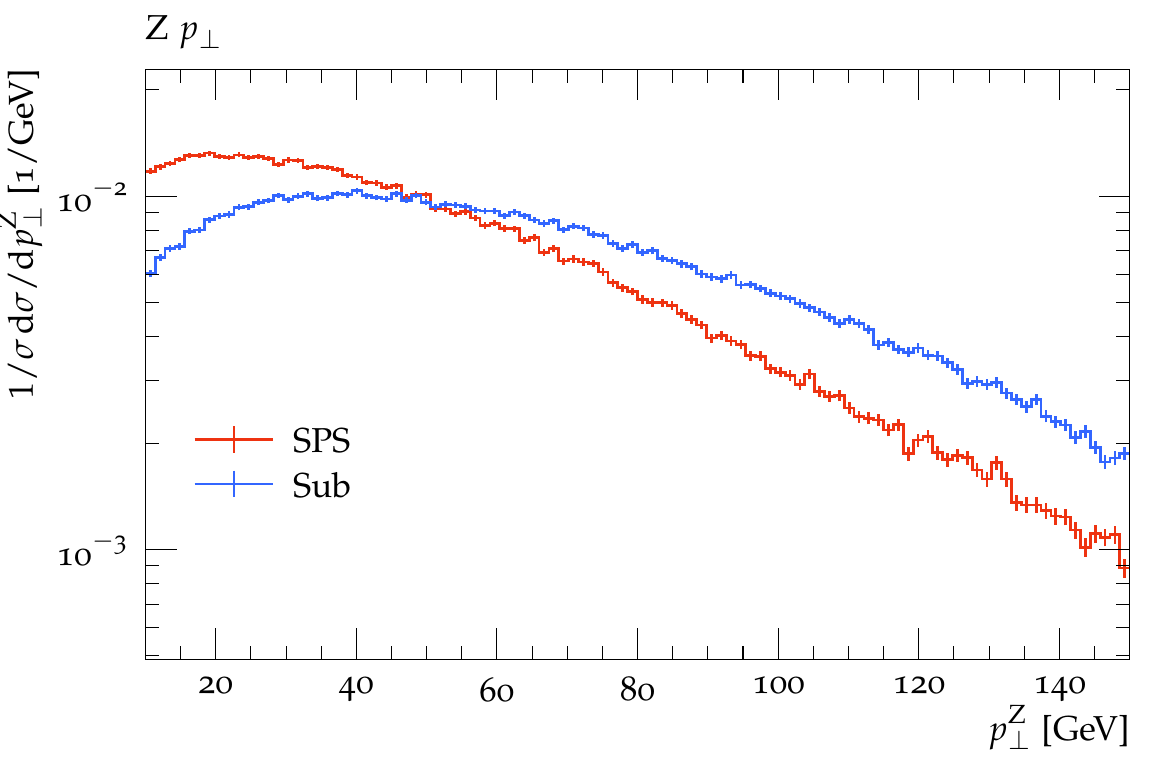}
\includegraphics[width=0.4\textwidth]{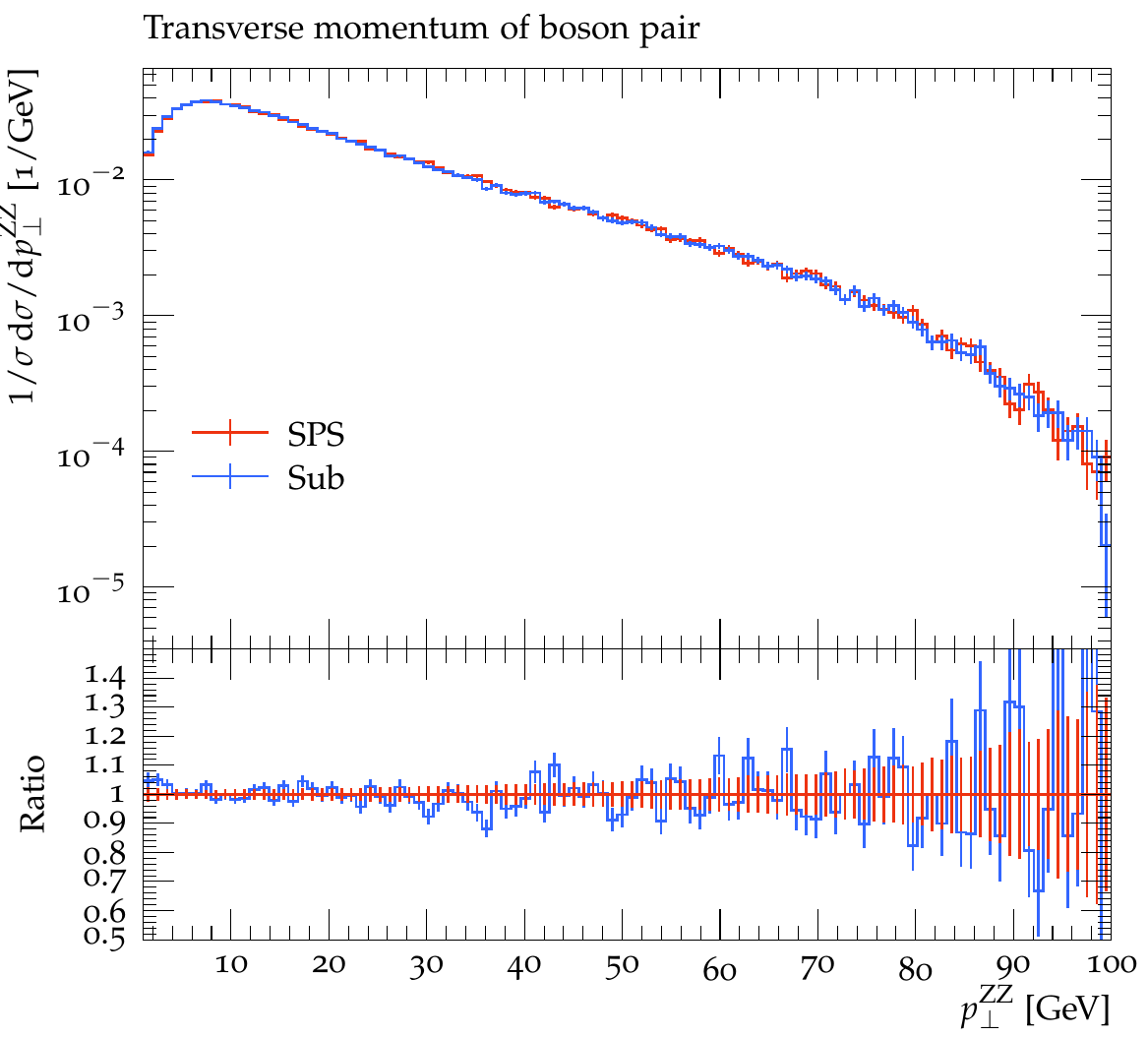} \\
(a) \hspace{230pt} (b)
\caption{(a) Transverse momenta of the $\Z$ bosons and (b) transverse momentum of the $\Z\Z$ pair as produced by the SPS and subtraction terms. The momentum fractions $X$ are set to be equal to the pre-kick fractions and $\mu=\mZ$. The SPS setup is the reference in the ratio plots. The histograms are normalised to unity.}
\label{Fig:yLarge}
\end{figure}

In $\Fig$ \ref{Fig:yLarge}, the subtraction term is compared to the SPS one. Here, the histograms are normalised to unity because we are mainly concerned with the shapes of the two different terms, as mentioned above. It can be seen that the $\pT$ spectra for the boson pair exactly match. This is because the $\pT$ spectrum of the $\Z\Z$ pair is mainly controlled by the shower algorithm used and the two terms are showered with the exact same algorithm $\sSh(\mZ)$. Nevertheless, the curves obtained for the $\pT$ spectrum of the $\Z$ bosons do not coincide. This is due to the fact that the $\Z$ $\pT$ is strongly determined by the cross section. The $\pT$ profile which was inserted in the expression of the subtraction term is the $\pT$ spectrum of a 1v1,pt event, and ensures an accurate subtraction with DPS in the region $y\sim 1/\nu$. However, this profile only approximates the $\pT$ spectrum of an SPS event and hence does not perfectly match the SPS cross section in the region $y\gg 1/\nu$. In particular, the small-$\pT$ behaviour obtained with the subtraction term is $\log(\pTs/\nu^2)$ instead of the $\log^2(\pTs/\nu^2)$ that can be extracted from the SPS cross section, recall $\Sec$ \ref{Sec:choicepT}. It will be seen in a later section how one can modify the transverse profile used in the merging kinematics to improve the matching between the SPS and the subtraction terms in the large-$y$ region.

\section{Numerical results}
\label{Sec:numerics}

In this last section, the results obtained from the numerical implementation of \Eqs (\ref{eq:sub}) and (\ref{eq:master}) are presented for $\Z\Z$ production via SPS and DPS at \mbox{$\sqrt{s}=13$ TeV}. The set of sPDFs, the running scheme for the strong coupling and the choices of scales and momentum fractions are identical to the ones mentioned in the previous section. In particular, the factorisation scales and the arguments of the couplings in all the cross sections as well as the starting scales of the showers are set to be equal to $\mZ$. The cross sections are computed either analytically or with \OL. As before, in this numerical study,  we will only include the loop-induced process in the SPS piece, although in principle one can also add other SPS processes on top of the loop-induced one (such as the $\q\qbar \to \Z \Z$ Born process).
 For the DPS cross section written in \Eq (\ref{eq:XSDPS}), the set of \mbox{$\y$-dependent} dPDFs that is used is the 3-flavour DGS set originally developed in \cite{Diehl:2017kgu} and improved in~\cite{Cabouat:2019gtm}. The results are presented at parton level, meaning that there is no hadronisation phase. In each event, there are at most two different hard scatters.

In this study we choose to rescale the SPS cross section by a factor $1/10$. This is to counteract the fact that the DPS cross section is power suppressed with respect to the SPS one \cite{Diehl:2011yj}. Such a rescaling is of course not physical, but is helpful in this proof-of-concept study to distinguish the DPS process from the SPS one in the histograms and to enhance the sensitivity to the $\nu$ variation. We recall here that the SPS term does not contain any dependence on the parameter $\nu$ and the cancellation of the dependence on this unphysical parameter only occurs between the subtraction term and the DPS one.

\subsection{Validation}
\label{Sec:val}

\begin{figure}[t!] 
\centering
\includegraphics[width=0.45\textwidth]{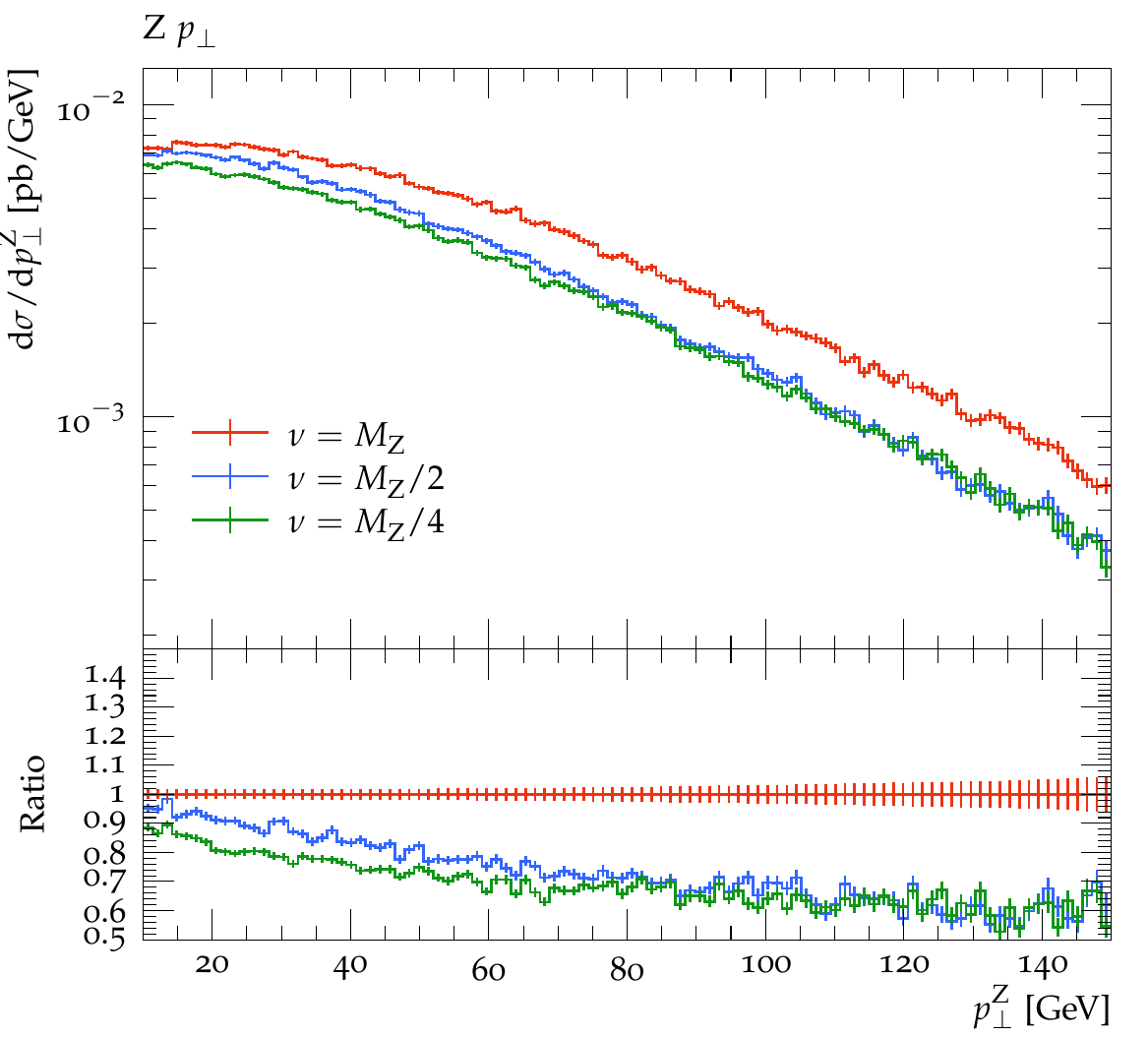}
\includegraphics[width=0.45\textwidth]{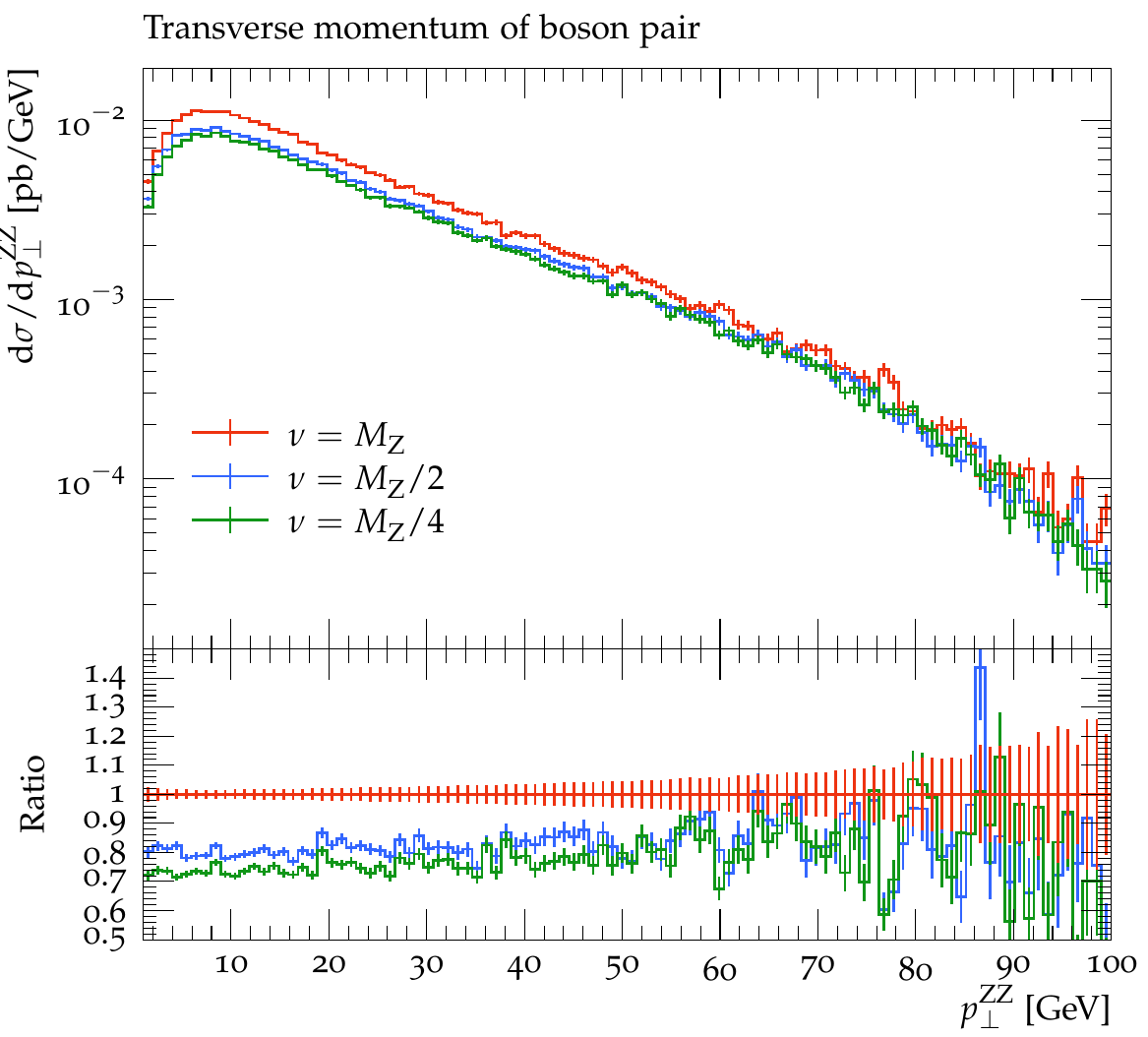} \\
(a) \hspace{180pt} (b)
\caption{(a) Transverse momenta of the $\Z$ bosons and (b) transverse momentum of the $\Z\Z$ pair as given by \Eq (\ref{eq:master}) with $\xsSub_\mathrm{(Z,Z)}=0$ and for three different values of $\nu$. The $\nu=\mZ$ setup is the reference in the ratio plots. The histograms are not normalised to unity.}
\label{Fig:subOff}
\end{figure}

\begin{figure}[t!] 
\centering
\includegraphics[width=0.45\textwidth]{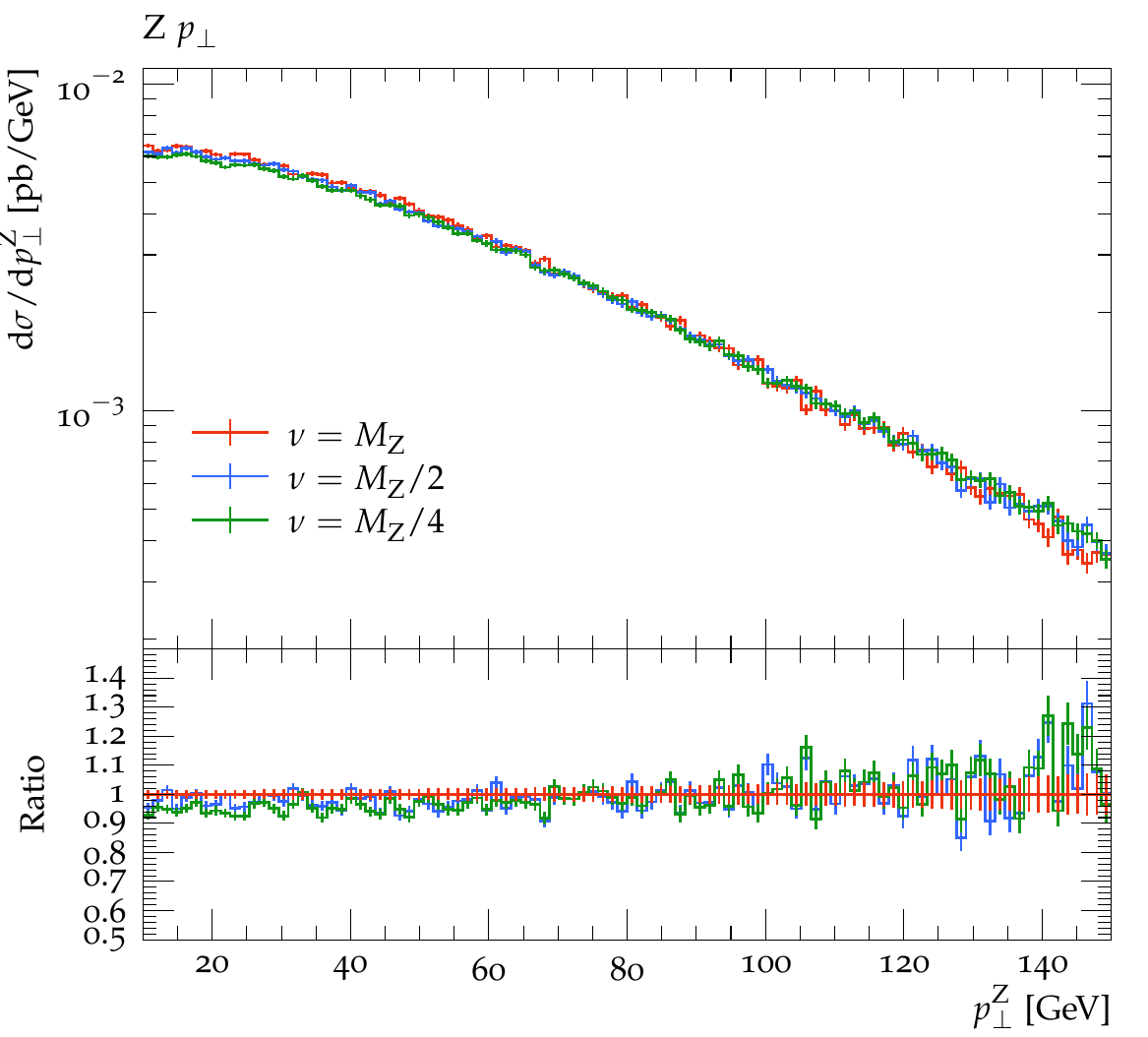}
\includegraphics[width=0.45\textwidth]{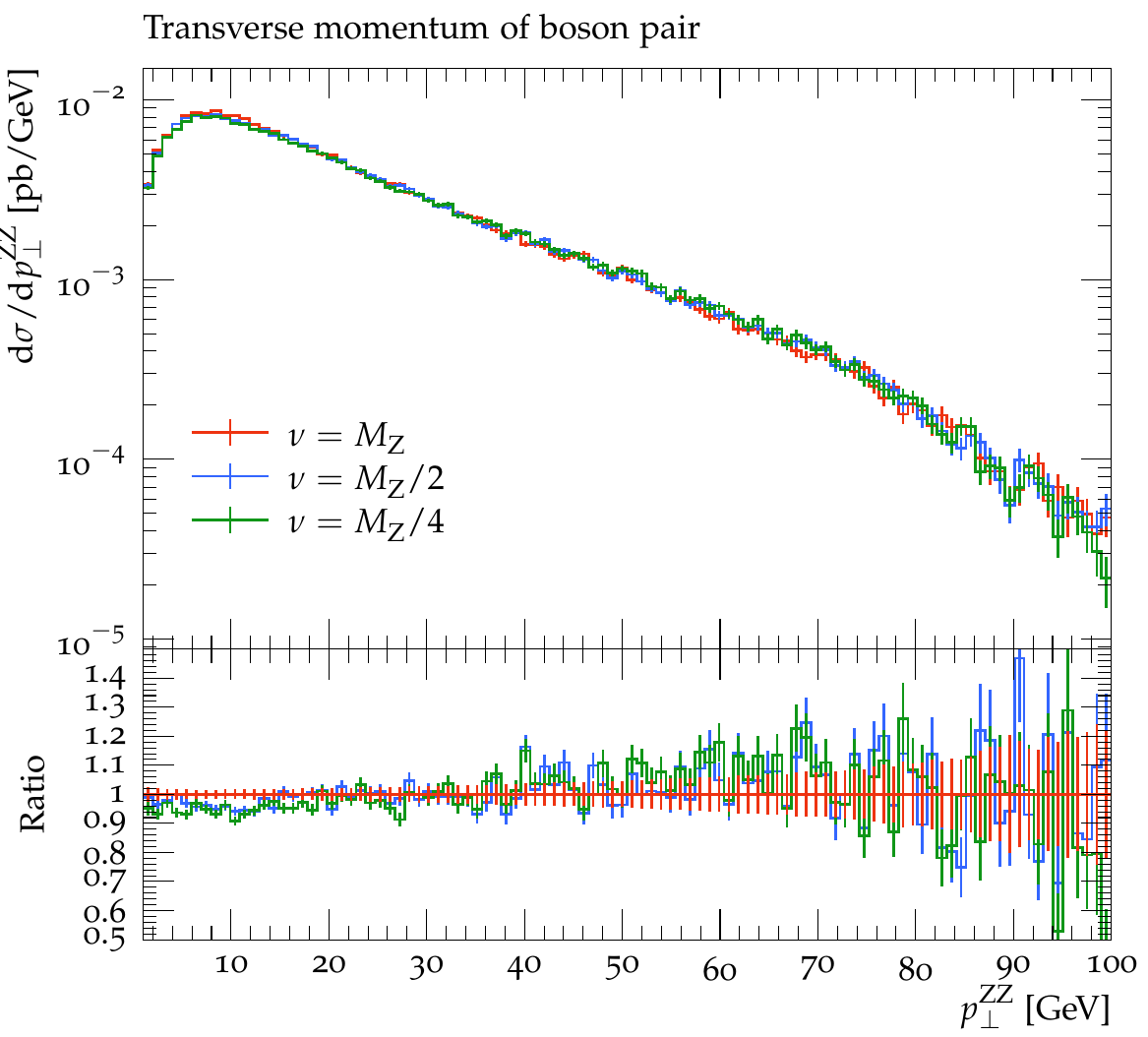} \\
(a) \hspace{180pt} (b)
\caption{(a) Transverse momenta of the $\Z$ bosons and (b) transverse momentum of the $\Z\Z$ pair as given by \Eq (\ref{eq:master}) for three different values of $\nu$. The $\nu=\mZ$ setup is the reference in the ratio plots. The histograms are not normalised to unity.}
\label{Fig:subOn}
\end{figure}

Let us start by studying the impact of the subtraction term. The histograms presented in $\Fig$ \ref{Fig:subOff} were produced setting $\xsSub_\mathrm{(Z,Z)}=0$, whereas the ones in $\Fig$ \ref{Fig:subOn} were obtained using all the terms present in \Eq (\ref{eq:master}). As expected, removing the subtraction term induces a strong dependence on the scale $\nu$ in the event shapes. The same effect can be observed for the total cross sections, see $\Tab$ \ref{tab:xs}.

\begin{table}[t!]
\centering
\begin{tabular}{|c|c|c|} 
\hline \multirow{2}{*}{Scale $\nu$} 
& \multirow{2}{*}{\begin{tabular}{c} With subtraction term \end{tabular}} 
& \multirow{2}{*}{\begin{tabular}{c} Without subtraction term \end{tabular}}
\\ & &  \\ \hline 
\multirow{2}{*}{\begin{tabular}{c} $\nu=\mZ$ \end{tabular}} &
\multirow{2}{*}{$0.222\pm0.002$} &
\multirow{2}{*}{$0.296\pm0.003$} 
\\ & &  \\ \hline 
\multirow{2}{*}{\begin{tabular}{c} $\nu=\mZ/2$ \end{tabular}} &
\multirow{2}{*}{$0.219\pm0.002$} &
\multirow{2}{*}{$0.240\pm0.002$} 
\\ & &  \\ \hline 
\multirow{2}{*}{\begin{tabular}{c} $\nu=\mZ/4$ \end{tabular}} &
\multirow{2}{*}{$0.216\pm0.001$} &
\multirow{2}{*}{$0.222\pm0.002$} 
\\ & &  \\ \hline 
\end{tabular}
\caption{Total cross section for pp $\to\Z\Z$ in picobarns [pb] at $\sqrt{s}=13$ TeV for different values of the scale $\nu$. The statistical error is given.}
\label{tab:xs}
\end{table}

In the case where the subtraction term is included, the fact that the event shapes are independent of $\nu$ (up to subleading terms in $\as$) can be understood as follows. As we increase $\nu$ from an initial value of the order\footnote{We remind the reader that in our current implementation we must have $\nu \leq Q_h$, although the general argument presented here also works for $\nu > Q_h$.} of $Q_h$, a positive contribution is added to the DPS term at small $y \sim 1/Q_h$. For the dominant 1v1 part of this, the double merging occurs very close to the two hard scatters. The additional 1v1 events hence develop a topology that is similar to the one usually associated to an SPS event. However, as we increase $\nu$, the subtraction term gets a nearly identical additional contribution at small $y$. This means that the term that is subtracted from the SPS cross section is larger, recall \Eqs (\ref{eq:master}) and (\ref{eq:subZZbis}), which implies fewer actual SPS events. The two mechanisms are designed to cancel each other. In practice, a slight dependence on $\nu$ may appear for some observables, however. This can be due to the fact that only the leading contributions were included in the definition of the DPS and subtraction terms. Adding higher-order corrections to both terms would reduce the residual $\nu$ dependence (a key result that is needed for this is obtained in \cite{Diehl:2019rdh}). In practice, the observables are even less sensitive to a $\nu$ variation than it appears in this proof-of-concept study because the DPS and subtraction terms are relatively small compared to the SPS one (recall the factor $1/10$ applied to the SPS cross section).

Let us briefly comment on the number of events with negative weights that are generated by our algorithm. The fraction of events that are accepted with a negative weight is rather small: $0.4\%$ for $\nu=\mZ$ and drops to $0\%$ for $\nu=\mZ/2$ and $\nu=\mZ/4$. Therefore, these events do not affect the efficiency of the algorithm. The fraction of events would be even smaller if the SPS cross section were not rescaled.

\subsection{Improving the matching at large $y$}

In $\Fig$ \ref{Fig:yLarge}a, it was observed that the shapes of the $\Z$ $\pT$ spectra produced by the SPS and subtraction terms do not coincide, even at small $\pT \ll Q_h$. This is due to a mismatch for large $y$ values between the $\pT$ profile of the subtraction term and the SPS cross section. 

It is actually possible to calculate the $\pT$ profile corresponding to the contribution to the SPS process which overlaps with DPS (i.e. the loop-induced process) in the large-$y$ region. This was achieved in \cite{Diehl:2011yj} and the $\pT$ profile of the unpolarised, colour-singlet contribution to SPS for large $y$ values can be approximated to be
\begin{equation} 
\label{eq:realprofileh}
h_{\mathrm{SPS}}(\pTv,y)=\frac{y^4}{(2\pi)^2}\int\frac{\d^2\z\,e^{i\z\cdot\pTv}}{\left(\y-\frac{1}{2}\z\right)^2\left(\y+\frac{1}{2}\z\right)^2}.
\end{equation}

\noindent The factor in front of the integral ensures that the profile is correctly normalised:
\begin{equation}
\begin{split}
\int h_{\mathrm{SPS}}(\pTv,y)\,\d^2\pTv
&=1.
\end{split}
\end{equation}

\noindent The $\pT$ profile in \Eq \eqref{eq:realprofileh} contains ultraviolet divergences at $\yp=0$ and $\ym=0$, where $\y_\pm = \y \pm \z /2$. However, no such divergences exist in the actual SPS cross section. This is because the integrand in \Eq \eqref{eq:realprofileh} is only valid in the region in which $|\y_\pm| \gg 1/Q_h \sim 1/\nu$, which is the region of the integral where a DPS description is most appropriate. The region in which one of $\y_\pm$ goes to zero whilst the other stays finite is the region of the integral where an SPS/DPS interference description is most appropriate. The ``DPS'' region $|\y_\pm| \gg 1/Q_h \sim 1/\nu$ ultimately yields the leading behaviour of the SPS cross section $\propto \log^2(\pT^2/\nu^2)$ (mentioned in $\Sec$ \ref{Sec:choicepT} and \cite{Nagy:2006xy, Gaunt:2011xd}), whilst the ``DPS/SPS interference'' region yields a subleading behaviour $\propto \log(\pT^2/\nu^2)$. Here, we are predominantly interested in the leading low-$\pT$ behaviour associated with the DPS region. To extract this behaviour, we can simply insert ultraviolet regulators in \Eq \eqref{eq:realprofileh} to cut off the integrand when $|\y_\pm| \sim 1/\nu$. In this work, we will regulate the ultraviolet divergences by adding a term $b_0^2/\nu^2$ to each denominator factor in \Eq \eqref{eq:realprofileh}, yielding:
\begin{equation}
\label{eq:massReg}
h_{\mathrm{SPS}}(\pTv,y;\nu)=\frac{y^4}{(2\pi)^2}\int\frac{\d^2\z\,e^{i\z\cdot\pTv}}{\left(\left(\y-\frac{1}{2}\z\right)^2+b_0^2/\nu^2\right)\left(\left(\y+\frac{1}{2}\z\right)^2+b_0^2/\nu^2\right)}.
\end{equation}

\noindent Integrating this profile over $y$ as in \Eq (\ref{eq:pTprofile}), one obtains
\begin{equation}
\label{eq:profileBessel}
\begin{split}
\int_0^{+\infty}\frac{\d^2 \y}{y^4}\,h_{\mathrm{SPS}}(\pTv,y;\nu)
&= \left[K_0\left(\frac{b_0\,\pT}{\nu}\right)\right]^2.
\end{split}
\end{equation}
%
The function $K_0(x)$ is one of the modified Bessel functions of the second kind and reads

\begin{equation}
K_0(x)=\int_0^{+\infty}\frac{\cos(x t)}{\sqrt{1+t^2}}\,\d t=\frac{1}{2}\int_{-\infty}^{+\infty}\frac{e^{ixt}}{\sqrt{1+t^2}}\,\d t.
\end{equation}

\noindent In the limit where $\pT\ll \nu$, one gets

\begin{equation} 
\label{eq:realprofint}
\int_0^{+\infty}\frac{\d^2 \y}{y^4}\,h_{\mathrm{SPS}}(\pTv,y;\nu)\sim\left(\frac{1}{2}\,\log\left(\frac{\pT^2}{\nu^2}\right)+\log\left(\frac{b_0}{2}\right)+\gamma_E\right)^2,
\end{equation}

\noindent which gives the leading $\log^2(\pT^2/\nu^2)$. Note that the regularisation in \Eq (\ref{eq:massReg}) changes the normalisation of the profile. This can be rectified by replacing the factor $y^4$ by $(y^2+b_0^2/\nu^2)^2$ in this same equation. This substitution then modifies the result obtained in \Eq (\ref{eq:profileBessel}) but does not change the leading $\log^2(\pT^2/\nu^2)$ behaviour that is extracted from this result for small values of $\pT$. Using another regularisation scheme has the same effect: it changes the subleading terms, but not the leading one. If the $\pT$ profile derived in \Eq (\ref{eq:profileBessel}) is then used to construct the subtraction term then the $\pT$ spectra obtained from the subtraction and SPS terms coincide in the small-$\pT$ region, up to corrections going like $\log(\pT^2/\nu^2)$ and terms which are not logarithmically enhanced.

The problem here is that the kinematics of the subtraction term must also match the one of a DPS 1v1,pt event in the small-$y$ region and it is cumbersome to design a transverse profile $g(\kTv,y)$ for the merging kinematics whose convolution with itself leads to a $\pT$ profile as given by \Eq (\ref{eq:profileBessel}) (recall \Eq (\ref{eq:ypTprofile})). This is the reason why the transverse profile $g(\kTv,y)$ was chosen to be Gaussian in this work, see \Eq (\ref{eq:gausskT}). Such a form leads to a resulting $\pT$ profile $h(\pTv,y)$ that can be analytically calculated and at least has a reasonably similar behaviour, once integrated over $y$, as the one given by \Eq (\ref{eq:profileBessel}) in the small-$\pT$ region. 

\begin{figure}[t!] 
\centering
\includegraphics[width=0.9\textwidth]{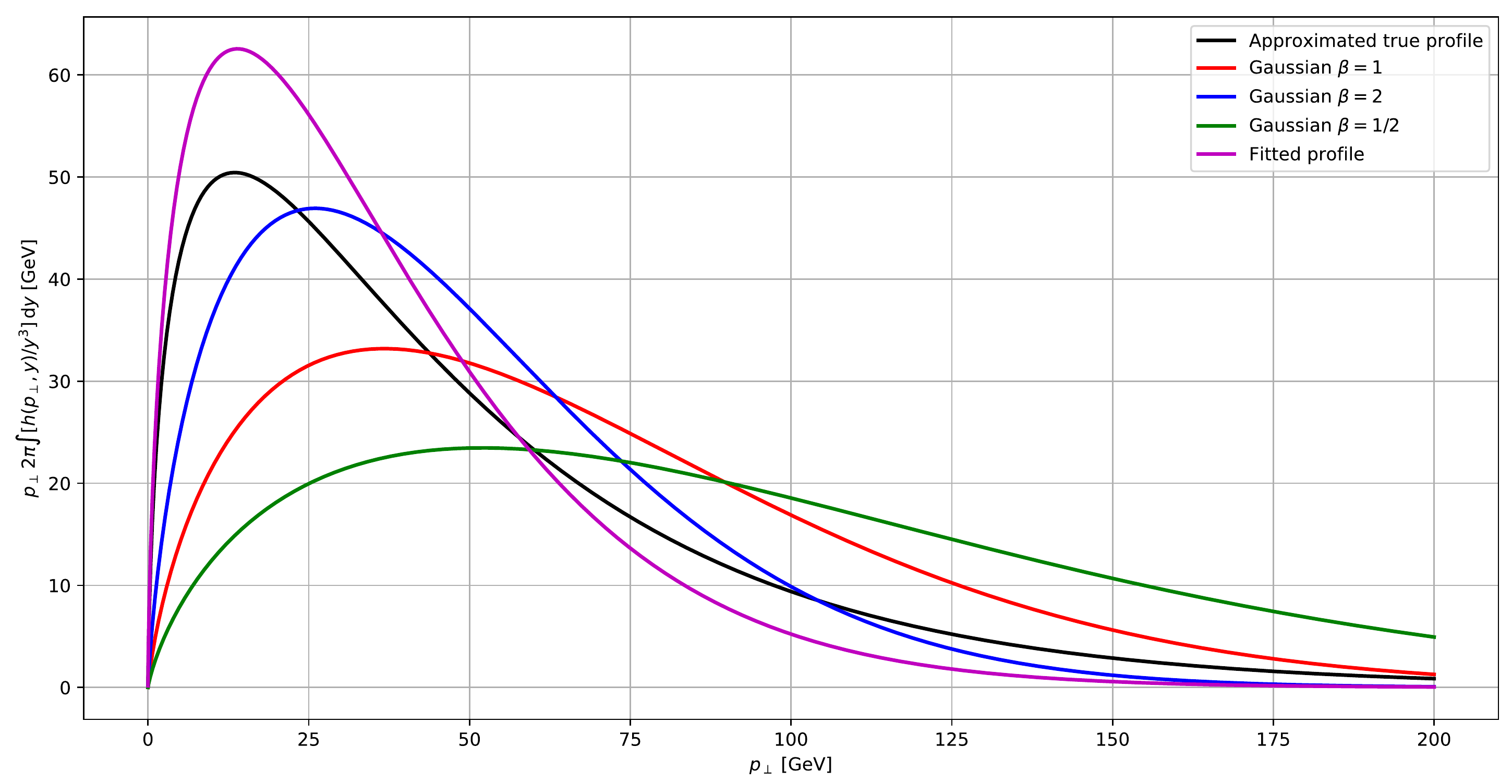}
\caption{Different $\pT$ profiles for the subtraction term for $\nu=\mZ$. The profile given by \Eq (\ref{eq:pTprofile}) which corresponds to a Gaussian distribution $g(\kTv,y)$ is represented for three values of $\beta$ (red, blue and green curves). The approximated ``true'' profile (black) and the fitted profile (magenta) are given respectively by \Eq (\ref{eq:profileBessel}) and \Eq (\ref{eq:fittedProfile}). The fitted profile corresponds to a decreasing Gaussian distribution $g(\kTv,y)$, as given by \Eq (\ref{eq:decreGausskT}). The area under each curve is equal to $\nu^2/(2b_0^2)$.}
\label{Fig:profileBeta}
\end{figure}

\begin{figure}[t!] 
\centering
\includegraphics[width=0.6\textwidth]{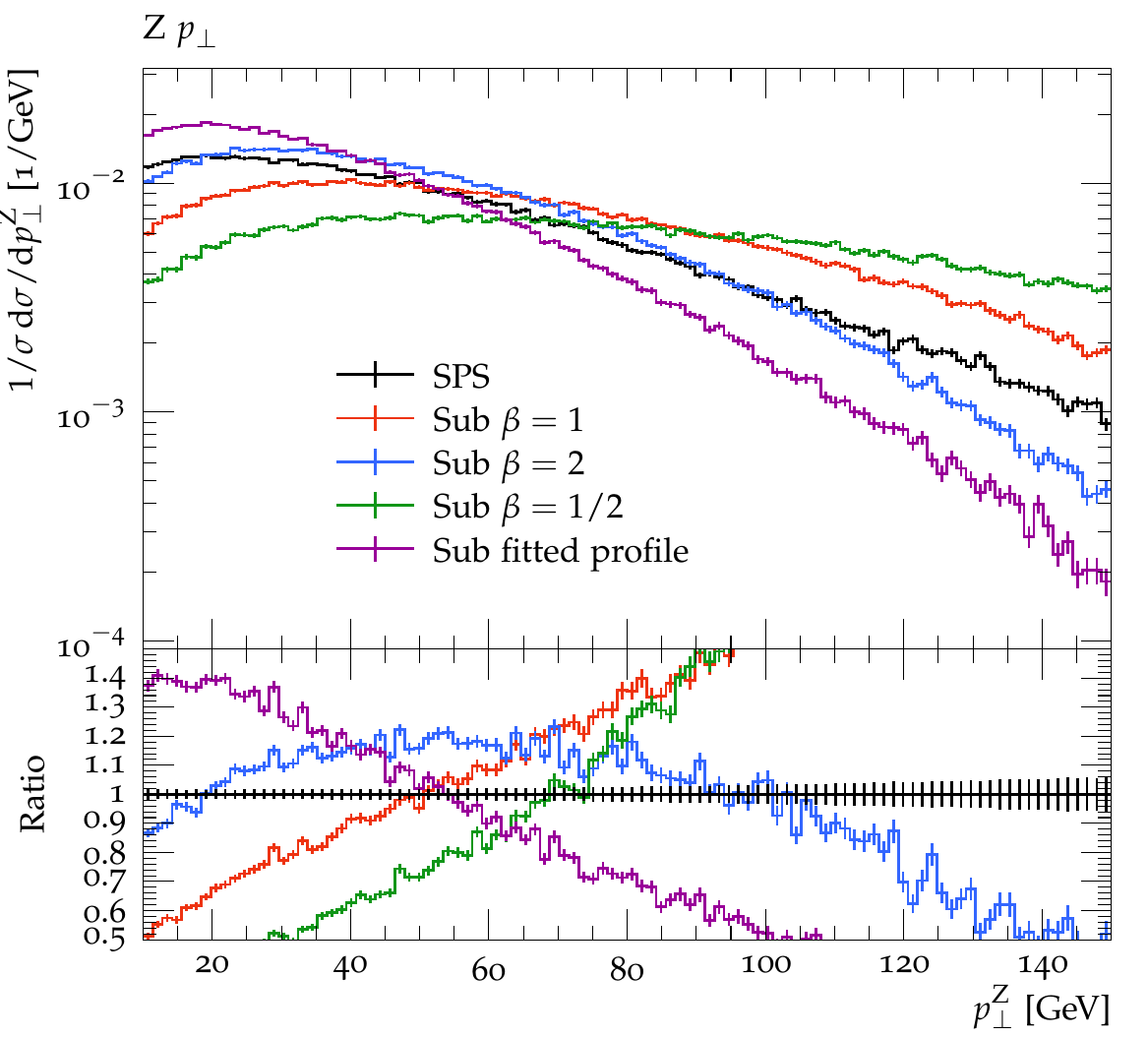}
\caption{Transverse momenta of the $\Z$ bosons as produced by the SPS and subtraction terms. The subtraction term corresponding to a Gaussian distribution $g(\kTv,y)$ is given for several values of $\beta$. The fitted profile corresponds to a decreasing Gaussian distribution $g(\kTv,y)$. The SPS setup is the reference in the ratio plot. The histograms are normalised to unity.}
\label{Fig:subBeta}
\end{figure}

In $\Fig$ \ref{Fig:profileBeta}, the approximated SPS $\pT$ profile given by \Eq (\ref{eq:profileBessel}) is compared to the one given by \Eq (\ref{eq:pTprofile}) for several values of $\beta$. This latter profile was obtained from a Gaussian distribution $g(\kTv,y)$. It can be observed that the shape of the SPS profile is best reproduced for $\beta=2$. This is confirmed in $\Fig$ \ref{Fig:subBeta} where the SPS term is compared to the subtraction term for several values of $\beta$. One observes in the plots that whatever the value of $\beta$ is, the shape of the subtraction term does not match that of the SPS term at the lowest $\pT$ values. This is due to the fact that changing the parameter $\beta$ cannot change the $\log(\pT^2/\nu^2)$ behaviour obtained from the resulting $\pT$ profile for small $\pT$ values, which does not match the SPS $\log^2(\pT^2/\nu^2)$. In this sense the Gaussian ansatz is not ideal. One has to keep in mind, however, that in fact the transverse profile $g(\kTv,y)$ of the $1\to 2$ splitting does not play a role at the leading-logarithmic level in the transverse-momentum distributions of the $\Z$ bosons, so these considerations are technically beyond our intended accuracy. The Gaussian ansatz implements in a simple way the physical intuition that the partons in the $1 \to 2$ splitting should be given a relative transverse momentum $\kT\sim1/y$.

\begin{figure}[t!] 
\centering
\includegraphics[width=0.45\textwidth]{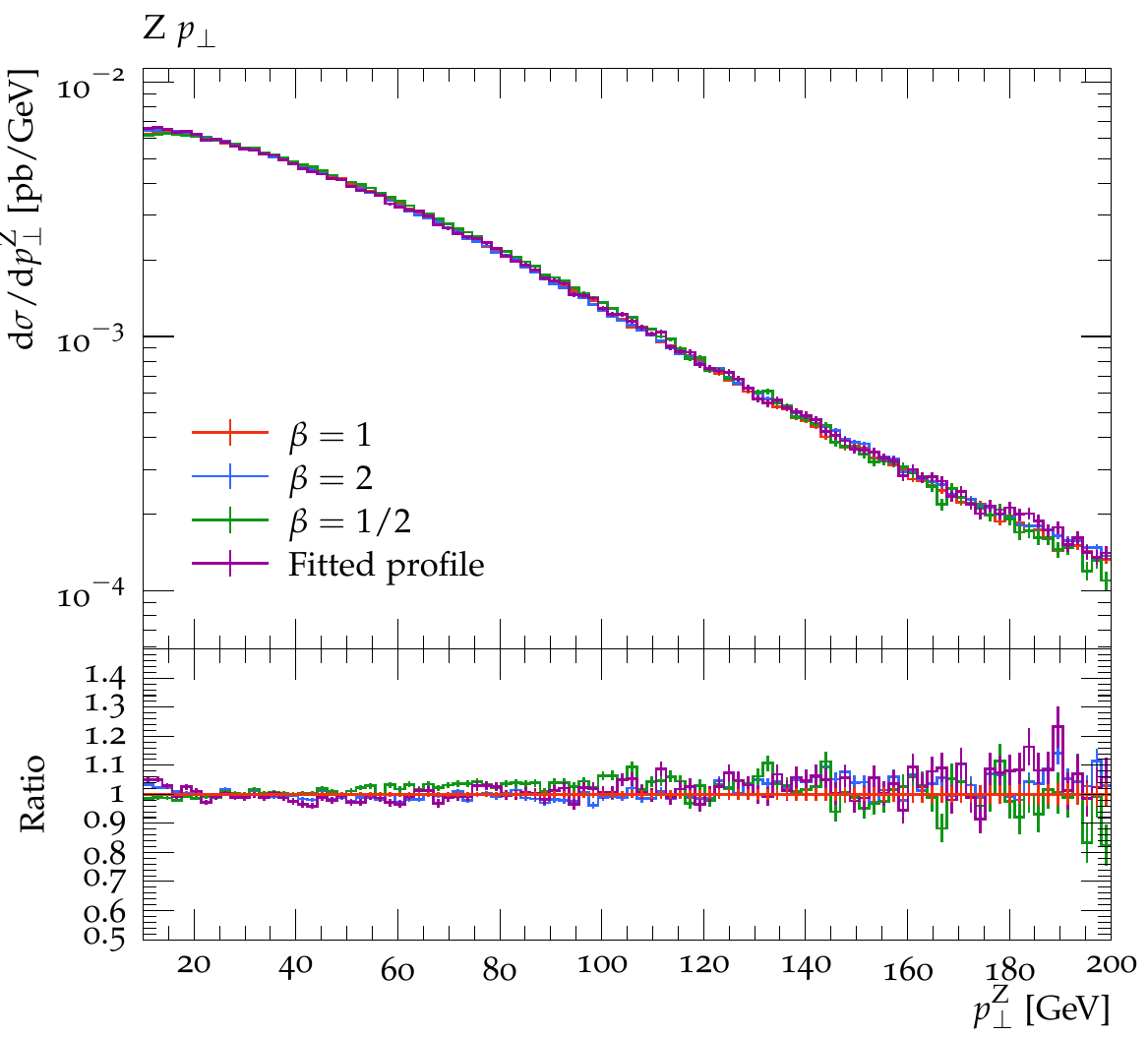}
\includegraphics[width=0.45\textwidth]{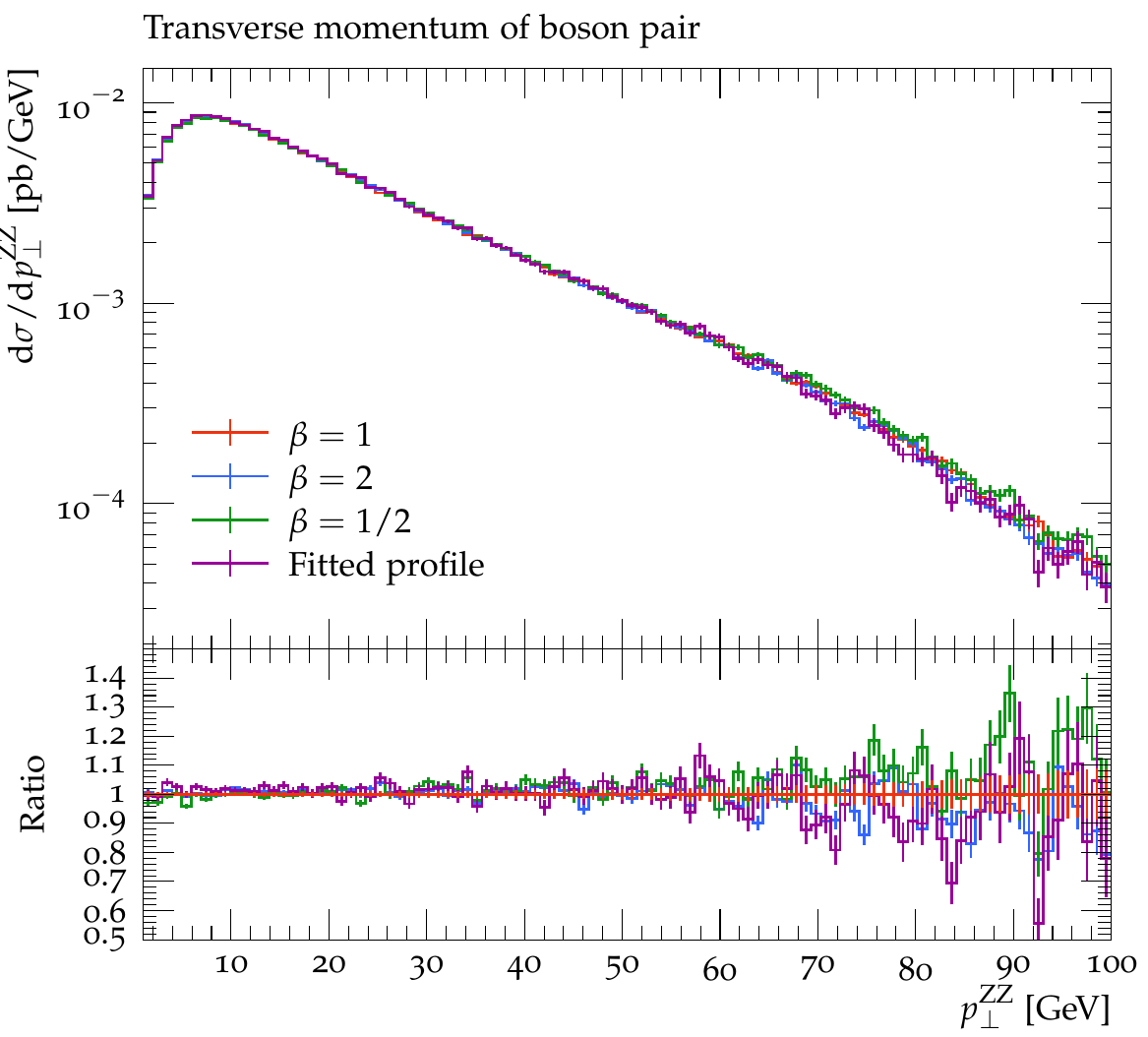} \\
(a) \hspace{180pt} (b)
\caption{(a) Transverse momenta of the $\Z$ bosons and (b) transverse momentum of the $\Z\Z$ pair as given by \Eq (\ref{eq:master}) for a Gaussian form of $g(\kTv,y)$ (with three different values of $\beta$) and for a decreasing Gaussian form (fitted profile). The $\beta=1$ setup is the reference in the ratio plots. The histograms are not normalised to unity.}
\label{Fig:betaNoCut}
\end{figure}

\begin{figure}[t!] 
\centering
\includegraphics[width=0.45\textwidth]{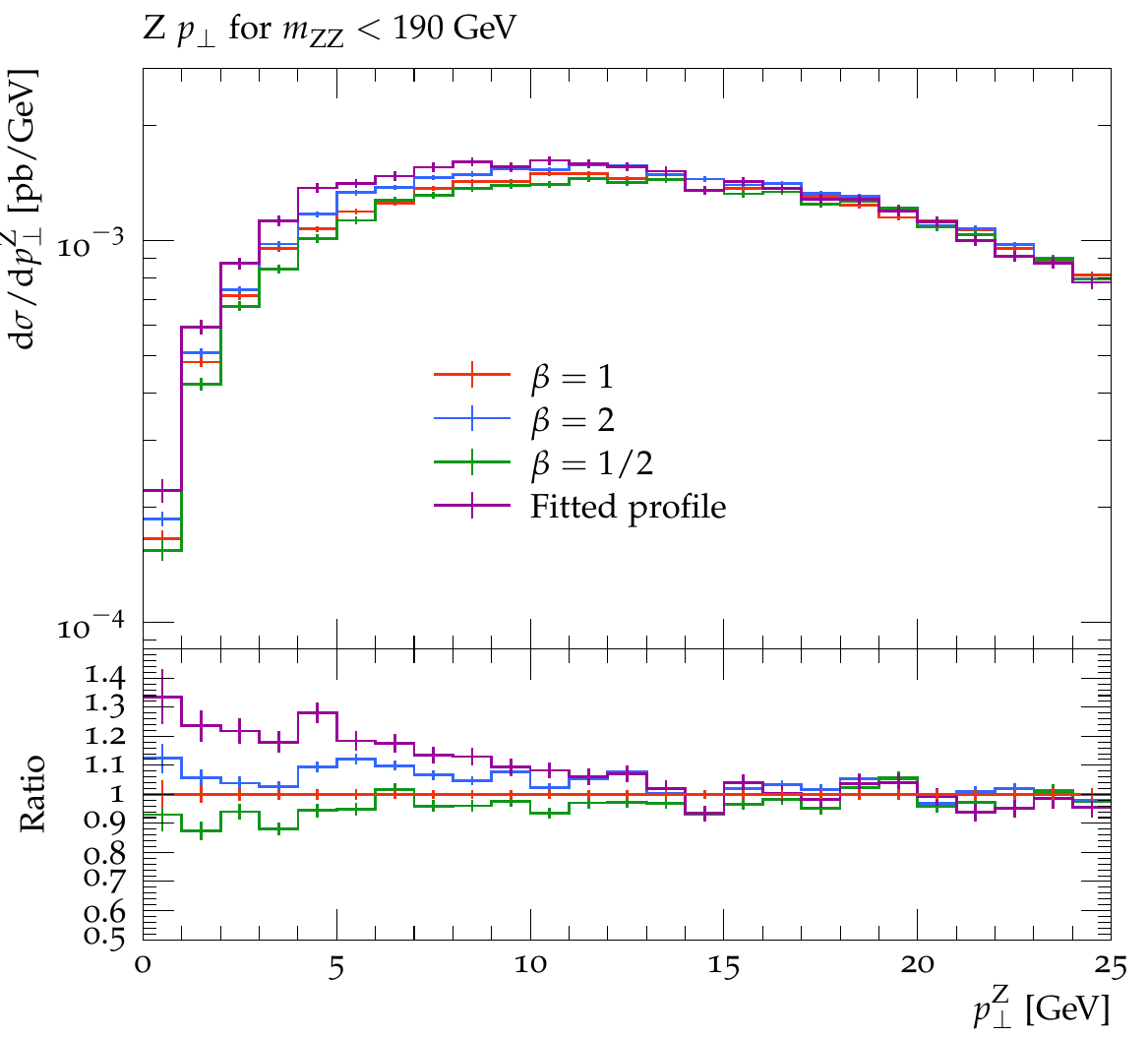}
\includegraphics[width=0.45\textwidth]{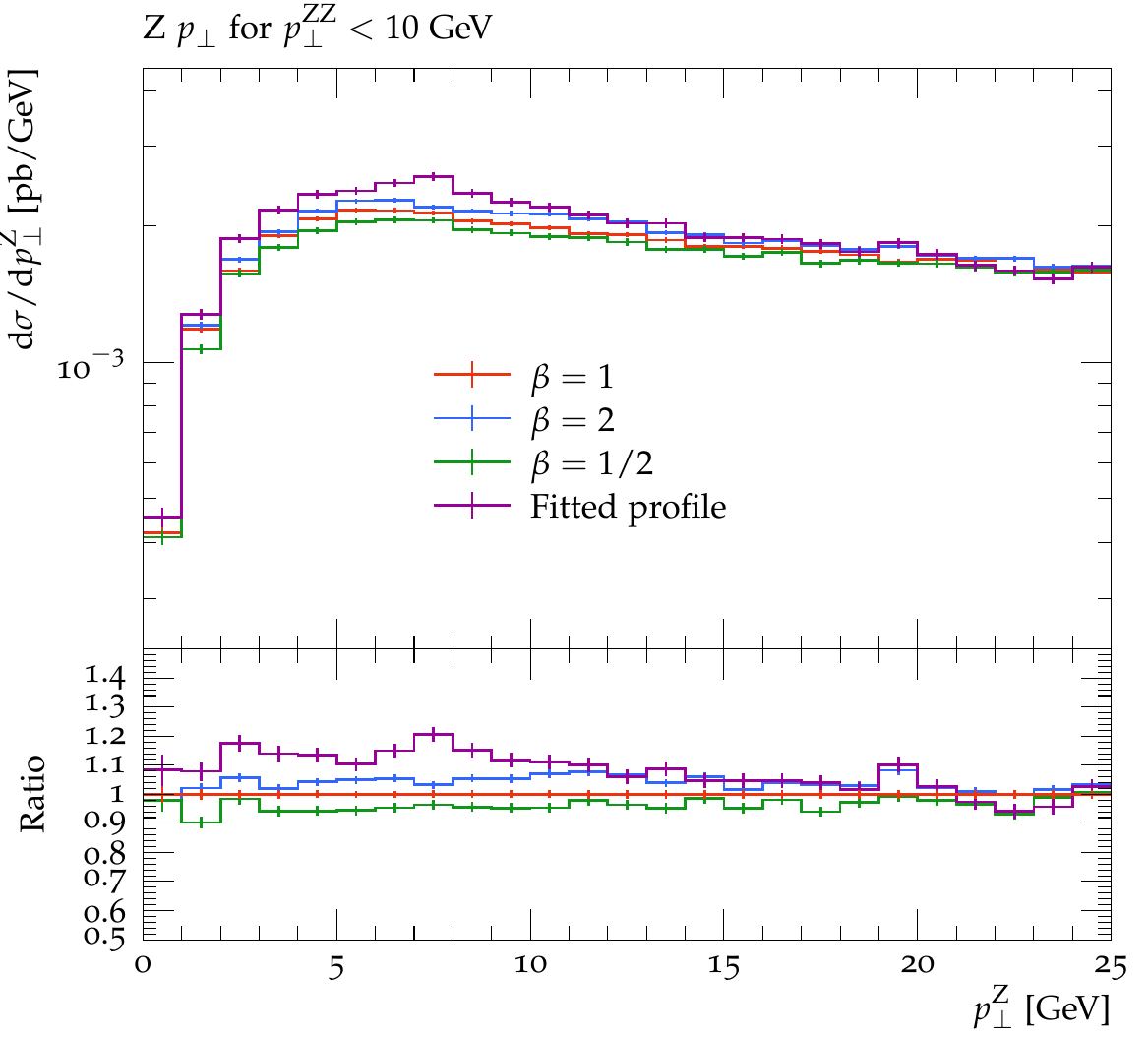} \\
(a) \hspace{180pt} (b)
\caption{Transverse momenta of the $\Z$ bosons with (a) a cut on the invariant mass of the $\Z\Z$ pair and (b) a cut on the transverse momentum of the pair. The results were produced using \Eq (\ref{eq:master}) for a Gaussian form of $g(\kTv,y)$ (with several values of $\beta$) and for a decreasing Gaussian form (fitted profile). The $\beta=1$ setup is the reference in the ratio plots. The histograms are not normalised to unity.}
\label{Fig:betaWithCut}
\end{figure}

In $\Figs$ \ref{Fig:betaNoCut} and \ref{Fig:betaWithCut}, the results obtained by combining all the contributions as described in \Eq (\ref{eq:master}) are given for several values of $\beta$. One can notice in $\Fig$ \ref{Fig:betaNoCut} that in general the value of $\beta$ does not affect too much the resulting kinematic distributions. In order to observe a discrepancy, one needs to study the small-$\pT$ region with extreme cuts on either the invariant mass or the transverse momentum of the $\Z\Z$ pair, see $\Fig$ \ref{Fig:betaWithCut}. The fact that the results do not depend strongly on the value of $\beta$ is expected: the discrepancy between the different choices is not a leading-logarithmic effect.

One may wonder whether it is possible to improve the Gaussian ansatz -- i.e. define a class of profiles $g(\kTv,y)$ such that the resulting $\pT$ profile behaves as $\log^2(\pT^2/\nu^2)$ in the small-$\pT$ region. To achieve such a goal, let us revisit the equations of $\Sec$ \ref{Sec:choicepT}. We recall that the small-$\pT$ behaviour of the loop-induced SPS term is dominated by contributions from the region $1/\nu\ll |\y_\pm| \ll 1/\pT$ (the logarithmic integrations for $\y_\pm$ are ``cut off'' at values of order $1/\pT$ by the exponential factor in \Eq \eqref{eq:massReg}). In a similar way, the dominant small-$\pT$ behaviour of the subtraction term under the Gaussian ansatz arises from the region $1/\nu\ll y\ll 1/\pT$ -- we have a logarithmic integration over $y$ that extends between $y \sim 1/\nu$ (where it is cut off by the factor $\Phi$) and $y \sim 1/\pT$ (where it is cut off by the Gaussian factor), recall \Eq \eqref{eq:pTprofile}. For the purposes of computing the leading low-$\pT$ behaviour, one can replace the Gaussian factor in \Eq \eqref{eq:ypTprofile} by a simple cut-off imposing $y\pT < 1$, yielding for the $\pT$ distribution:
\begin{equation} 
\label{eq:GaussianLL}
\int_{b_0/\nu}^{1/\pT}\frac{\d^2\y}{y^4}\,\left(\frac{\beta}{2\pi}\,y^2\right)=\pi\int_{b_0^2/\nu^2}^{1/\pT^2}\frac{\d y^2}{y^2}\,\frac{\beta}{2\pi}=-\frac{\beta}{2}\log\left(\frac{b_0^2\,\pT^2}{\nu^2}\right).
\end{equation}
This agrees with \Eq \eqref{eq:DPSsing} at the leading-logarithmic level.

This insight allows us to design an $h(\pTv,y)$ that yields a double logarithmic behaviour in the small-$\pT$ limit. We need an expression which is strongly suppressed for $y\pT > 1$, as for the Gaussian ansatz, but which is proportional to $-y^2\,\log(y\pT)$ in the limit $y\pT \ll 1$ rather than $y^2$. Then, the leading low-$\pT$ behaviour will be proportional to (recall \Eq \eqref{eq:GaussianLL})
\begin{equation}
\label{eq:decGaussianLL}
\int_{b_0/\nu}^{1/\pT}\frac{\d^2\y}{y^4}\,\left(-y^2\,\log(y\pT)\right)=\pi\int_{b_0^2/\nu^2}^{1/\pT^2}\frac{\d y^2}{y^2}\,(-\log(y\pT))=\frac{\pi}{4}\log^2\left(\frac{b_0^2\,\pT^2}{\nu^2}\right).
\end{equation}
Such a profile $h(\pTv,y)$ can be obtained for example from the following form for $g(\kTv,y)$:
\begin{equation}
\label{eq:decreGausskT}
g(\kTv,y)=\frac{1}{\pi\sqrt{2}}\,\frac{y}{\kT}\,\exp\left(-\frac{\pi}{2}\,y^2\kT^2\right).
\end{equation}
The width of the Gaussian in this expression has been chosen such that when this profile is used to construct the subtraction term, the coefficient of the $\log^2(\pT^2/\nu^2)$ term in the $\pT$ distribution is the same as the corresponding coefficient in \Eq \eqref{eq:realprofint}. 

Unfortunately, we were not able to obtain the $\pT$ profile of the subtraction term corresponding to \Eq \eqref{eq:decreGausskT} analytically. However, one can perform a fit of this profile, using the following functional form:
\begin{equation}
\label{eq:fittedProfile}
\int_{b_0/\nu}^{+\infty}\frac{\d^2 \y}{y^4}\,h(\pTv,y)\simeq e^{-\beta_0 z^{\alpha_0}}\left(\gamma+\beta_1 z^{-\alpha_1}+\beta_2\log(z)+\log^2(z)\right),
\end{equation}

\noindent with $z=b_0\,\pT/\nu$. The result of the fit gives the coefficients\footnote{Note that technically the leading small-$z$ behaviour of the right-hand side is $z^{-\alpha_1}$ -- however, the size of this term only actually overtakes the $\log^2(z)$ one once $z \lesssim 10^{-9}$, which is not practically relevant.} $\beta_0=3.58$, $\alpha_0=1.16$, $\gamma=1.18$, $\beta_1=3.58$, $\alpha_1=0.23$ and $\beta_2=1.42$.

In $\Fig$ \ref{Fig:profileBeta}, the fit of the $\pT$ profile is compared to the approximated SPS profile given by \Eq (\ref{eq:profileBessel}) and to profiles corresponding to a Gaussian $g(\kTv,y)$. One can see that this fitted profile more closely approximates the shape of the SPS profile than the other ones for small values of $\pT$. This is due to the fact that the two profiles have the same double-logarithmic behaviour in the small-$\pT$ region. 

\begin{figure}[t!] 
\centering
\includegraphics[width=0.6\textwidth]{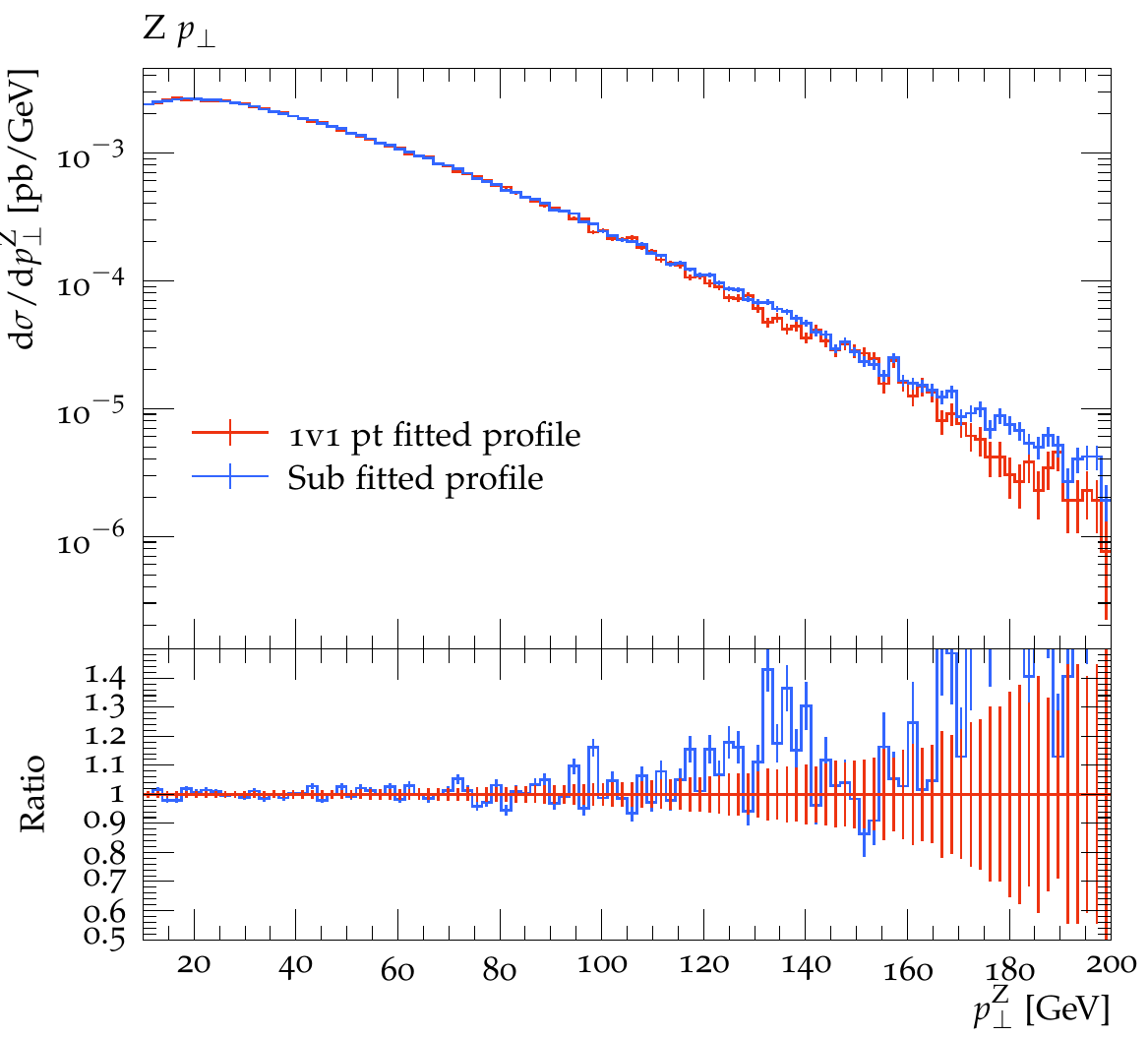}
\caption{Transverse momenta of the $\Z$ bosons as produced by the 1v1,pt and subtraction terms for a decreasing Gaussian form of $g(\kTv,y)$. The 1v1,pt setup is the reference in the ratio plot and is defined as in $\Sec$ \ref{Sec:subTerm}. The histograms are not normalised to unity.}
\label{Fig:subFit}
\end{figure}

Using an approximation of the $\pT$ profile instead of the exact expression does mean that the matching between 1v1,pt events and the subtraction term is to some extent degraded. In $\Fig$ \ref{Fig:subFit}, the subtraction term corresponding to the fitted profile given in \Eq (\ref{eq:fittedProfile}) is compared to the 1v1,pt DPS term, as defined in $\Sec$ \ref{Sec:subTerm}. As a reminder, the transverse momenta $\kTv$ of the merging partons in a 1v1,pt event are selected according to $g(\kTv,y)$, which is here the ``decreasing Gaussian'' given by \Eq (\ref{eq:decreGausskT}). In this figure, it can be observed that the two terms start to disagree at large $\pT$ values. This is in contrast with the case where $g(\kTv,y)$ is a bare Gaussian, where the $\pT$ profile of the subtraction term can be analytically calculated. Indeed, it was noticed in $\Fig$ \ref{Fig:subFrac} that the 1v1,pt and subtraction terms overlap perfectly in this instance. 

The mismatch at large $\Z$ $\pT$ leads to an imperfect subtraction between the DPS and subtraction terms at small $y$ and large $\pT$. However, one notes that when all contributions are combined, the use of a fitted $\pT$ profile instead of an analytical result does not have a strong impact on the kinematic distributions, including the $\Z$ $\pT$ -- see $\Fig$ \ref{Fig:betaNoCut}, where the fitted-profile result agrees well with the Gaussian-ansatz results, even at large $\pT$. This is because the subtraction term for the decreasing Gaussian ansatz falls more steeply than the SPS term, such that it is much smaller than SPS at large $\pT$ -- see $\Fig$ \ref{Fig:subBeta}. Since the large-$\pT$ region is dominated by contributions from the  small-$y$ region, the DPS term should also be much smaller than the SPS term at large $\pT$. The mis-cancellation seen in $\Fig$ \ref{Fig:subFit} is then numerically unimportant in the combination.

Both the Gaussian ansatz (with adjustable $\beta$) and the decreasing Gaussian ansatz (using the fitted profile of \Eq \eqref{eq:fittedProfile} in the subtraction term) are available as options in the code.

\subsection{Distinguishing DPS from SPS}

As previously mentioned, we do not aim here at a full phenomenological analysis of DPS in the $\Z\Z$ production process. However, even in the context of our toy set-up where we only have the loop-induced process in the SPS piece, and this is multiplied by $1/10$, it is interesting to investigate in what kinematic regions we can observe the largest impact from the DPS process.

We recall from $\Sec$ \ref{Sec:DGS} that the DPS cross section is generically not well-defined on its own, since it depends on the unphysical parameter $\nu$, and that the well-defined combination is the total cross section SPS+DPS-sub. How can we then define a separation of SPS and DPS? Note that, from a theoretical point of view, the SPS cross section for \mbox{pp $\to\Z\Z$} is perfectly defined on its own. Therefore, we can compare the signal produced by the SPS process on its own to the one obtained when combining SPS and DPS. Any discrepancy between the two we attribute to DPS. In this way we effectively define the quantity ``DPS-sub'' to be the DPS contribution, putting the large-$y$ parts of 1v1 loops that are not already described by the SPS term into the DPS contribution.

\begin{figure}[t!] 
\centering
\includegraphics[width=0.45\textwidth]{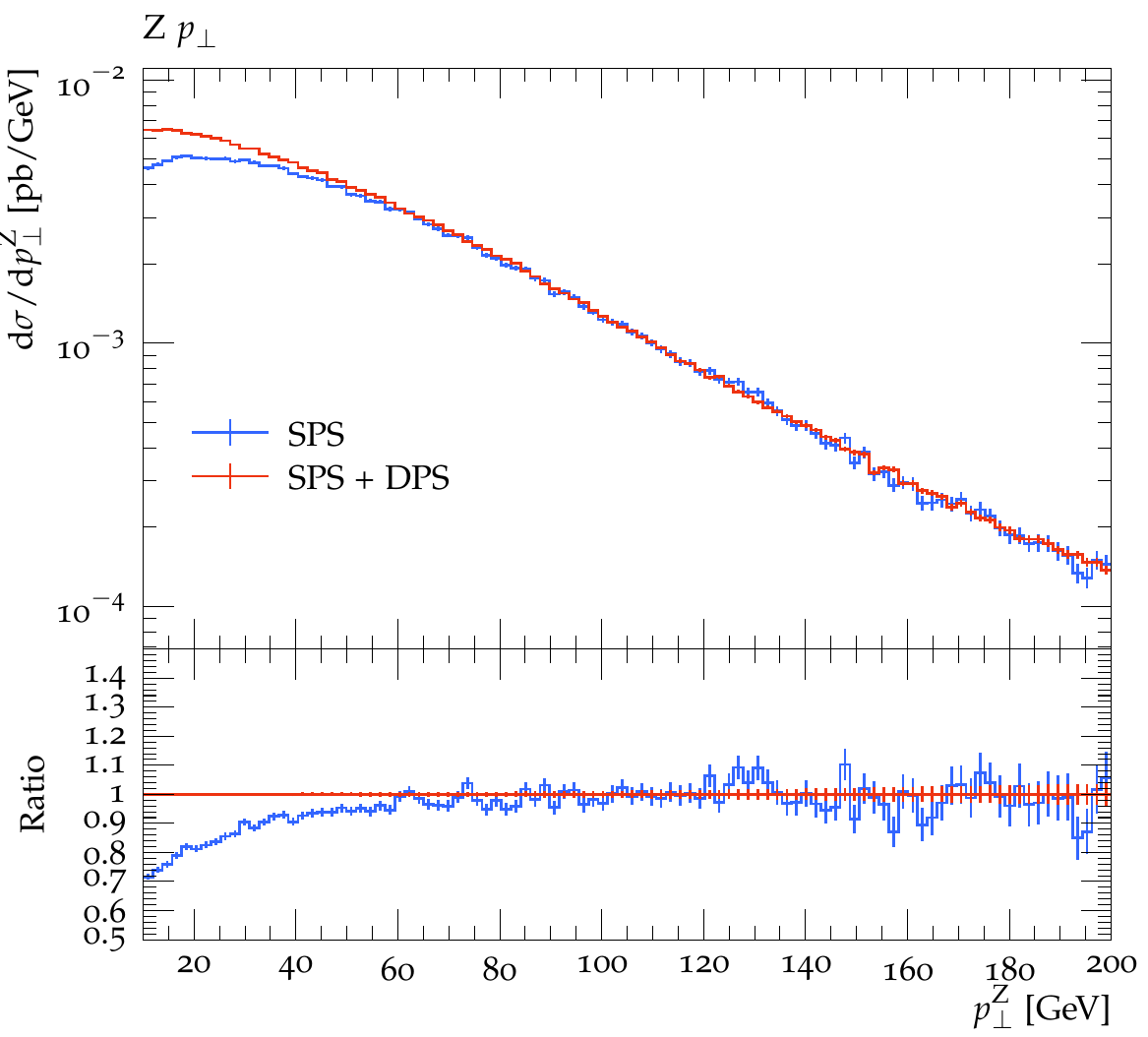}
\includegraphics[width=0.45\textwidth]{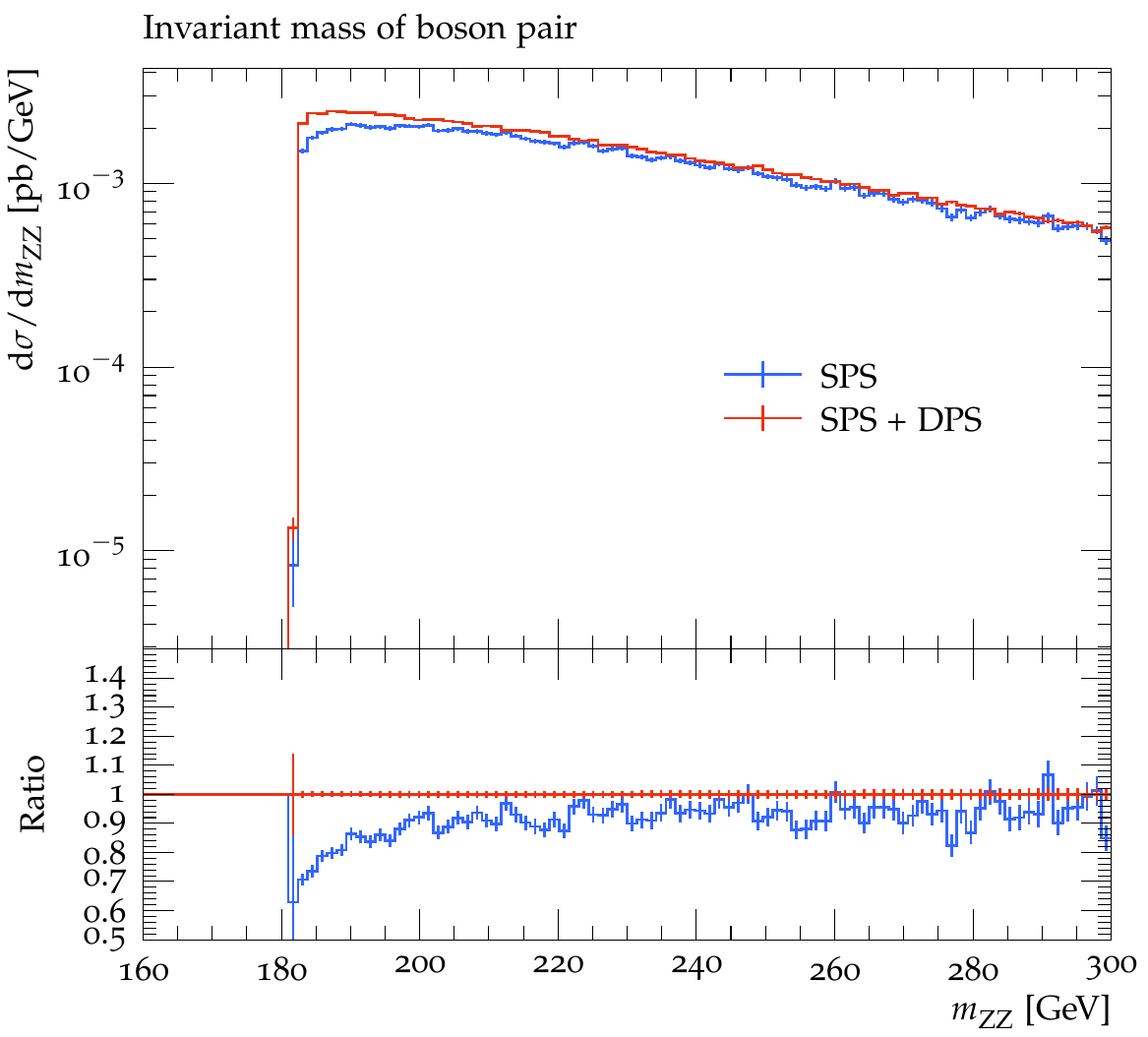} \\
(a) \hspace{180pt} (b)
\caption{(a) Transverse momenta of the $\Z$ bosons and (b) invariant mass of the $\Z\Z$ pair for the production via SPS only and via SPS combined with DPS. The SPS+DPS setup is the reference in the ratio plots. The histograms are not normalised to unity.}
\label{Fig:comp1}
\end{figure}

\begin{figure}[t!] 
\centering
\includegraphics[width=0.45\textwidth]{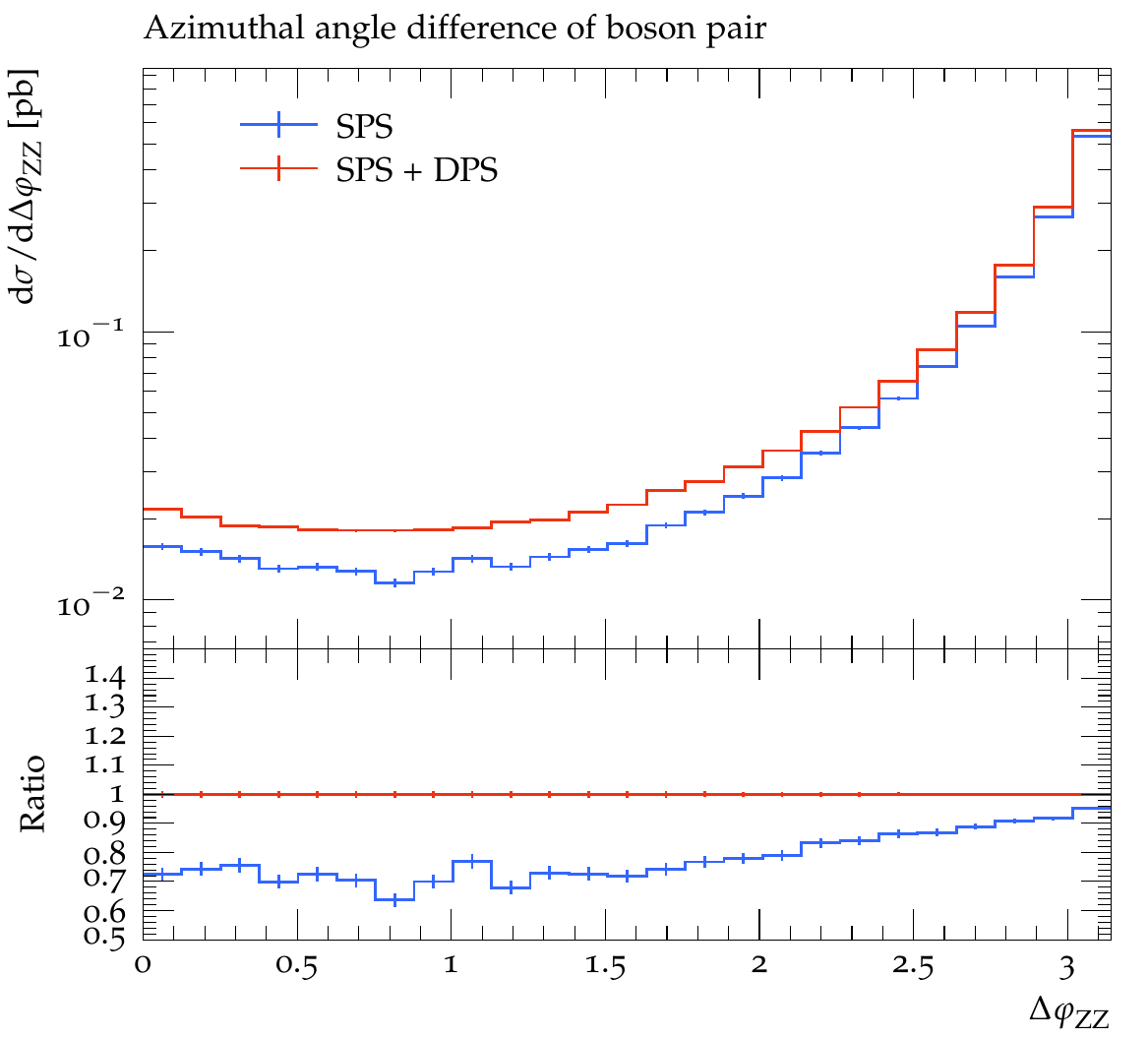}
\includegraphics[width=0.45\textwidth]{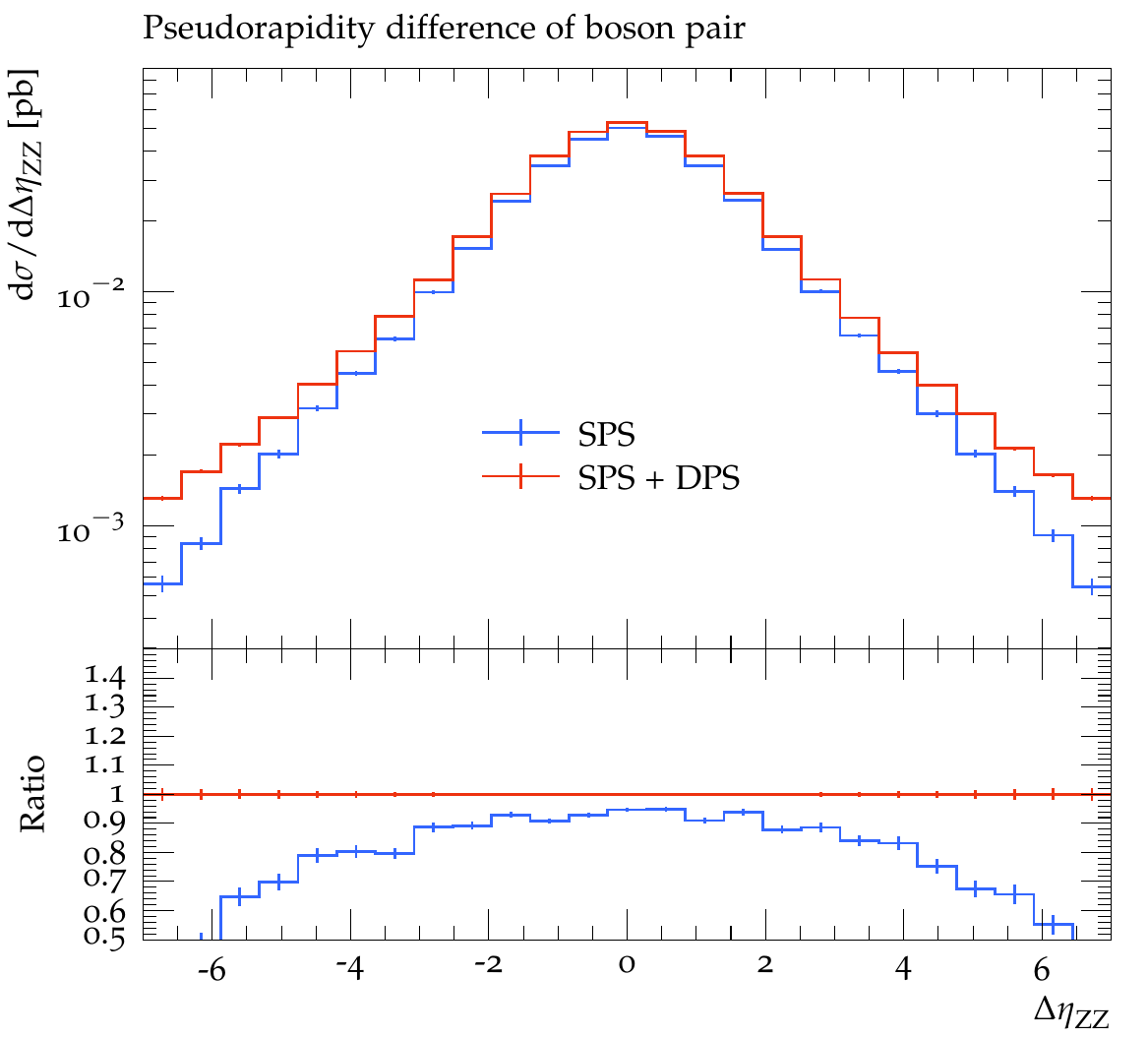} \\
(a) \hspace{180pt} (b)
\caption{(a) Difference of the azimuthal angles of the two $\Z$ bosons and (b) difference of the pseudorapidities of the two $\Z$ bosons for the production via SPS only and via SPS combined with DPS. The SPS+DPS setup is the reference in the ratio plots. The histograms are not normalised to unity.}
\label{Fig:comp2}
\end{figure}

\begin{figure}[t!] 
\centering
\includegraphics[width=0.45\textwidth]{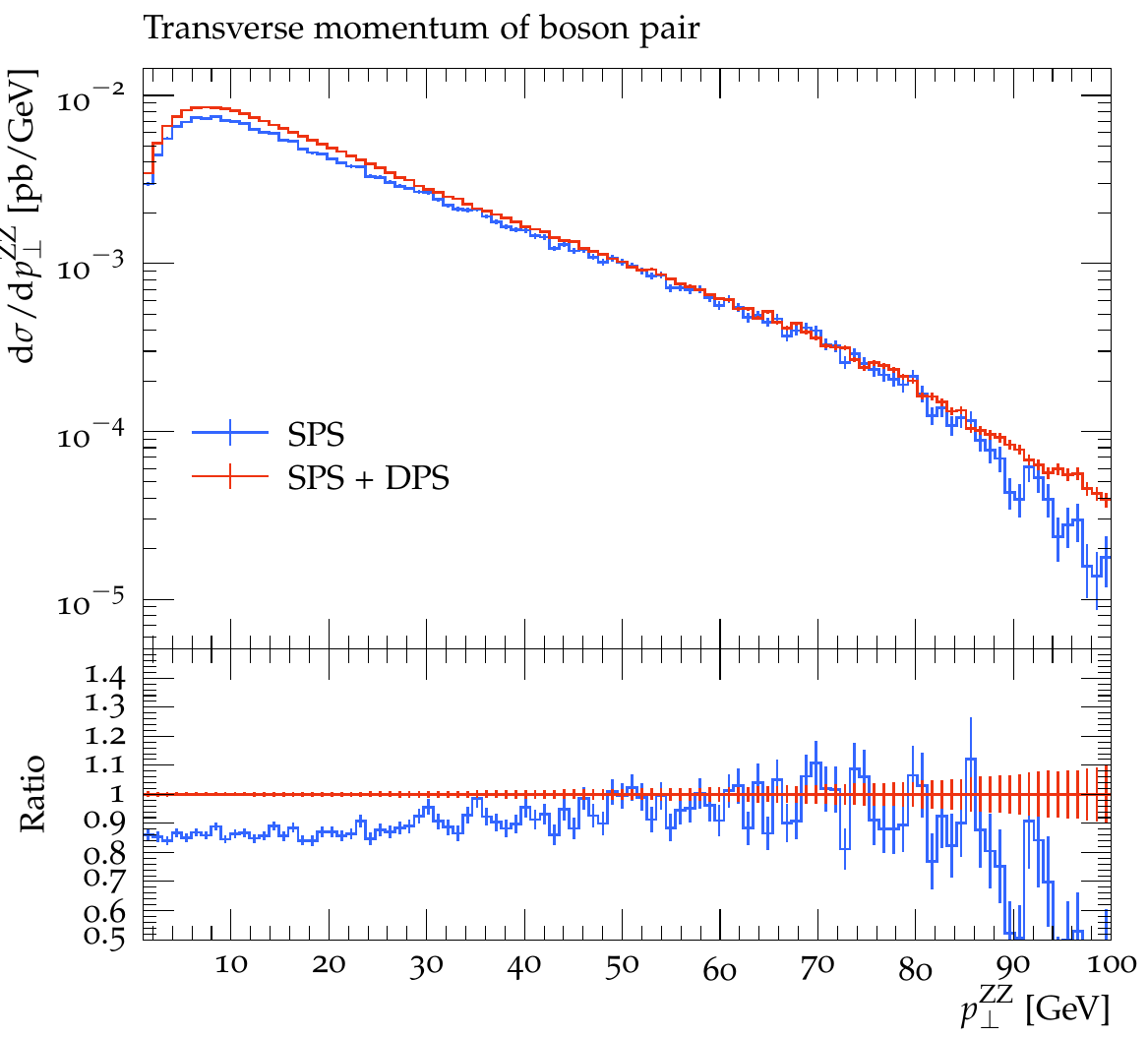}
\includegraphics[width=0.45\textwidth]{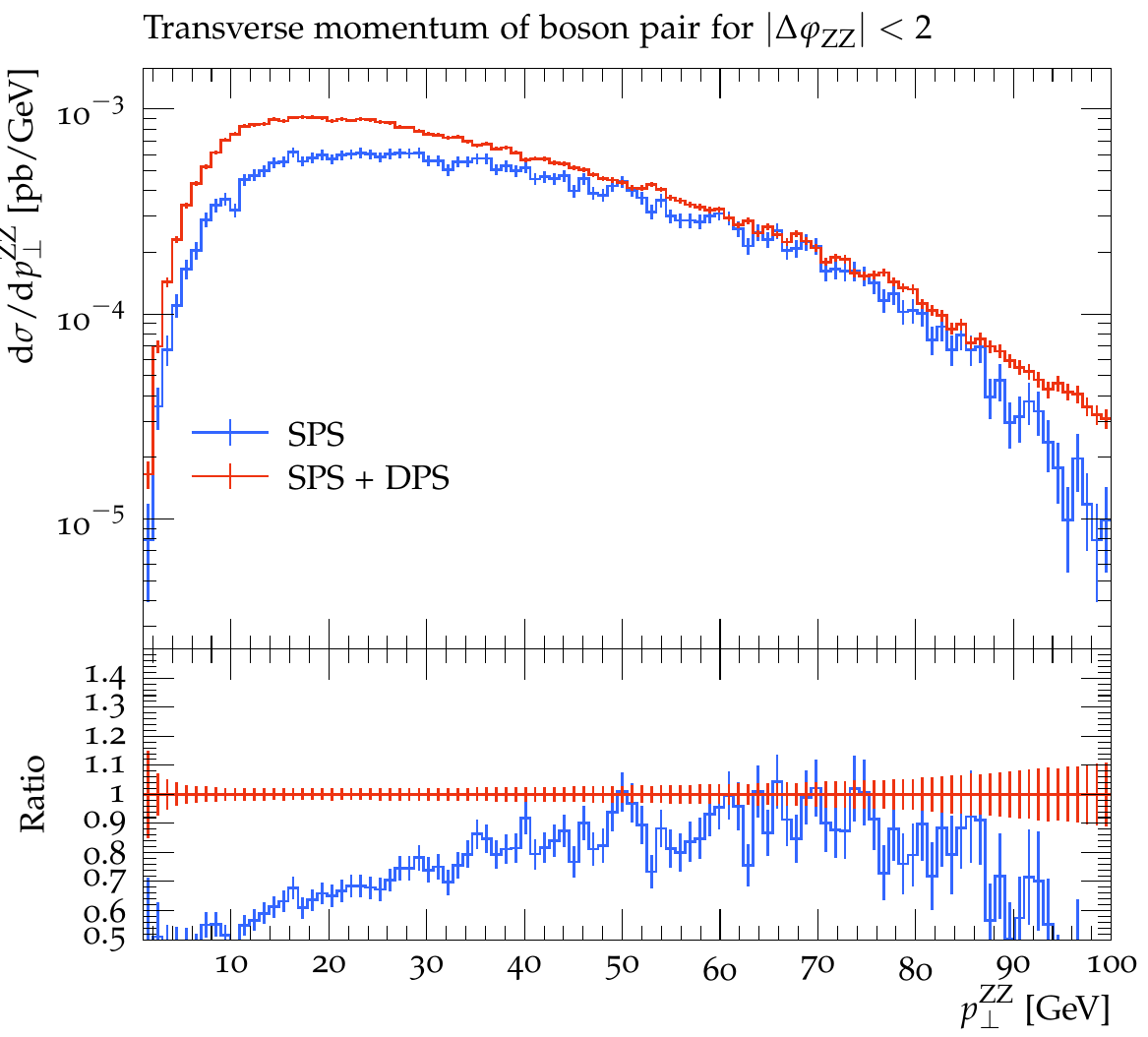} \\
(a) \hspace{180pt} (b)
\caption{(a) Transverse momentum of the $\Z\Z$ pair and (b) transverse momentum of the pair with a cut on the azimuthal difference for the production via SPS only and via SPS combined with DPS. The SPS+DPS setup is the reference in the ratio plots. The histograms are not normalised to unity.}
\label{Fig:comp3}
\end{figure}

In $\Figs$ \ref{Fig:comp1}, \ref{Fig:comp2} and \ref{Fig:comp3}, some event shapes are given. The setups of the simulations are the same as before. More precisely, the label ``SPS+DPS'' refers to the results obtained using \Eq (\ref{eq:master}) for $\nu=\mZ$ and $\beta=2$ i.e. by combining SPS and DPS. The ``SPS'' curves were again produced with the loop-induced process only, with the cross section multiplied by a factor $1/10$. The comparison shows that the inclusion of DPS leads to more events in the regions of small transverse momenta and small invariant masses. It is natural that DPS should be concentrated in this region since, at LO, the bosons are produced with zero transverse momenta in the DPS process. Combining the DPS process with the SPS one should then add to the SPS cross section a contribution that is peaked at zero transverse momentum and at an invariant mass of $2\mZ$, recall \Eq (\ref{eq:mZZ}). This leads us to propose an upper cut on either the transverse momenta of the bosons (or of the pair) or the invariant mass of the pair as a useful cut to distinguish DPS from SPS. Moreover, the results presented in $\Fig$ \ref{Fig:comp2} seem to advocate an upper cut on the difference in azimuthal angles $\Delta\varphi_\mathrm{ZZ}$ and a lower cut on the absolute value of the difference in pseudorapidities $\Delta\eta_\mathrm{ZZ}$ of the bosons as discriminating cuts. For instance, in $\Fig$ \ref{Fig:comp3}b, the $\pT$ spectrum of the pair was produced for both setups by only accepting the events that satisfy $\Delta\varphi_\mathrm{ZZ}<2$. This seems to enhance the discrepancy between the two setups, especially in the region of small transverse momenta which is the region where the DPS contribution is expected to be important.

Removing the factor of $1/10$ in the SPS piece will reduce the differences that can be observed between the SPS and SPS+DPS curves. Including the other contributions to the SPS process may affect the event shapes observed for $\Z\Z$ production, which may lead to  different discriminating cuts being appropriate. However, this is probably not the case since our reasoning uses rather general distinguishing characteristics of the DPS signal. Moreover, the proposed cuts are used in many phenomenological and experimental analyses to distinguish the DPS signal from the background SPS signal. For instance, similar cuts were already proposed in the context of a phenomenological study of $\Z\Z$ production in \cite{Kom:2011nu} and for the CMS extraction of DPS in same-sign WW production, where discriminating variables of the kind we discussed were used to train boosted decision trees \cite{Sirunyan:2019zox}. For an extensive review of experimental extractions of DPS, where in many places such variables are used to discriminate DPS and SPS, see Chapters 6-8 of \cite{Bartalini:2017jkk}.

\section{Summary}
\label{Sec:summary}

In this work, the Monte-Carlo simulation of DPS $\dShower$ introduced in \cite{Cabouat:2019gtm} has been augmented such that SPS and DPS processes can be combined in a consistent manner for the first time. This is a non-trivial task; simply adding up SPS and DPS leads to a double-counting issue both at the inclusive and differential levels. At the inclusive level, the problem of combining DPS and SPS without double counting was solved in \cite{Diehl:2017kgu}, via the inclusion of a subtraction term.
The objective of this work was to extend this subtraction scheme to the differential level in such a way that it can be implemented within a probabilistic  parton-shower algorithm. 

This required several steps. First of all, the kinematics of the $1\to2$ splittings was modified such that a relative transverse momentum $\kT\sim 1/y$ was generated between the daughter partons (with $y$ the partonic transverse separation). In the original $\dShower$ algorithm \cite{Cabouat:2019gtm}, the daughter partons were produced with zero relative $\kT$. This new kinematics is more realistic, and ensures that the kinematics of ``1v1,pt'' DPS events (in which $1\to2$ splittings occur in both protons and there are no QCD emissions above the characteristic scale of the $1\to2$ splittings) mimic more closely at large $y$ the kinematics of an SPS event, whose topology at such $y$ values is equivalent to the 1v1,pt one (see $\Fig$ \ref{Fig:ZZprod}). 

Then, a subtraction term was introduced, whose kinematics was chosen to be the one generated by the shower algorithm for a 1v1,pt DPS event. With such a choice (and with the modification to the DPS algorithm just described), the kinematics of the subtraction term matches the DPS one at small $y$ by definition, and approximately matches the SPS one at large $y$, thus extending the subtraction scheme at the differential level. Finally, each term was combined with a shower algorithm, such that event shapes corresponding to the production of a given final state via both SPS and DPS could be simulated without double counting. The overall design of the subtraction scheme in the shower is to a certain extent similar to techniques used in the matching of NLO computations to the parton shower \cite{Bengtsson:1986hr, Seymour:1994we, Seymour:1994df, Miu:1998ju, Lonnblad:1995ex, Frixione:2007vw, Nason:2004rx, Catani:2001cc, Lonnblad:2001iq, Mrenna:2003if, Frixione:2002ik}.

This subtraction scheme was implemented in the new version of the $\dShower$ simulation, thus allowing the combination of SPS and DPS. The implementation was numerically validated at parton level in the context of $\Z\Z$ production. In our proof-of-concept study, the SPS term was the loop-induced process initiated by a pair of gluons since it is the only contribution that overlaps with the DPS process and that has a large-$y$ tail. This SPS term was divided by 10, to boost the visibility of the DPS contribution and reduce the required statistics. We studied the dependence of the algorithm on the quantity $\nu$, an unphysical parameter that effectively demarcates SPS and DPS.  Once the subtraction term is included, the results show a rather small dependence of the cross section and event shapes on this scale, as should be the case. We also investigated several different sensible choices for the $\kT$ profile $g(\kTv,y)$ in the $1\to2$ splitting process and subtraction term, including an ``optimal'' choice for which the behaviour of the subtraction term matches that of the SPS loop-induced term at small $\pT$. For many distributions, almost no difference was observed between the different choices, with a small difference being observed in the region of phase space where the transverse momenta of both bosons are small. The implementation of this subtraction scheme generates some counter-events that contribute to the histograms with a negative weight (as is also encountered in NLO+shower matching schemes such as \textsc{MC@NLO} \cite{Frixione:2002ik, Frixione:2010wd, Frederix:2012ps, Frederix:2020trv}). However, it was shown that it is possible to limit the fraction of events with negative weights to a few percent if one couples the SPS cross section and the subtraction term to the exact same shower algorithm.

Using the toy set-up described above, we also studied in what kinematic regions the inclusion of DPS has an observable impact. Our results indicate that upper cuts on $\pT^\mathrm{Z}$, $\pT^\mathrm{ZZ}$, $m_\mathrm{ZZ}$ and $\Delta\varphi_\mathrm{ZZ}$ as well as a lower cut on $|\Delta\eta_\mathrm{ZZ}|$ will lead to an enhanced DPS contribution. This is consistent with previous experimental and phenomenological studies of DPS.

In the future, it would be interesting to use this algorithm to make a proper phenomenological analysis of $\Z\Z$ production and other processes of interest such as $\W^+\W^-$ production. For such studies it would be desirable to include at least the Born SPS process in addition to the loop-induced one, massive quark flavours, decays of the bosons, and hadronisation of the low-scale partons. It would also be interesting to study the effects of different sets of sPDFs and dPDFs in the simulation, or to adapt the algorithm such that it can handle unequal-scale dPDFs. The new PDF interpolation library \textsc{ChiliPDF} \cite{Nagar:2019njl} could help to achieve such goals. The first aspect would help to assess the uncertainties related to the PDFs, whereas the second one would be relevant for DPS processes that involve hard scatters characterised by two different scales such as four-jet or W$+ 2$ jet production.

\section*{Acknowledgements}

BC would like to thank Simon-Luca Villani for helpful discussions regarding loop-induced processes. This work has received funding from the European Union's Horizon 2020 research and innovation programme as part of the Marie Sk\l{}odowska-Curie Innovative Training Network MCnetITN3 (grant agreement no. 722104). The histograms were produced with \textsc{Rivet} \cite{Buckley:2010ar, Bierlich:2019rhm} and the sketches with Axodraw \cite{Collins:2016aya}.

\clearpage

\bibliographystyle{utphys}
\bibliography{Paper_ZZ}

\end{document}